\documentclass[a4paper,11pt]{article}
\pdfoutput=1 % if your are submitting a pdflatex (i.e. if you have
\usepackage{jheppub} % for details on the use of the package, please
\usepackage[T1]{fontenc}
\usepackage[italian,english]{babel}
\usepackage{ifpdf}
\usepackage{subfigure}

\usepackage{amssymb}
\usepackage{amsfonts}
\usepackage{epsf}
\usepackage{rotating}
\usepackage{graphicx}
\usepackage{amsmath}
\usepackage{fancyhdr}
\usepackage{lineno}
\usepackage{babel}
\usepackage{graphics}
\usepackage{pstricks}
\usepackage{color}
\usepackage{multirow}
\usepackage{nicefrac}
\usepackage{bm}
\usepackage{slashed}
\usepackage{float}
\usepackage{array}
\usepackage{boldline}

\usepackage{mathrsfs}
\usepackage{stackengine}
\usepackage{xcolor}

\usepackage{hyperref}
\hypersetup{colorlinks=true,linkcolor=blue,citecolor=red,urlcolor=lightseagreen}

\definecolor{lightseagreen}{rgb}{0.13, 0.7, 0.67}

\usepackage[numbers,sort&compress]{natbib}
\usepackage{tikz}
\usepackage{tikz-feynman}
\tikzfeynmanset{compat=1.0.0}

\DeclareFontFamily{OT1}{pzc}{}
\DeclareFontShape{OT1}{pzc}{m}{it}{<-> s * [1.2] pzcmi7t}{}
\DeclareMathAlphabet{\mathpzc}{OT1}{pzc}{m}{it}

\def\fbi{$\text{fb}^{-1}$ }

\usepackage{braket}
\usepackage{mathrsfs}
\usepackage{multicol,caption}
\usepackage{chngcntr}
\usepackage{multirow}
\usepackage{balance}

\def\tev{\ensuremath{\mathrm{\,Te\kern -0.1em V}}}
\def\gev{\ensuremath{\mathrm{\,Ge\kern -0.1em V}}}
\def\mev{\ensuremath{\mathrm{\,Me\kern -0.1em V}}}

\newcommand{\gsim}{\raisebox{-0.13cm}{~\shortstack{$>$ \\[-0.07cm]
			$\sim$}}~}

\newcommand*\xbar[1]{%
	\hbox{%
		\vbox{%
			\hrule height 0.65pt % The actual bar
			\kern0.4ex%         % Distance between bar and symbol
			\hbox{%
				\kern-0.05em%      % Shortening on the left side
				\ensuremath{#1}%
				\kern0.0em%      % Shortening on the right side
			}%
		}%
	}%
}

\DeclareFontFamily{OT1}{pzc}{}
\DeclareFontShape{OT1}{pzc}{m}{it}{<-> s * [1.2] pzcmi7t}{}
\DeclareMathAlphabet{\mathpzc}{OT1}{pzc}{m}{it}
\newcommand{\Q}{\bm{Q}}
\newcommand{\LL}{\bm{L}}
\newcommand{\cl}{\mathpzc{l}}
\newcommand{\uq}{\mathpzc{u}}
\newcommand{\dq}{\mathpzc{d}}
\newcommand{\be}{\begin{equation}}
\newcommand{\ee}{\end{equation}}
\newcommand{\bea}{\begin{eqnarray}}
\newcommand{\eea}{\end{eqnarray}}
\newcommand{\LQ}{Leptoquark\, }
\newcommand{\LQs}{Leptoquarks\, }
\allowdisplaybreaks

\newcommand{\bh}{\hat\beta}
\newcommand{\cts}{\cos^2\theta}
\newcommand{\Mp}{M_{\phi_v}}
\begin{document}

%\begin{frontmatter}

%%%%%%%%%%%%%%%%%%%%%%%%%%%%%%%%%%%%%%%%%%%%%
\title{Distinguishing Leptoquarks at the LHC/FCC}

\author[a]{Priyotosh Bandyopadhyay,}
\author[a]{Saunak Dutta,}
\author[b]{Mahesh Jakkapu,}
\author[a]{Anirban Karan.}
\affiliation[a]{Indian Institute of Technology Hyderabad, Kandi,  Sangareddy-502285, Telengana, India}

\affiliation[b]{Graduate University for Advanced Studies(SOKENDAI), Shonan Village, Hayama, Kanagawa 240-0193 Japan}

\emailAdd{bpriyo@phy.iith.ac.in}
\emailAdd{ph17resch11002@iith.ac.in}
\emailAdd{maheshj@post.kek.jp}
\emailAdd{kanirban@iith.ac.in}

\preprint{ IITH-PH-0005/20}
	
%%%%%%%%%%%%%%%%%%%%%%%%%%%%%%%%%%%%%%%%%%%%%
%%%%%%%%%%%%%%%%%%%%%%%%%%%%%%%%%%%%%%%%%%%%%
	\abstract{In this article, we deal with how to distinguish the signatures of different \LQs at the LHC/FCC if all of them lie within similar mass and coupling range and can be produced at present and future colliders. It has been found that  hard scattering cross-sections and angular distributions can be used to differentiate scalar and vector Leptoquarks. On the other hand, final state topology and determination of jet charge can separate \LQs with same spin even from same $SU(2)_L$ multiplet. We performed a PYTHIA8 based analysis considering all the dominant Standard Model (SM) backgrounds at the LHC/FCC with centre of mass energies of 14, 27 and 100 TeV for scalar ($S_1$) and vector ($\widetilde{U}_{1\mu}$) Leptoquarks. We see that confirming evidence of scalar Leptoquark at 14 TeV requires 1000 fb$^{-1}$ of integrated luminosity, whereas the vector Leptoquark can be probed with very early data. But, at 100 TeV with 1000 fb$^{-1}$ of integrated luminosity, scalar Leptoquark of mass 3.5 TeV and vector Leptoquark of mass more than 5 TeV can be probed easily. }

\maketitle
\flushbottom
%%%%%%%%%%%%%%%%%%%%%%%%%%%%%%%%%%%%%%%%%%%%%%%%
%%%%%%%%%%%%%%%%%%%%%%%%%%%%%%%%%%%%%%%%%%%%%%%%

\section{Introduction}

\LQs are  special kind of beyond Standard Model (BSM) particles carrying both non-zero lepton and baryon numbers \cite{Hewett:1997ce,pdg,Dorsner:2016wpm}. Therefore, they can interact with quarks and leptons simultaneously.  They are colour triplet (fundamental or anti-fundamental) as well as electromagnetically charged. However, under $SU(2)_L$ gauge representation, they could be singlet, doublet or triplet. Moreover, according to Lorentz transformation, they might be scalar (spin 0) and vector (spin 1) as well. The notion of \LQ has been there in literature for more than forty years.
They appear naturally in various BSM scenarios involving higher gauge representations that unify the matter fields \cite{Pati:1973uk,Pati:1974yy,Georgi:1974my,Georgi:1974sy,Dimopoulos:1979es,Farhi:1980xs,Schrempp:1984nj,Wudka:1985ef,Nilles:1983ge,Haber:1984rc}. They are also quite useful in explaining various experimental and theoretical anomalies \cite{Crivellin:2021egp,Azizi:2021lnb,Faisel:2020php,Azuelos:2020gvi,Hati:2020cyn,Iguro:2020keo,Crivellin:2020ukd,Bordone:2020lnb,Babu:2020hun,Kumbhakar:2020okw,Dorsner:2020aaz,Crivellin:2020mjs,Dev:2020qet,Bigaran:2020jil,Altmannshofer:2020ywf,Crivellin:2019dwb,Dorsner:2019vgp,Hou:2019wiu,Mandal:2019gff,sLQ1,Dorsner:2019itg,Davidson:2010uu,Leurer:1993em,Deshpande:1994vf}.  Their signatures at different colliders have been also studied widely for several phenomenological interests  \cite{sLQ,S2,S3,LQS1,LQS4,Bhaskar:2021pml,Haisch:2020xjd,Buonocore:2020erb,Bhaskar:2020gkk,Borschensky:2020hot,Allanach:2019zfr,Alves:2018krf,Mandal:2018qpg,Padhan:2019dcp,Baker:2019sli,Blumlein:1996qp,Abreu:1998fw,Chekanov:2003af,Alitti:1991dn,Abe:1995fj,Abe:1996dn,Bandyopadhyay:2020klr,Bandyopadhyay:2020jez,Aaboud:2019jcc,Aaboud:2019bye,Sirunyan:2018vhk,Sirunyan:2018ryt,Sirunyan:2018nkj,Sirunyan:2018kzh,Sirunyan:2018btu,Buchmuller:1986zs,Hewett:1987bh,Belyaev:2005ew,Bhattacharyya:1994ig,Hewett:1987yg,Plehn:1997az,Kramer:1997hh,Cuypers:1995ax,Eboli:1993qx,Nadeau:1993zv,Atag:1994hk,Gunion:1987ge,Ilyin:1995jv,Atag:1994np,Blumlein:1994qd,Bhaskar:2020kdr}. However, no conclusive evidence for there existence has been found yet.

In this paper, we investigate how to differentiate the signatures of scalar and vector \LQs at the LHC/FCC. For this, we have to first assume that both type of \LQs exist in nature with such masses and couplings that they can be produced at present and future colliders like LHC/FCC. In a PYTHIA8 \cite{Pythia8} based analysis, we have looked into hard scattering cross-section, angular distribution, jet charge determination, transverse momenta of jet and lepton and few other aspects  for this distinction. It turns out that total cross-section and angular distribution can be used to separate scalar and vector \LQs at the LHC/FCC, whereas final state topology and determination of jet charge become instrumental in distinguishing \LQs with same spin as well as \LQs belinging to same $SU(2)_L$ multiplet. 

A detailed PYTHIA8 \cite{Pythia8} based simulation with all dominant SM backgrounds have been carried out which shows that the events number for vector \LQ could be a few times larger than the scalar one for the same choices of mass of \LQs viz. $\widetilde{U}_{1\mu}$ and $S_1$. This attributes to the fact of more spin degrees of freedom for the former one and also due to higher branching for the allowed benchmark points. Reconstruction of \LQ mass along with the angular distribution in the centre of mass(CM) frame enable us to distinguish the spin representations of such Leptoquarks.  We also comment on the possibility of different final states from a Leptoquakrs in higher gauge representation. Their decays can be explored by the study of different final state topologies as well as the construction of the jet charges coming from the Leptoquark decays. 

The paper is organised in the following way. We briefly describe all the scalar and vector \LQs in the next section (Section \ref{sec:LQs}). Section \ref{sec:BP} deals with current experimental bounds on the masses and couplings of \LQs and choice of benchmark points for simulation. The theoretical aspects for pair production of scalar and vector \LQs at proton-proton collider have been illustrated in Section \ref{sec:theory}. In Section \ref{sec:dist}, we have discussed how to identify the signatures of scalar and vector \LQs as well as the same of different excitations lying in the same $SU(2)_L$ multiplet. Finally, we conclude our study in Section \ref{sec:conc}.

\section{Scalar and vector \LQs}
\label{sec:LQs}

\begin{table}[h!]
	\renewcommand{\arraystretch}{0.98}
	\begin{tabular*}{\textwidth}{|@{\hspace*{4mm}\extracolsep{\fill}}cccccc|}
		\hline
		&&&&&\\[-3mm]
		$\phi$ & $SU(3)$ & $\mathtt Y_\phi$ & $T_3$ & $Q_{\phi}$  &  Interaction (+ h.c.) \\[1mm]
		\hline
        &&&&&\\[-3mm]
		\multicolumn{5}{|l}{\textbf{Scalar Leptoquarks} $\bm{\phi_s}$}&\\
		\multirow{2}{*}{$S_1$} & \multirow{2}{*}{$\overline 3$} &\multirow{2}{*}{$\nicefrac{2}{3}$} & \multirow{2}{*}{0} & \multirow{2}{*}{$\nicefrac{1}{3}$}   & $Y_L\,\xbar \Q_L^c\, \left(i\sigma^2 \, S_1\right)\LL_L$\\
		&&&&&$+Y_R \,\xbar \uq^c_R \,S_1\,\cl_R$\\[2mm]
		
		$\widetilde S_1$ & $\overline 3$& $\nicefrac{8}{3}$ & 0 & $\nicefrac{4}{3}$  & $ Y_R\, \xbar{\dq}^c_R\, \widetilde{S}_1\,\cl_R $ \\[2mm]
		
		\multirow{2}{*}{$R_2$} & \multirow{2}{*}{3} & \multirow{2}{*}{$\nicefrac{7}{3}$} & $\nicefrac{1}{2}$ & $\nicefrac{5}{3}$ & $ Y_L\, \xbar{\uq}_R \left(i\sigma^2 R_2\right)^T \LL_L$  \\
		&  & & $\nicefrac{-1}{2}$ &$\nicefrac{2}{3}$   &  $+ Y_R\, \xbar{\Q}_L\, R_2\, \cl_R$ \\[2mm]
		
		\multirow{2}{*}{$\widetilde{R}_2$} & \multirow{2}{*}{3} & \multirow{2}{*}{\nicefrac{1}{3}} & $\nicefrac{1}{2}$ &\nicefrac{2}{3} & \multirow{2}{*}{$Y_L\,\xbar\dq_R\left(i\sigma^2\widetilde{R}_2\right)^T\LL_L$} \\
		&  & & $\nicefrac{-1}{2}$ & $\nicefrac{-1}{3}$   &  \\[2mm]
		
		\multirow{3}{*}{$\vec{S}_3$} &\multirow{3}{*}{$\overline 3$}& \multirow{3}{*}{$\nicefrac{2}{3}$} & 1 & $\nicefrac{4}{3}$    & \multirow{3}{*}{$ Y_L\,\xbar{\Q}_L^c \left(i\sigma^2 \,S_3^{adj}\right) \LL_L $} \\
		&   & & 0 & $\nicefrac{1}{3}$  &   \\
		&  & & $-1$ &$\nicefrac{-2}{3}$   & \\[2mm]
		
		\hline
		&&&&&\\[-3mm]
		\multicolumn{5}{|l}{\textbf{Vector Leptoquarks} $\bm{\phi_v}$}&\\
		
		\multirow{2}{*}{$U_{1\mu}$} & \multirow{2}{*}{3} & \multirow{2}{*}{$\nicefrac{4}{3}$} & \multirow{2}{*}{0} & \multirow{2}{*}{$\nicefrac{2}{3}$}  & $Y_L\,\xbar\Q_L\gamma^\mu\,U_{1\mu}\LL_L$\\
		&&&&&$+Y_R\,\xbar\dq_R\gamma^\mu\,U_{1\mu}\cl_R$	\\[2mm]
		
		$\widetilde U_{1\mu}$ &3& $\nicefrac{10}{3}$ & 0 & $\nicefrac{5}{3}$ & $Y_R\,\xbar{\uq}_R\, \gamma^\mu \,\widetilde{U}_{1\mu}\,\cl_R $ \\[2mm]
		
		\multirow{2}{*}{$V_{2\mu}$} &\multirow{2}{*}{$\overline 3$}& \multirow{2}{*}{$\nicefrac{5}{3}$} & $\nicefrac{1}{2}$ & $\nicefrac{4}{3}$ & $Y_L\,\xbar{\dq}_R^c\, \gamma^\mu \left(i\sigma^2 V_{2\mu}\right)^T \LL_L $ \\
		& & & $\nicefrac{-1}{2}$ &$\nicefrac{1}{3}$  & $+ Y_R\,\xbar{\Q}_L^c \,\gamma^\mu \left(i\sigma^2 V_{2\mu}\right)\cl_R$\\[2mm]
		
		\multirow{2}{*}{$\widetilde V_{2\mu}$} &\multirow{3}{*}{$\overline 3$} & \multirow{2}{*}{$\nicefrac{-1}{3}$} & $\nicefrac{1}{2}$ & $\nicefrac{1}{3}$  & \multirow{2}{*}{$Y_L\,\xbar\uq_R^c\,\gamma^\mu\left(i\sigma^2\widetilde{V}_{2\mu}\right)^T\LL_L$}\\
		&  & &$\nicefrac{-1}{2}$ &$\nicefrac{-2}{3}$ & \\[2mm]
		
		\multirow{3}{*}{$\vec{U}_{3\mu}$}& \multirow{3}{*}{$ 3$}& \multirow{3}{*}{$\nicefrac{4}{3}$} & 1 & $\nicefrac{5}{3}$ & \multirow{3}{*}{$Y_L\,\xbar{\Q}_L \,\gamma^\mu\, U_{3\mu}^{adj}\, \LL_L $}  \\
		&  & &0 & $\nicefrac{2}{3}$ & \\
		&  & &$-1$ &$\nicefrac{-1}{3}$  & \\[1mm]
		\hline
	\end{tabular*}
	\caption{Specification of scalar and vector Leptoquarks. The $S_3^{adj}$ and $U_{3\mu}^{adj}$ are the scalar and vector triplet \LQs in adjoint representation.}
	\label{tab:LQ}
\end{table}

 In this section, we discuss different scalar and vector Leptoquarks, their nomenclature and quantum numbers and interaction terms in brief. We have generically denoted the \LQs  as $\phi_{s,v}$ indicating the scalar and vector types respectively. In Table \ref{tab:LQ}, we summarize all the scalar and vector Leptoquarks.
 We follow the similar notations of Refs.\cite{Hewett:1997ce,Dorsner:2016wpm,Buchmuller:1986zs,Hewett:1987yg,Belyaev:2005ew,Davidson:1993qk} for the nomenclature of the Leptoquarks. The vector ones are indicated by the Lorentz index $\mu$ in the subscript of their names. Additionally, the subscripts 1, 2 and 3 in the names of \LQs signify singlet, doublet and triplet \LQs under $SU(2)_L$ gauge group. Thus we have five scalar ($S_1\,,\widetilde{S}_1\,,R_2\,,\widetilde{
	R}_2$ and $\vec S_3$) and five vector ($U_{1\mu}\,,\widetilde{U}_{1\mu}\,,V_{2\mu}\,,\widetilde{V}_{2\mu}$ and $\vec U_{3\mu}$) \LQs and in each set there are two singlets, two doublets and one triplet. Different quantum numbers like $SU(3)$ behaviour, weak hypercharge $(\mathtt Y_\phi)$, the third component of weak isospin $(T_3)$, electromagnetic charge $(Q_{\phi})$ as well as their interactions with quarks and leptons are also mentioned in Table \ref{tab:LQ}. Here, $\Q_L$ and $\LL_L$ are $SU(2)_L$ doublets for quarks and leptons given by $\Q_L = (\uq_L, \dq_L)^T$ and $\LL_L = (\nu_L, \cl_L)^T$ respectively, whereas $\uq_R$, $\dq_R$ and $\cl_R$ represent all the three generations of right-handed $SU(2)_L$ singlets for up type quark, down type quark and charged lepton, respectively   (the generation and colour indices are suppressed). The superscript ``c'' in the interaction terms indicates charge conjugate of a field. It is interesting to notice that for scalar Leptoquarks, only the doublets ($R_2$ and $\widetilde{R}_2$) are in fundamental representation of $SU(3)$ and the rest are in anti-fundamental representation whereas the scenario becomes reverse for vector Leptoquarks.

\section{Experimental bounds and benchmark points}
\label{sec:BP}

\begin{figure}

	\ContinuedFloat*
	\includegraphics[height=0.23\textheight,width=0.48\textwidth]{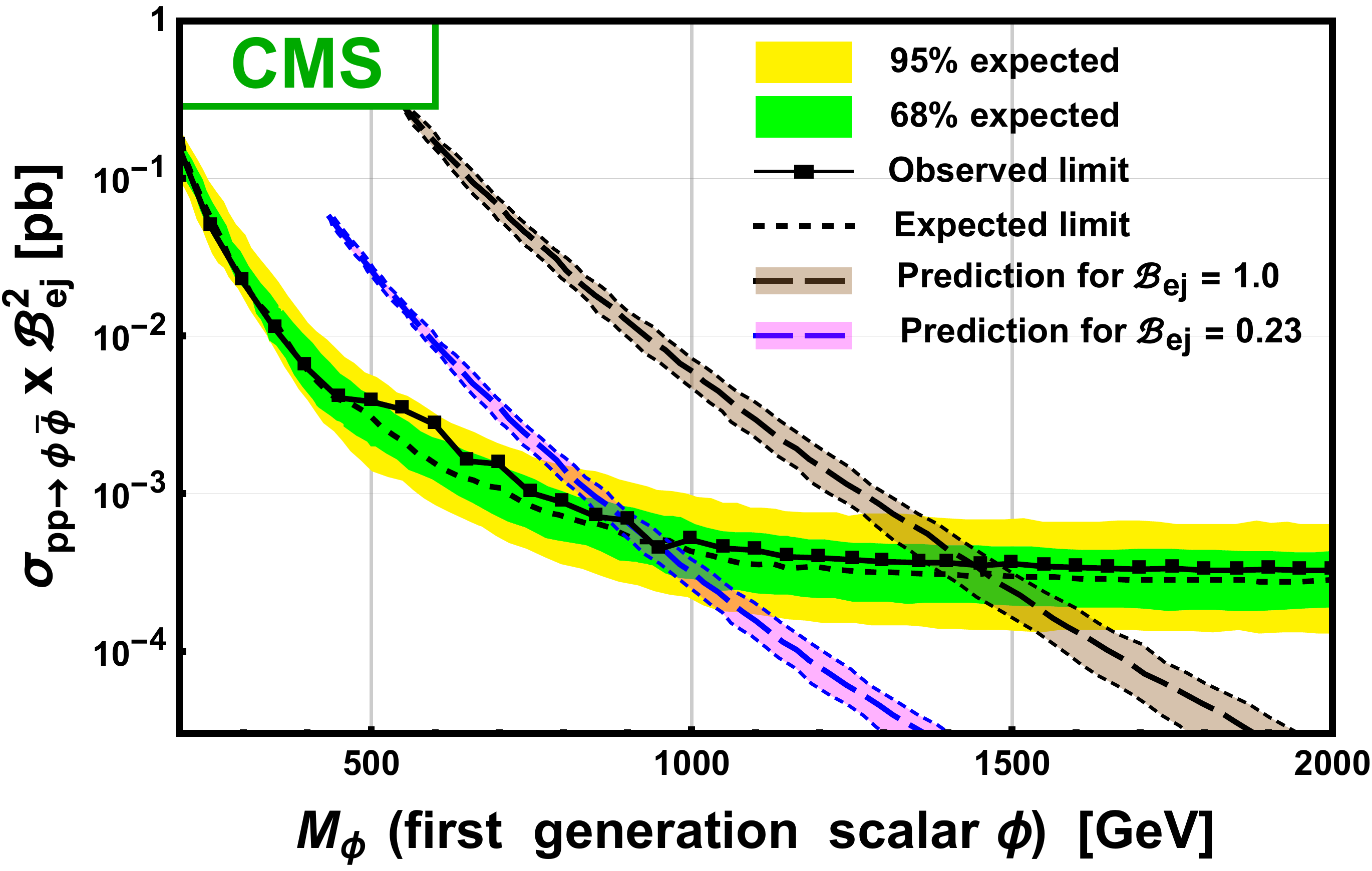}
	\hfil
	\includegraphics[height=0.23\textheight,width=0.48\textwidth]{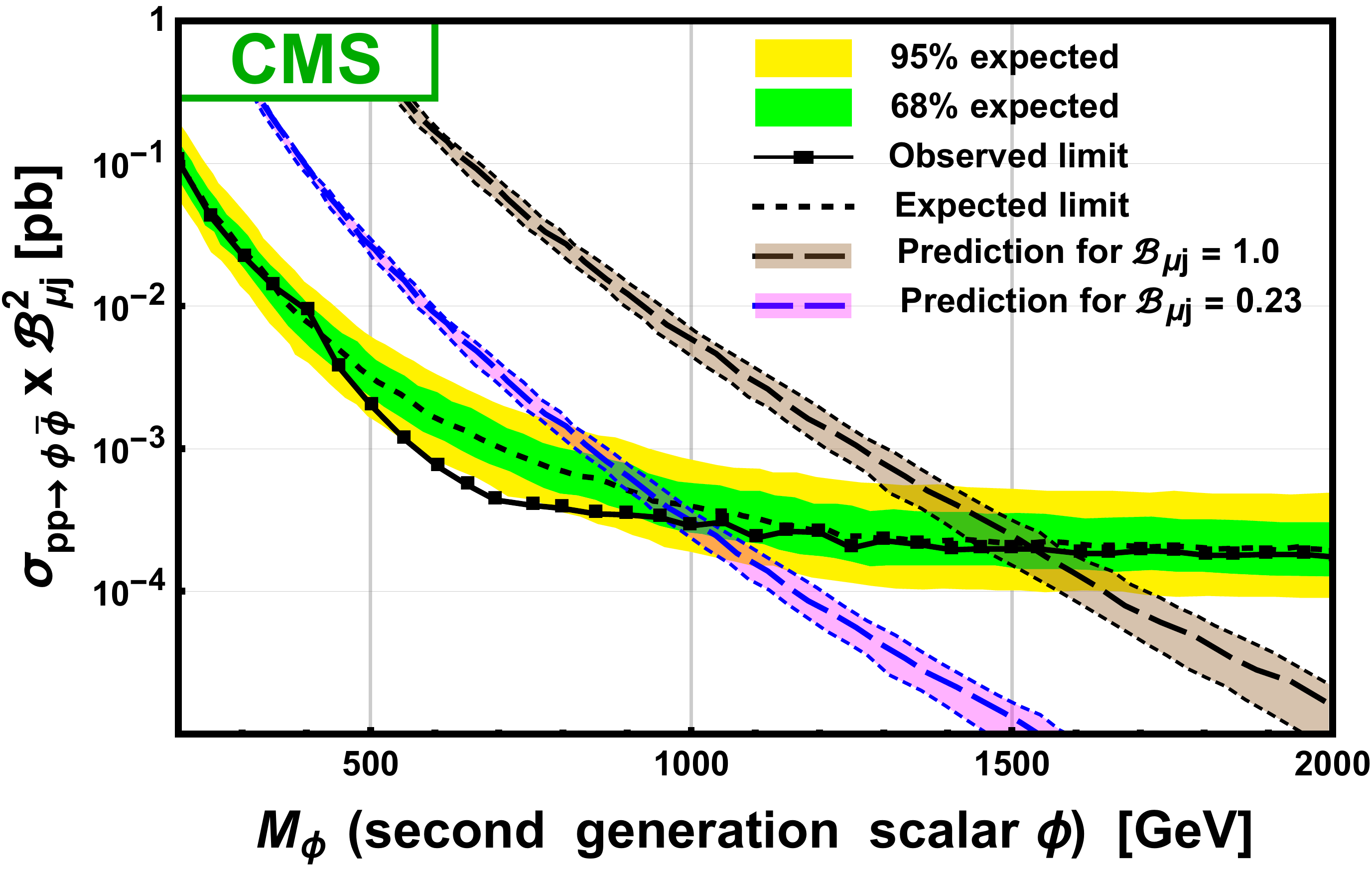}\\[2mm]
	\hspace*{0.2cm}\includegraphics[height=0.23\textheight,width=0.47\textwidth]{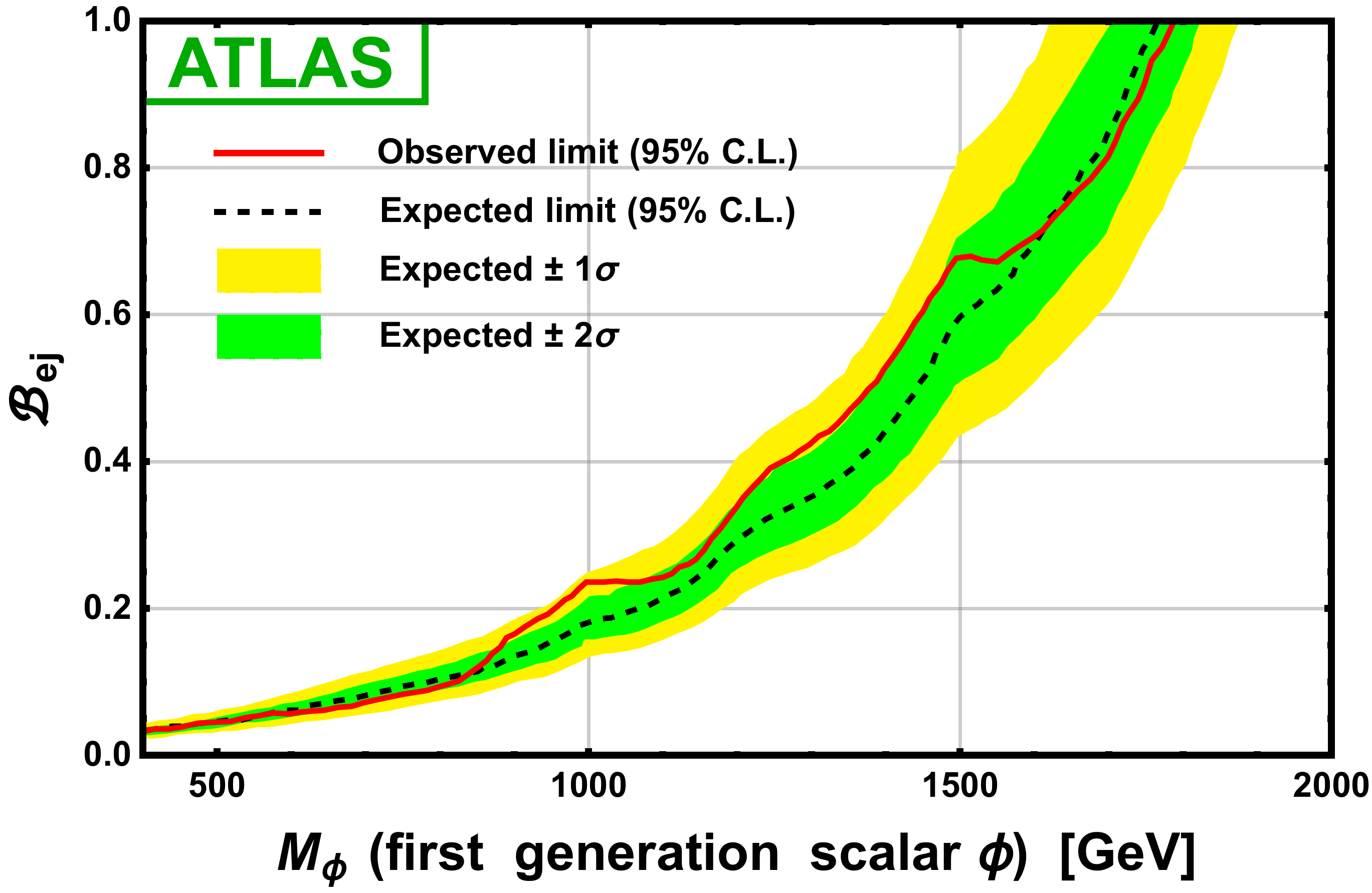}
	\hfil
	\includegraphics[height=0.23\textheight,width=0.47\textwidth]{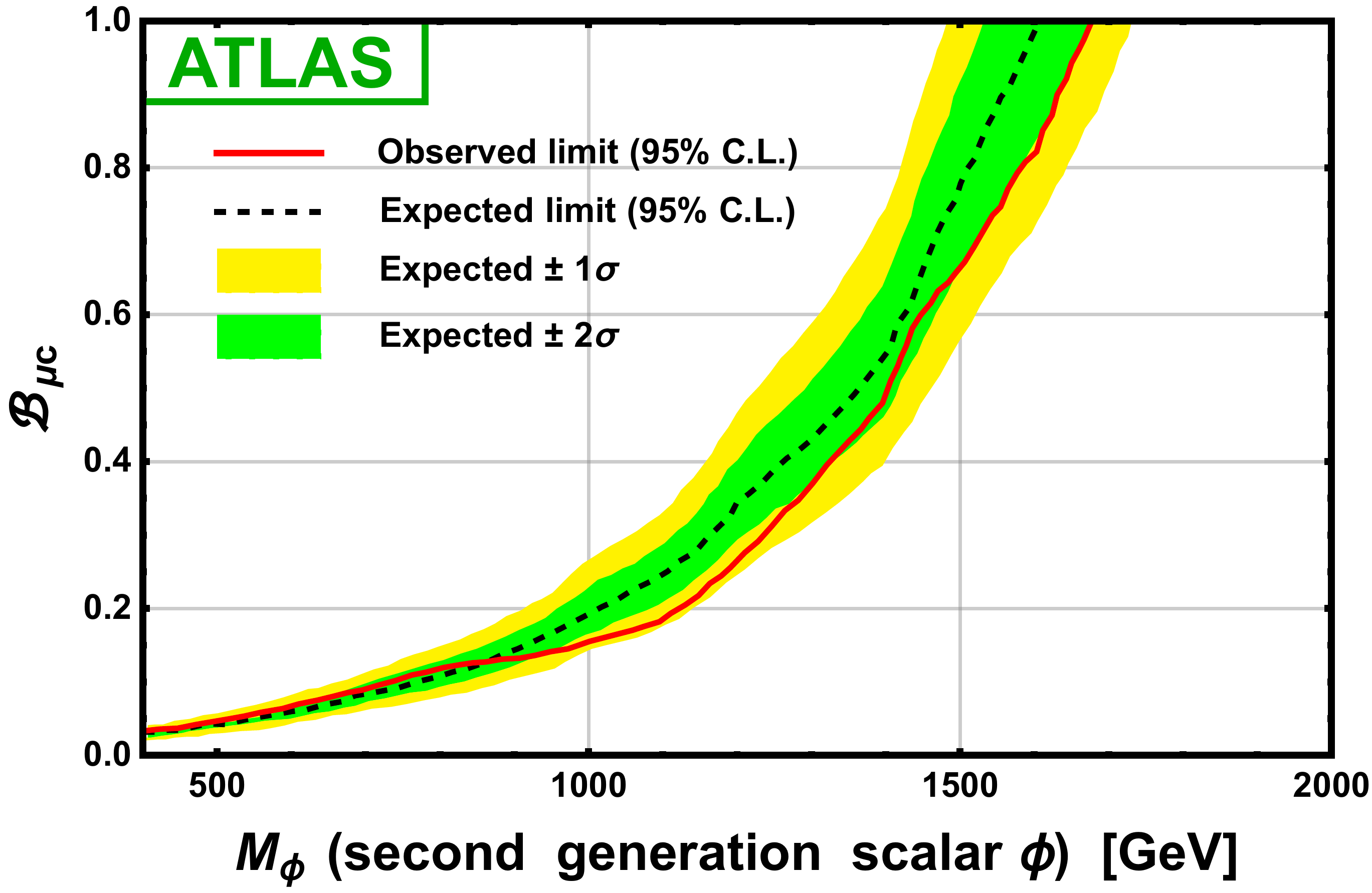}
	
		\caption{LHC bounds on first and second generations of scalar Leptoquarks. First and second plots show the constraints from CMS on $\sigma_{pp\to\phi\bar\phi}\times \mathcal B^2$ for first and second generations of scalar Leptoquarks decaying to a jet and charged lepton \cite{Sirunyan:2018btu,Sirunyan:2018ryt}. Third and fourth plots illustrate the bounds from ATLAS on the branching fractions of the same \LQs to a jet and electron or $c-$jet and muon \cite{Aad:2020iuy}. The black dotted lines indicate the expected limits whereas the black solid lines with small squares (or the solid red lines) depict the observed limits. The green and yellow bands describe the $1\sigma$ and $2\sigma$ regions respectively over the expected limits. The black and blue dashed curve (along with brown and magenta shades on them) signify the theoretical prediction for the pair production of \LQs (with theoretical uncertainties) with branching 100\% and 23\% respectively to a particular mode.}\label{fig:bounds1}
\end{figure}

\begin{figure}
	\centering
	\ContinuedFloat
	\includegraphics[height=0.23\textheight,width=0.48\textwidth]{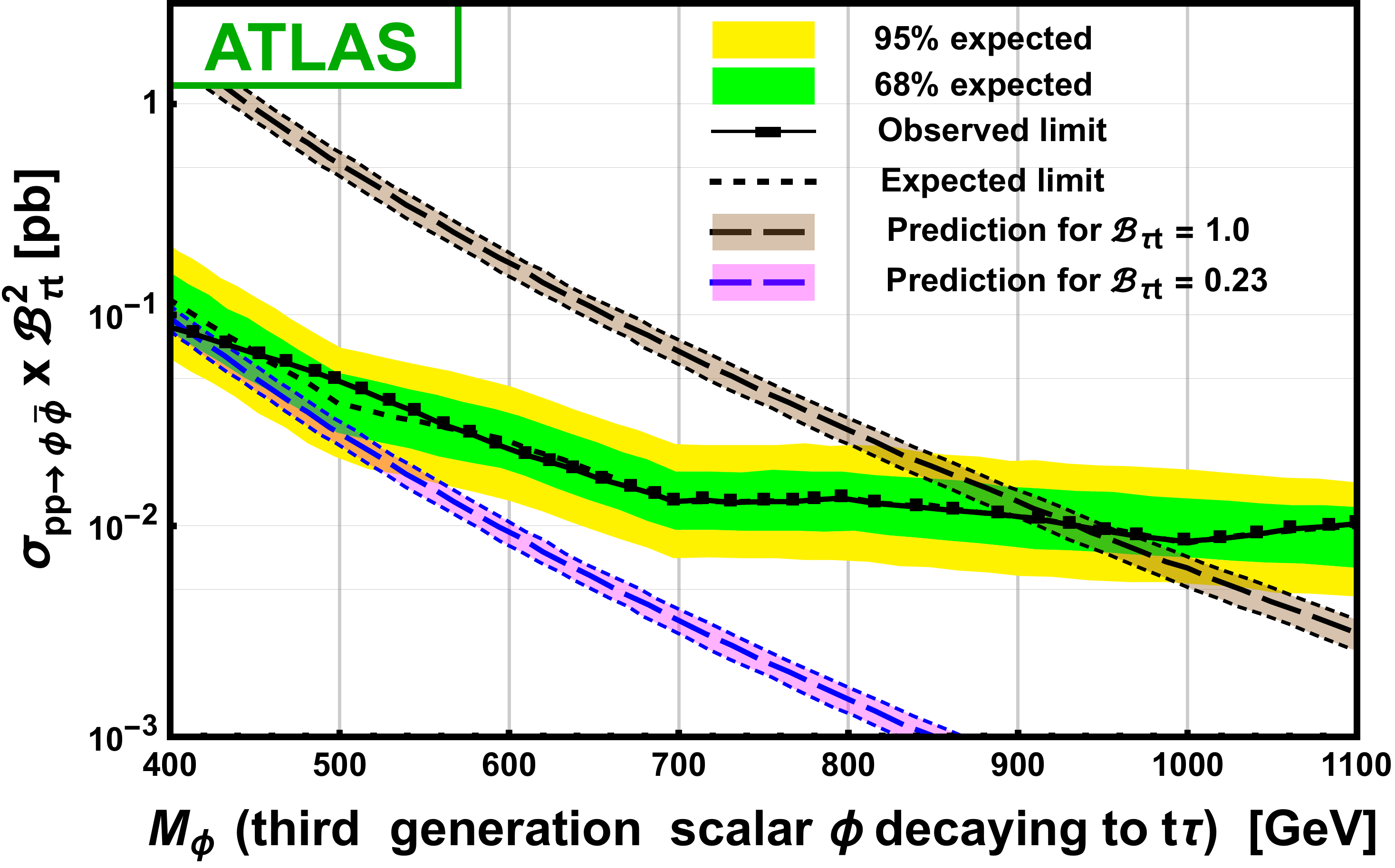}
	\hfil
	\includegraphics[height=0.23\textheight,width=0.48\textwidth]{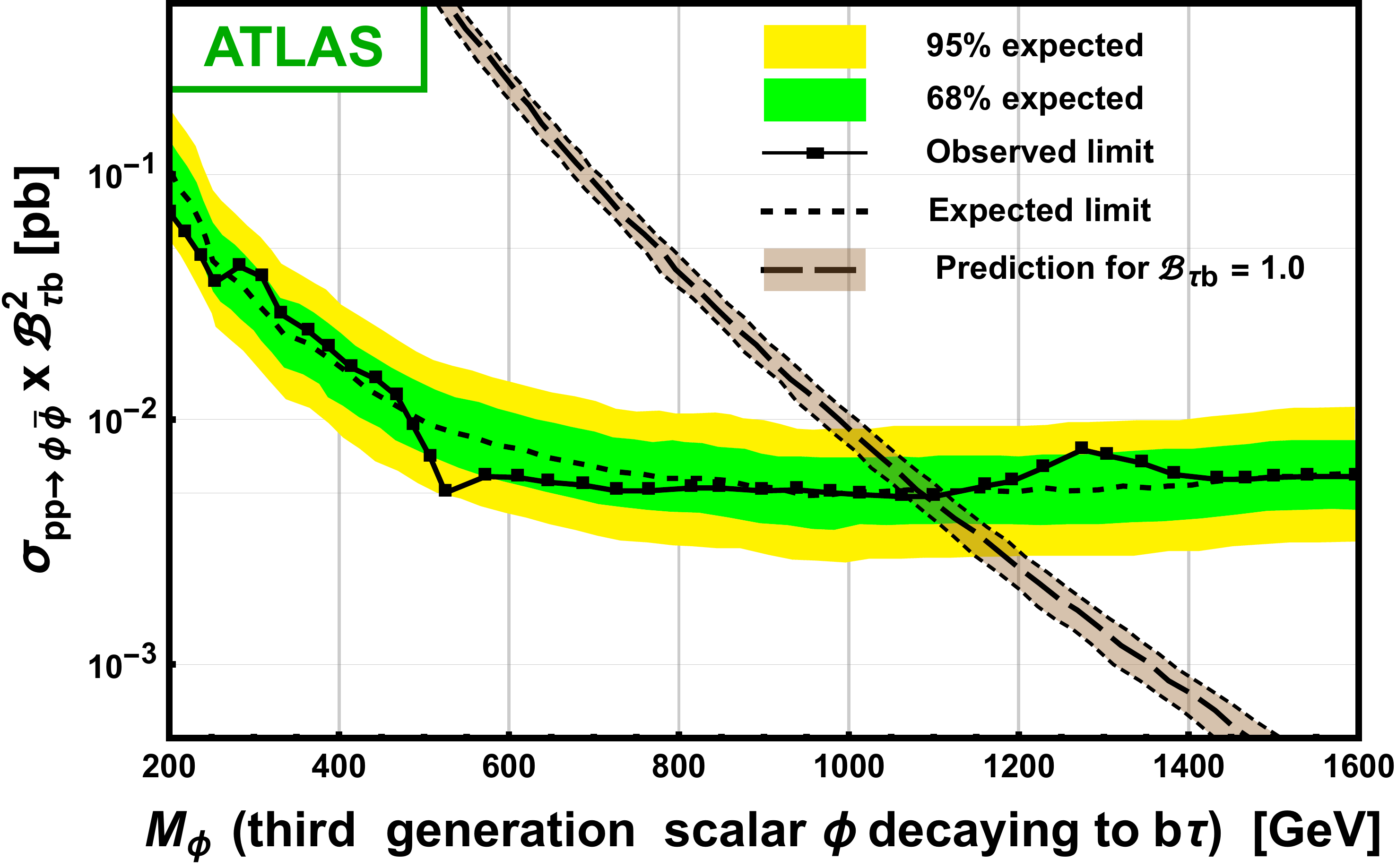}\\[2mm]
	\includegraphics[height=0.23\textheight,width=0.48\textwidth]{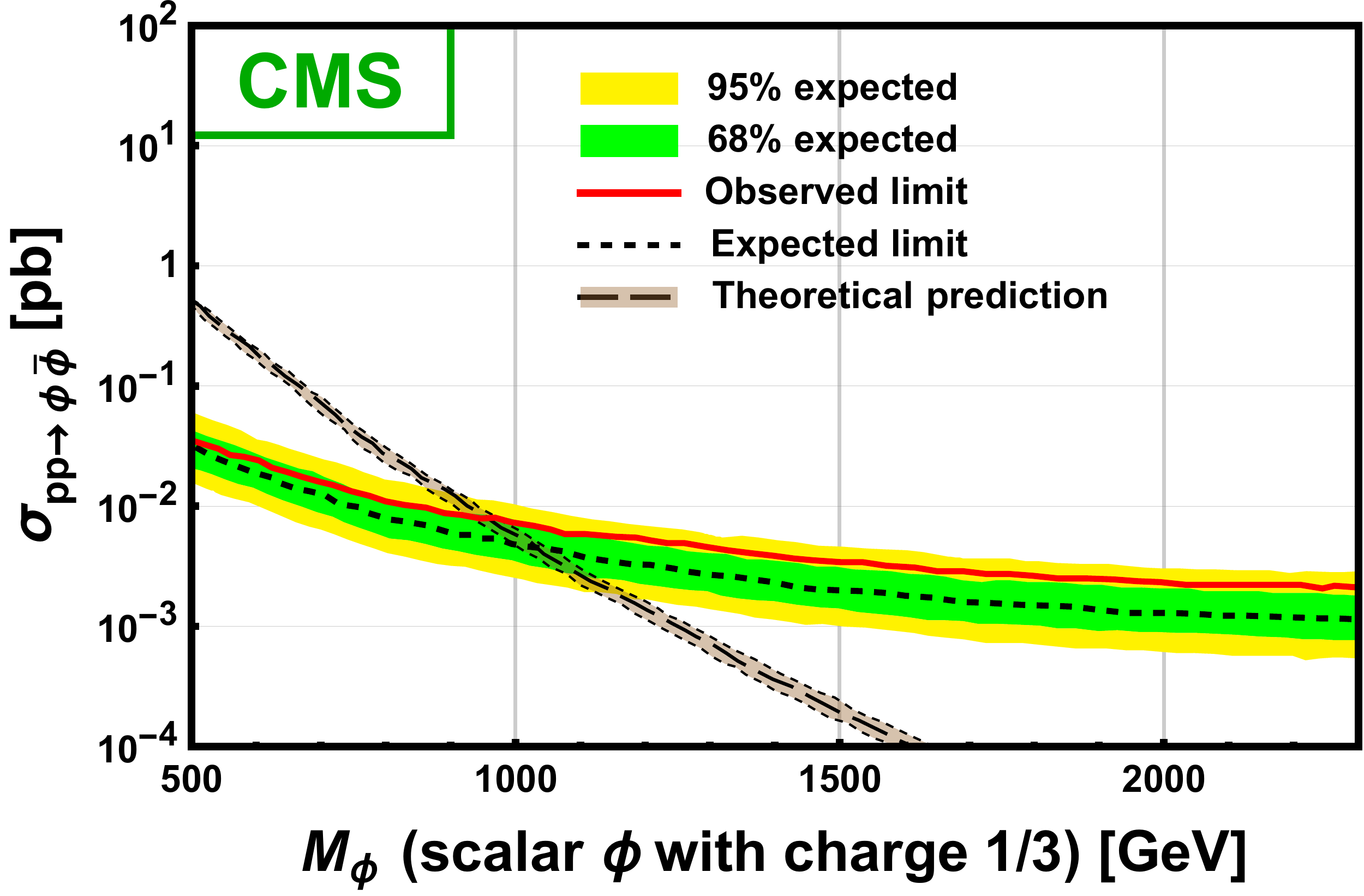}
	
	\caption{LHC bounds on third generation of scalar Leptoquarks. First and second plots show the constraints from ATLAS on $\sigma_{pp\to\phi\bar\phi}\times \mathcal B^2$ for third generation scalar Leptoquarks decaying to $t\tau$ or $b\tau$ \cite{Aaboud:2019bye}. Third one illustrates the bound from CMS on the cross-section for pair production of scalar \LQ with charge 1/3 at LHC considering $b\nu$ and $t\tau$ modes only \cite{Sirunyan:2020zbk}. The black dotted lines indicate the expected limits whereas the black solid lines with small squares (or the solid red lines) depict the observed limits. The green and yellow bands describe the $1\sigma$ and $2\sigma$ regions respectively over the expected limits. The black and blue dashed curve (along with brown and magenta shades on them) signify the theoretical prediction for the pair production of \LQs (with theoretical uncertainties) with branching 100\% and 23\% respectively to a particular mode.}\label{fig:bounds2}
	
\end{figure}

\begin{figure}
	\ContinuedFloat
	\centering
    \includegraphics[height=0.23\textheight,width=0.48\textwidth]{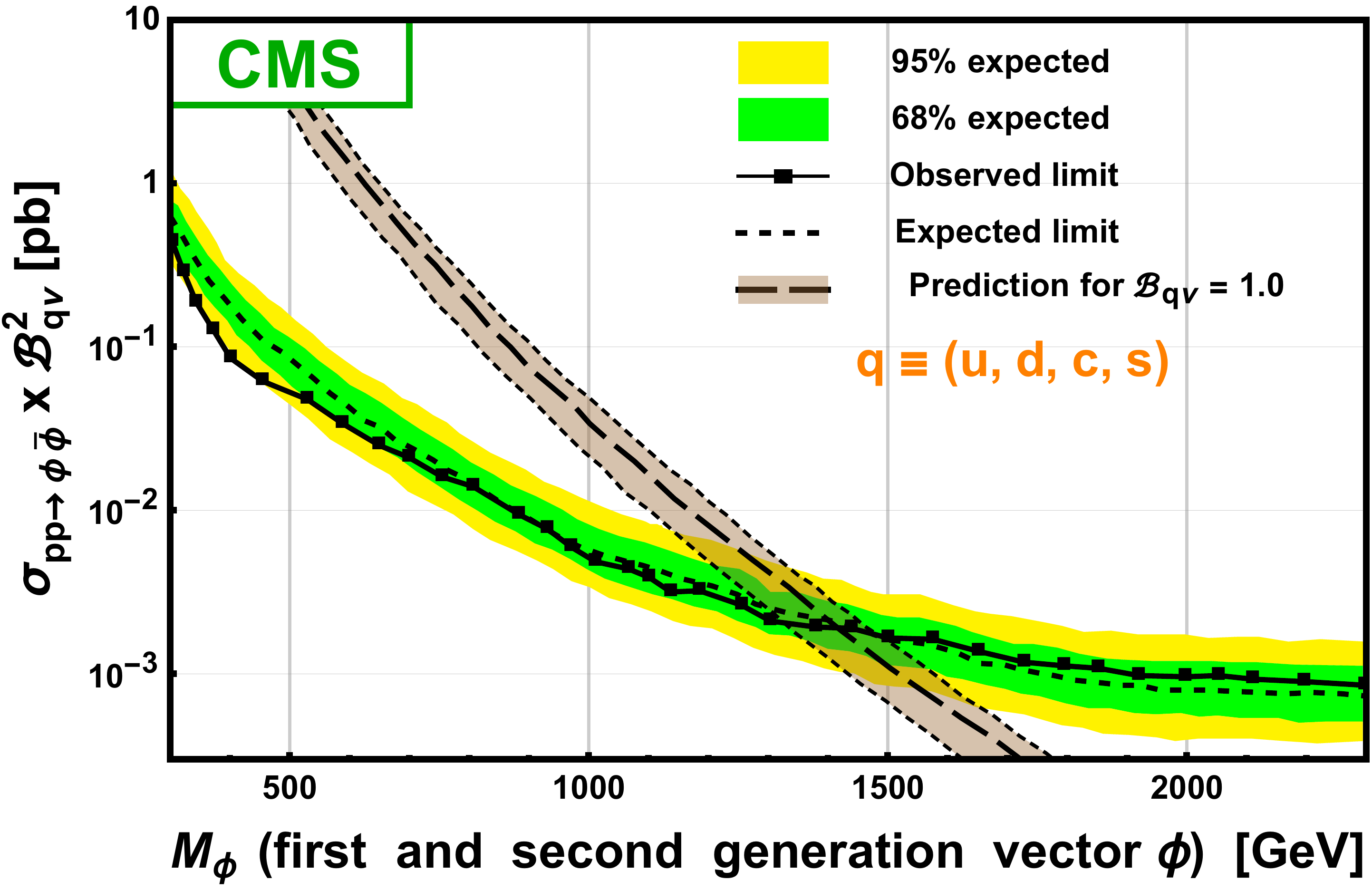}
    \hfil
	\includegraphics[height=0.23\textheight,width=0.48\textwidth]{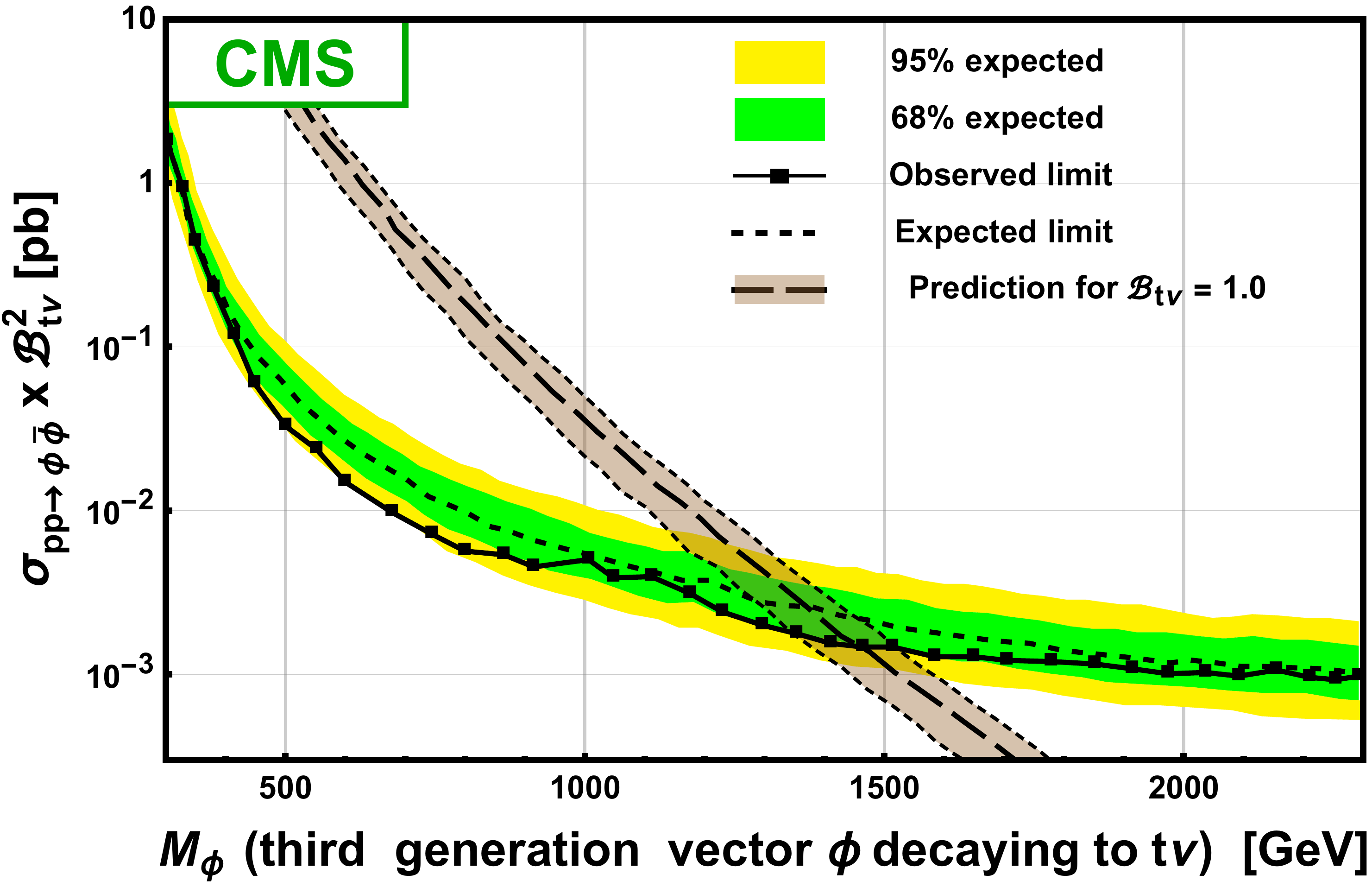}\\[2mm]
	\includegraphics[height=0.23\textheight,width=0.48\textwidth]{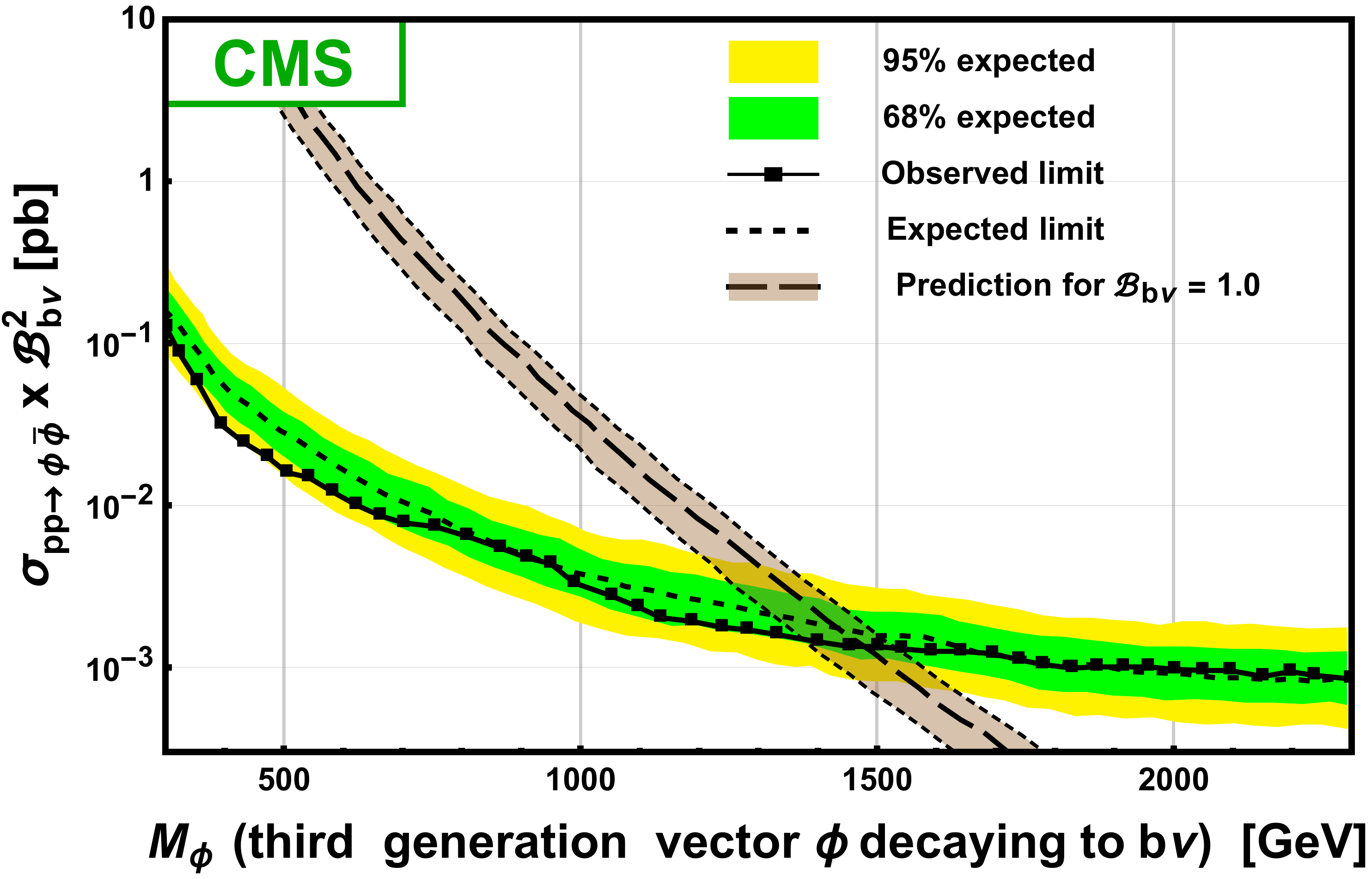}
	
	\caption{LHC bounds on vector Leptoquarks \cite{Sirunyan:2018vhk} from CMS considering the neutrino decay modes. While the first plot shows constraints  on first two generations of vector \LQs, the second and third plots depict the same for third generation. The black dotted lines indicate the expected limits whereas the black solid lines with small squares illustrate the observed limits for pair production of \LQs at LHC. The green and yellow bands describe the $1\sigma$ and $2\sigma$ regions respectively over the expected limits. The black dashed curve (along with brown shade) depict the theoretical prediction for the pair production of \LQs (with theoretical uncertainties) with branching 100\% to a particular mode.}
	\label{fig:bounds3}
\end{figure}

Before we choose benchmark points for collider simulation let us first summarize various bounds on the parameter-space of Leptoquarks. Several direct and indirect constraints on the masses and couplings of \LQs have been studied in literature from different perspectives. Results from low energy experiments help to restrict the Leptoquark-induced four-fermion interactions which provide indirect bound on the parameter-space of the Leptoquarks. Refs. \cite{Davidson:1993qk,Leurer:1993em,Leurer:1993qx,Carpentier:2010ue,Mandal:2019gff} deal with the indirect constraints on \LQs in a quite extensive manner. All the indirect bounds on Leptoquarks are listed in the ``Indirect Limits for Leptoquarks'' section of Ref. \cite{pdg}. However, we mainly focus on the direct bounds on \LQs coming from the chance for them to be detected at various high energy colliders.

\begin{figure}[h!]
	\begin{tabular}{p{5.9cm}p{1cm}p{5.9cm}p{0.5cm}}
	\vspace{0pt}\includegraphics[scale=0.32]{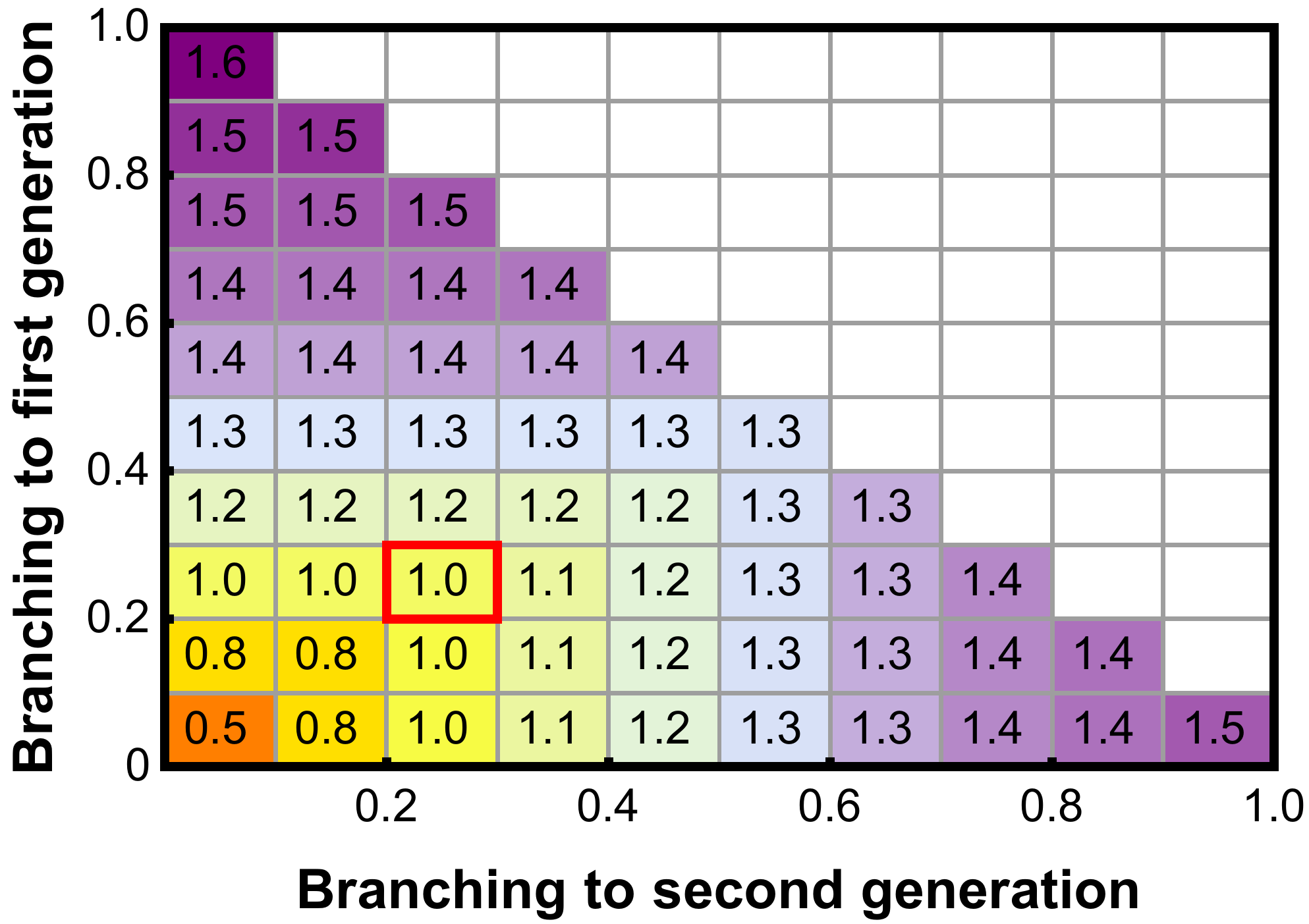}&\vspace{0pt}\includegraphics[scale=0.263]{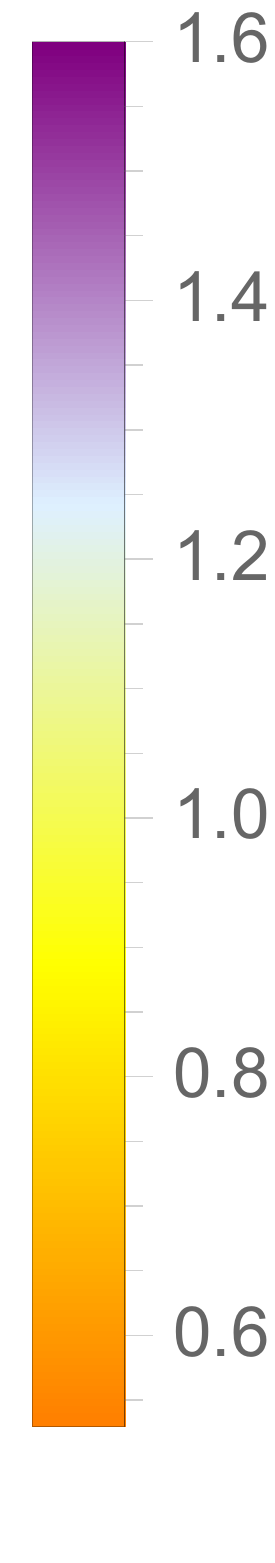}&
	\vspace{0pt}\includegraphics[scale=0.32]{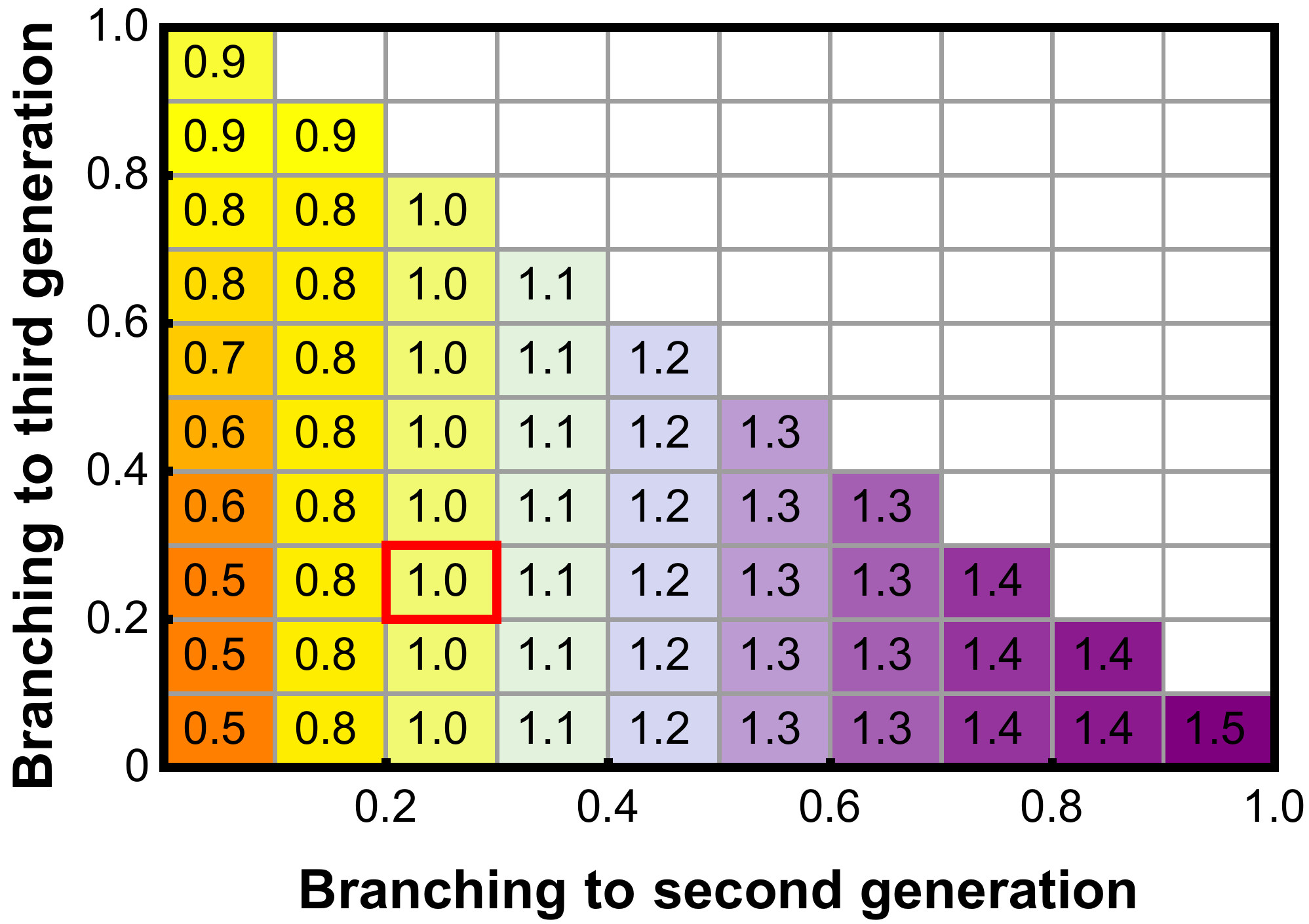}&\vspace{0pt}\includegraphics[scale=0.263]{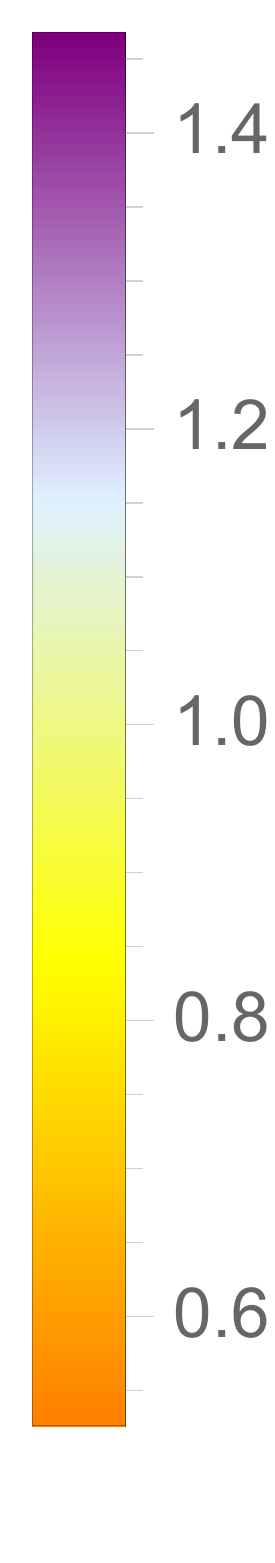}
	\end{tabular}
\caption{Combined lower mass bound (in TeV) from ATLAS and CMS on scalar leptoquark decaying to all three generations of quarks and leptons using the charged lepton modes.}\label{fig:new}	
\end{figure}

The experimental hunt for \LQs started around thirty five years ago. The CELLO \cite{Behrend:1986jz} and JADE \cite{Bartel:1987de} Collaborations were the first to search for \LQs at the PETRA through their pair production in $e^+e^-$ collision. After that the AMY Collaboration \cite{Kim:1989qz} at TRISTAN, the ALEPH \cite{Decamp:1991uy}, L3 \cite{Adriani:1993gk}, OPAL \cite {Abbiendi:2003iv} and DELPHI \cite{Abreu:1998fw} Collaborations at LEP, the H1 \cite{Collaboration:2011qaa} and ZEUS \cite{Abramowicz:2012tg,Abramowicz:2019uti} Collaborations at HERA, the UA2 Collaboration \cite{Alitti:1991dn} at CERN, the CDF \cite{Acosta:2005ge,Abulencia:2005ua,Aaltonen:2007rb} and D$\slashed {\text{O}}$ \cite{Abazov:2008np,Abazov:2010wq,Abazov:2011qj} Collaborations at Fermilab Tevatron have done exhaustive work to get the first evidence of Leptoquark. But none of them succeeded in discovering \LQ and thus the direct bound on the parameter-space of \LQs arise. After each experiment the allowed mass for \LQ gets higher than the previous analysis.

The strongest constraints till now on \LQs come from the ATLAS and CMS Collaborations at the LHC. The ATLAS Collaboration has looked for pair production of first and second generation \LQs with the $\mathcal{L}_{int}=36.1$ \fbi data set of the LHC at $\sqrt s=13$ TeV. However, they cannot find any conspicuous signal over SM background, and hence they rule out first and second generation of scalar \LQs with masses below 1400 GeV (1290 GeV) and 1560 GeV (1230 GeV) at 95\% C.L. assuming branching $\beta=1\,(0.5)\,$ \cite{Aaboud:2019jcc}. Using same data set they have also excluded third generation scalar \LQ lighter than 800 GeV irrespective of any branching fraction \cite{Aaboud:2019bye}. The CMS Collaboration has also performed similar analysis taking the LHC data set at $\mathcal{L}_{int}=35.9$ \fbi and $\sqrt s=13$ TeV. They put lower bounds on the masses of first and second generation scalar \LQs to be 1435 GeV (1270 GeV) and 1530 GeV (1285 GeV) respectively for $\beta=1\,(0.5)$ at 95\% C.L. \cite{Sirunyan:2018btu,Sirunyan:2018ryt}. About third generation scalar \LQs, CMS Collaboration has reported at 95\% C.L. that they should be heavier than 900 GeV and 1020 GeV, if they decay to  top-quark plus $\tau$-lepton  and bottom-quark plus $\tau$-lepton respectively with $\beta=1$ \cite{Sirunyan:2018kzh,Sirunyan:2018nkj}. On the other hand, bounds on vector \LQs have been drawn from neutrino decay channels only. Results from CMS Collaboration \cite{Sirunyan:2018vhk} states that if a vector \LQ decays to $t\nu$ and $b\tau$ channels with 50\% branching fractions in each, then it should have mass larger than 1530 GeV (1115 GeV) is excluded for $\kappa=1 \, (\kappa=0)$. At this point, it is worth mentioning that  $\kappa(\equiv 1- \kappa_G^{})$ is a dimensionless parameter related to the \textit{anomalous chromo-magnetic moment} and \textit{anomalous chromo-electric dipole moment} of the vector Leptoquarks. The interactions of gluons with vector \LQs depend on this parameter \cite{Blumlein:1996qp}. The case with $\kappa=1$ is usually termed as \textit{Yang-Mills} coupling whereas the the scenario with $\kappa=0$ is called \textit{minimal} coupling. In Figures \ref{fig:bounds1}, \ref{fig:bounds2} and \ref{fig:bounds3} we have illustrated all the current bounds on scalar and vector Leptoquarks. While Figure \ref{fig:bounds1} shows ATLAS and CMS bounds on first and second generations of scalar Leptoquarks, Figure \ref{fig:bounds2}
indicates the same on third generation scalar Leptoquarks. On the other hand, Figure \ref{fig:bounds3} depicts different bounds from CMS on vector \LQs through the decay modes involving neutrinos.

\begin{table*}[t!]
	\centering
	\renewcommand{\arraystretch}{1.1}
	\begin{tabular*}{\textwidth}{|c@{\extracolsep{\fill}}cccccccc|}
		\hline
		$\phi$  & BP & $M_\phi$ (GeV) & $Y_L^{11}$& $Y_L^{22}$& $Y_L^{33}$& $Y_R^{11}$& $Y_R^{22}$& $Y_R^{33}$ \\
		\hline
		
		\multirow{3}{*}{$S_1$} & BP1 &1000 &\multirow{3}{*}{0.2} &\multirow{3}{*}{0.2}&\multirow{3}{*}{0.2}&\multirow{3}{*}{0.2}&\multirow{3}{*}{0.2}&\multirow{3}{*}{0.2}\\
		
		& BP2 &1500 &&&&&&\\
		
		& BP3 &2000 &&&&&&\\
		\hline
		
		\multirow{3}{*}{$\widetilde{U}_{1\mu}$} & BP1 &1000 &\multirow{3}{*}{---} &\multirow{3}{*}{---}&\multirow{3}{*}{---}&\multirow{3}{*}{0.2}&\multirow{3}{*}{0.2}&\multirow{3}{*}{0.2}\\
		
		& BP2 &1500 &&&&&&\\
		
		& BP3 &2000 &&&&&&\\
		\hline
	\end{tabular*}
	\caption{Benchmark points for \LQs $S_1$ and $\widetilde{U}_{1\mu}$.}
	\label{tab:BP}
\end{table*}

\begin{table*}[h!]
	\renewcommand{\arraystretch}{1.1}
	\centering
	\begin{tabular*}{\textwidth}{|l@{\extracolsep{\fill}}ccc|lccc|}
		\hline
		Modes &  BP1 & BP2 & BP3 & Modes &  BP1 & BP2 & BP3\\
		\hline
		\textbf{\LQ $\bm {S_1}$}&&&& \textbf{\LQ $\bm{\widetilde{U}_{1\mu}}$}&&&\\
		$u\,e$ &0.225&0.223&0.223&$e^+\,u$&0.338&0.336&0.335\\
		$c\, \mu$&0.225&0.223&0.223&$\mu^+\,c$&0.338&0.336&0.335\\
		$t\, \tau$&0.212&0.218&0.221&$t^+\,\tau$&0.323&0.329&0.331\\
		$d\, \nu_e$&0.113&0.112&0.111&---&---&---&---\\
		$s\, \nu_\mu$&0.113&0.112&0.111&---&---&---&---\\
		$b\, \nu_\tau$&0.113&0.112&0.111&---&---&---&---\\
		\hline
	\end{tabular*}
	\caption{Branching fractions of \LQs $S_1$ and $\widetilde{U}_{1\mu}$ for the benchmark points specified in Table \ref{tab:BP}.}
	\label{tab:BF}
\end{table*}

The theoretically predicted curves, shown by brown bands in Figures \ref{fig:bounds1}, \ref{fig:bounds2} and \ref{fig:bounds3}, consider a \LQ to couple to a single generation of quark and lepton only indicating 100\% branching fraction to a particular mode. But if the \LQs are assumed to interact with all generations of quarks and leptons, the branching ratio in each generation diminishes. Consequently, the brown banded curves will now get scaled down as square of branching fraction, and they will intersect the experimental bands at lower masses than the earlier scenarios. For example, the magenta shaded curves signify the branching fraction to a particular mode to be 23\% which obviously hit the experimental curves at lower masses than the brown strips. Additionally, the third and fourth plots of Figure \ref{fig:bounds1} confirms that adjustment in the branching fractions could allow us to work with \LQs of a bit lower mass. Similarly, the plots in Figure \ref{fig:bounds2} signify that scalar \LQs with masses above 1000 GeV are allowed with any branching fraction to third generation of quarks and leptons.  We combine the CMS and ATLAS results from charged lepton modes in Figure \ref{fig:new} to show the lower bounds (in TeV) on the mass of scalar leptoquark considering its decay to all three generations of quarks and leptons. On the other hand, bounds on vector Leptoquarks, shown in \ref{fig:bounds3}, are  on the invisible decay modes only.

 For our analysis, we have taken scalar singlet \LQ $S_1$ and vector singlet \LQ $\widetilde U_{1\mu}$ with masses 1 TeV, 1.5 TeV and 2 TeV respectively. The couplings ($Y_L$ and $Y_R$) of these \LQs with different generations of quarks and leptons are taken to be diagonal $3\times3$ matrices with entries 0.2, as shown in Table \ref{tab:BP}. It should be noted that both the couplings $Y_L$ and $Y_R$ exist for \LQ $S_1$, but there exists $Y_R$ only for \LQ $\widetilde U_{1\mu}$ which can easily be seen from Table \ref{tab:LQ}. It should also be noticed that our couplings are less than the electromagnetic coupling constant. The branching fractions of these \LQs to different decay modes are listed in Table \ref{tab:BF}. The \LQ $S_1$ has around 23\% branching to each of the charged lepton decay mode. It is assured from Figure \ref{fig:bounds1} that scalar \LQs having 23\% of branching fraction to first and second generations of quarks and leptons are allowed for all the three masses as considered in case of three BPs. Moreover, Figure \ref{fig:bounds2} indicates that these BPs are permitted while considering the bounds on scalar \LQs that couple to third generation of quarks and leptons. One can also observe from Figure \ref{fig:new} that scalar leptoquarks with branching 20\% to 30\% in each of the charged lepton modes, as depicted by the red box, should not be lighter than 1 TeV. On the other hand, the vector \LQ $\widetilde U_{1\mu}$ does not have any invisible decay mode, as shown in Figure \ref{fig:bounds3}, and hence there is no such direct bound on its mass.

\section{Cross-section and angular distribution}
\label{sec:theory}
In this section, we briefly discuss the theoretical aspects to determine the angular distribution as well as the total cross-section for the pair production of scalar and vector \LQs in proton-proton collision \cite{Blumlein:1996qp}. Though these modes are accessible through photon and $Z$-boson mediated electroweak channels as well as the lepton mediated $t$-channel Feynman diagrams, for the sake of simplicity regarding theoretical calculations we neglect them. This presumption is justified since at very high energy the pair production of \LQ will be mostly QCD dominated\footnote{For our simulation, we ignored the $s$-channel mediated electroweak processes, however, we do include the lepton mediated $t$-channel diagrams.}. The Lagrangian related to the mass and kinetic part (QCD) of the scalar and vector \LQs can be expressed as:
\begin{gather}
\mathcal L_s=\Big(D_{ij}^\mu\,\phi_s^j\Big)^\dagger \Big(D^{ij}_\mu\,\phi_{s,j}\Big)-M_{\phi_s}^2\, \phi_s^{i\,\dagger}\,\phi_{s,i}~,\\
\mathcal L_v=-\frac{1}{2}\,G^{i\,\dagger}_{\mu\nu}\,G_{i}^{\mu\nu}+M_{\phi_v}^2\, \phi_{v,\mu}^{i\,\dagger}\,\phi_{v,i}^\mu-ig_s\bigg[(1-\kappa_G)\,\phi_{v,\mu}^{i\,\dagger}\,T^a_{ij}\,\phi_{v,\nu}^j\,\mathcal G^{\mu\nu}_a+\frac{\lambda_G}{M_{\phi_v}^2}\,G^{i\,\dagger}_{\sigma\mu}\,T^a_{ij}\,G_{\nu}^{j\mu}\,\mathcal G^{\nu\sigma}_a\bigg]~,
\end{gather}  
where $\phi_{s,v}$ are scalar and vector \LQs with masses $M_{\phi_{s,v}}$, $\kappa_G$ and $\lambda_G$ are anomalous couplings, $g_s$ is the strong coupling constant and $T^a$ are the generators of $SU(3)$ colour gauge group. The covariant derivative as well as the field strength tensors for gluon $(\mathcal A_\mu)$ and vector \LQ are given by:
\begin{gather}
D_\mu^{ij}=\partial_\mu\,\delta^{ij}-ig_s\,T_a^{ij}\mathcal{A}_\mu^{a}~,\\
\mathcal G_{\mu\nu}^a=\partial_\mu \,\mathcal{A}^a_\nu-\partial_\nu\, \mathcal{A}^a_\mu+g_s\,f^{abc}\, \mathcal{A}_{\mu b}\, \mathcal{A}_{\nu c}~,\\
G_{\mu\nu}^i=D_\mu^{ik}\,\phi_{v,\nu k}-D_\nu^{ik}\,\phi_{v,\mu k}~.
\end{gather}

 Assuming all the quarks to be massless, the differential and integral partonic cross-sections for the pair production of scalar \LQ from $gg$ and $q\bar q$ fusion becomes:   
\begin{gather}
\hspace{-4cm}\frac{d\,\hat\sigma^{gg}_s}{d\cos\theta}=\frac{\pi\,\alpha_s^2\,\hat \beta}{6\,\hat s}\,\bigg[\frac{1}{32}\,\Big(25-18\hat \beta^2+9\hat \beta^2\cos^2\theta\Big)\nonumber\\
\hspace*{5cm}-\,\frac{1}{16}\,\bigg(\frac{25-34\hat \beta^2+9\hat \beta^4}{1-\hat \beta^2\cos^2\theta}\bigg)+\bigg(\frac{1-\hat \beta^2}{1-\hat \beta^2\cos^2\theta}\bigg)^{\hspace*{-0.5mm}2}\;\bigg]~,\\
\hat{\sigma}_s^{gg}=\frac{\pi\,\alpha_s^2}{96\,\hat s}\,\bigg[\hat \beta\,\big(41-31\hat \beta^2\,\big)-\big(17-18\hat \beta^2+\hat \beta^4\big)\,\log\Big|\frac{1+\hat \beta}{1-\hat \beta}\Big|\,\bigg]~,\\
\frac{d\,\hat\sigma^{q\bar q}_s}{d\cos\theta}=\frac{\pi\,\alpha_s^2}{18\,\hat s}\,\hat \beta^3\sin^2\theta\quad\text{and}\quad\hat\sigma_s^{q\bar q}=\frac{2\pi\,\alpha_s^2}{27\,\hat s}\,\hat \beta^3,
\end{gather}
where $\hat \beta=\sqrt{1-4\,M_{\phi_s}^2/\hat s}$ and $\alpha_s=g_s^2/4\pi$ with $\hat s$ being the centre of mass energy and $\theta$ being the \LQ scattering angle in partonic CM frame.

 However, the expression for pair production of vector \LQs is not very simple and the angular distribution depends on the anomalous couplings $\kappa_G$ and $\lambda_G$ too. In this case, the angular distribution can be expanded in terms of polynomials of $\kappa_G$ and $\lambda_G$, and the coefficients for the polynomial expansion can be expressed as functions of $s/\Mp^2$, $\hat \beta$ and $\theta$. Thus, the differential and integral partonic cross-sections for the pair production of vector \LQ from $gg$ and $q\bar q$ fusion can be written as:
\begin{gather}
\frac{d\,\hat\sigma^{gg}_v}{d\cos\theta}=\frac{\pi\,\alpha_s^2\,\hat \beta}{192\,\hat s}\,\sum_{i=0}^{14}\chi_i^g(\kappa_G,\lambda_G)\,\frac{F_i(\hat s,\hat \beta,\cos\theta)}{(1-\hat\beta^2\cos^2\theta)^2}~,\\
\hat\sigma^{gg}_v=\frac{\pi\,\alpha_s^2}{96\,M_{\phi_v}^2}\,\sum_{i=0}^{14}\chi_i^g(\kappa_G,\lambda_G)\,\widetilde F_i(\hat s,\hat \beta)~,\\
\frac{d\,\hat\sigma^{q\bar q}_v}{d\cos\theta}=\frac{2\pi\,\alpha_s^2\,\hat \beta^3}{9\,M_{\phi_v}^2}\,\sum_{i=0}^{5}\chi_i^q(\kappa_G,\lambda_G)\,G_i(\hat s,\hat \beta,\cos\theta)~,\\
\hat\sigma^{q\bar q}_v=\frac{4\pi\,\alpha_s^2\,\hat \beta^3}{9\,M_{\phi_v}^2}\,\sum_{i=0}^{5}\chi_i^q(\kappa_G,\lambda_G)\,\widetilde G_i(\hat s,\hat \beta)~,\\
\text{with}\quad \widetilde{F}_i=\frac{M_{\phi_v}^2}{\hat s}\int_0^{\hat\beta}d\xi\, \frac{F_i(\xi=\hat \beta\cos\theta)}{(1-\xi^2)^2} \quad \text{and}\quad \widetilde{G}_i=\int_0^1d\cos\theta\, G_i(\hat s,\hat \beta,\cos\theta)~.
\end{gather}
However, it is important to mention that this expansion is model dependent and applicable to Leptoquarks with mass range few hundred GeV to few TeV. Now, for minimal coupling scenario $(\kappa_G=1,\, \lambda_G=0)$, we have:
\begin{align}
\sum_{i=0}^{14}F_i\,\chi_i^g(\kappa_G&=1,\, \lambda_G=0)=F_0+F_1+F_3+F_6+F_{10}~,\\
\sum_{i=0}^{5}G_i\,\chi_i^g(\kappa_G&=1,\, \lambda_G=0)=G_0+G_1+G_3~.
\end{align}
The relevant $F_i$ and $G_i$ functions are listed in \autoref{app}. Finally, wrapping each partonic cross-section by corresponding parton distribution function (PDF) and summing over all such contributions, the total cross-section for pair production of \LQ at proton-proton collider is achieved. At this point, it is important to mention that the terms linear in $\kappa_G$ and $\lambda_G$ do not contain the unitarity violating factor $s/M_\phi^2$ for gluon fusion channel; however, to restore the unitarity in quark fusion mode, the lepton exchanging $t$-channel diagrams must be included.  On the other hand, if the energy is very high the Lagrangian for vector \LQ also needs to be corrected appropriately  \cite{Blumlein:1996qp}.

\section{Distinguishing features of \LQs}
\label{sec:dist}

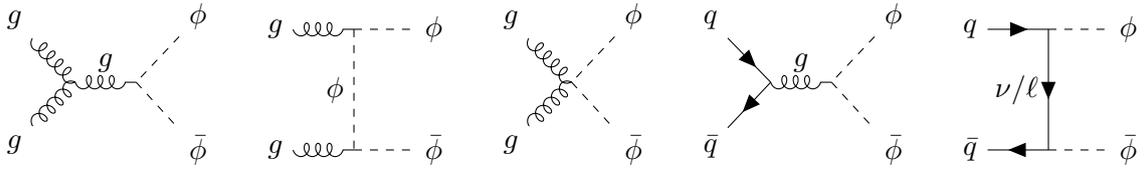
\begin{figure}[h!]
	\centering
	\begin{tikzpicture}
	\begin{feynman}
	\vertex (a2);
	\vertex [above left=0.8cm of a2] (z1){\( g \)} ;
	\vertex [below left=0.8cm of a2] (b1){\( g \)};
	\vertex [right=0.8cm of a2] (a3);
	\vertex [above right=0.8cm of a3] (z4){\( \phi \)};
	\vertex [below right=0.8cm of a3] (b4){{\( \bar{\phi} \)}};
	\diagram{(z1)--[gluon](a2)--[gluon,edge label=\( g\)](a3),
		(a2)--[gluon](b1),
		(z4)--[scalar](a3)--[scalar](b4)};
	\end{feynman}
	\end{tikzpicture}
	\hfill
	\begin{tikzpicture}
	\begin{feynman}
	\vertex (a2);
	\vertex [left=0.8cm of a2] (a1){\( g \)} ;
	\vertex [right=0.8cm of a2] (a3){\( \phi \)};
	\vertex [below=1.6cm of a2] (b2);
	\vertex [left=0.8cm of b2] (b1){\( g \)};
	\vertex [right=0.8cm of b2] (b3){\( \bar{\phi} \)};
	\diagram{(a1)--[gluon](a2)--[scalar](a3),
		(a2)--[scalar, edge label'=\( \phi \)](b2),
		(b2)--[scalar](b3),
		(b1)--[gluon](b2)};
	\end{feynman}
	\end{tikzpicture}
	\hfill
	\begin{tikzpicture}
	\begin{feynman}
	\vertex (a2);
	\vertex [above left=0.8cm of a2] (z1){\( g \)} ;
	\vertex [below left=0.8cm of a2] (b1){\( g \)};
	\vertex [above right=0.8cm of a2] (z4){\( \phi \)};
	\vertex [below right=0.8cm of a2] (b4){{\( \bar{\phi} \)}};
	\diagram{(z1)--[gluon](a2)--[gluon](b1),
		(z4)--[scalar](a2)--[scalar](b4)};
	\end{feynman}
	\end{tikzpicture}
	\hfill
	\begin{tikzpicture}
	\begin{feynman}
	\vertex (a2);
	\vertex [above left=0.8cm of a2] (z1){\( q \)} ;
	\vertex [below left=0.8cm of a2] (b1){\( \bar{q} \)};
	\vertex [right=0.8cm of a2] (a3);
	\vertex [above right=0.8cm of a3] (z4){\( \phi \)};
	\vertex [below right=0.8cm of a3] (b4){{\( \bar{\phi} \)}};
	\diagram{(z1)--[fermion](a2)--[gluon,edge label=\( g \)](a3),
		(a2)--[fermion](b1),
		(z4)--[scalar](a3)--[scalar](b4)};
	\end{feynman}
	\end{tikzpicture}
		\hfill
	\begin{tikzpicture}
	\begin{feynman}
	\vertex (a2);
	\vertex [left=0.8cm of a2] (a1){\( q \)} ;
	\vertex [right=0.8cm of a2] (a3){\( \phi \)};
	\vertex [below=1.6cm of a2] (b2);
	\vertex [left=0.8cm of b2] (b1){\( \bar{q} \)};
	\vertex [right=0.8cm of b2] (b3){\( \bar{\phi} \)};
	\diagram{(a1)--[fermion](a2)--[scalar](a3),
		(a2)--[fermion, edge label'=\( \nu / \ell \)](b2),
		(b2)--[scalar](b3),
		(b2)--[fermion](b1)};
	\end{feynman}
	\end{tikzpicture}
	\caption{Feynman diagrams for Leptoquark pair production at LHC. The photon and Z mediated diagrams have been ignored due to very small contribution.}
	\label{FDLQ}
\end{figure}

This section deals with distinguishing the features of different Leptoquarks from one another at LHC. In Figure \ref{FDLQ}, we have shown the dominant Feynman diagrams for the pair production of \LQ at proton-proton collision. As expected, this process is mainly dominated by QCD. Hence, the tiny contributions from photon and Z mediated diagrams have been ignored. 

\subsection{Separating scalar and vector \LQs}

In this section, we focus on distinguishing the scalar Leptoquarks from their spin-1 vector counterparts. Since the pair-production of \LQ at LHC is QCD dominated, the production cross-section and the angular distribution remain practically independent of the gauge representation and electromagnetic charge of the Leptoquarks but depends on the spins. For convenience, we choose scalar singlet \LQs $S_1$ and vector singlet \LQ $\widetilde U_{1\mu}$ with the masses and couplings specified in Table \ref{tab:BP} and perform a PYTHIA based analysis. In order to study the angular distribution of the scattered Leptoquarks, we first reconstruct them from decay products (i.e. a charged and a quark) and then boost the whole system back in the rest frame of interaction.

\subsubsection{Event rates of hard scattering cross-section:} \label{EvRat}

\begin{figure}[h!]
	\includegraphics[height=0.21\textheight]{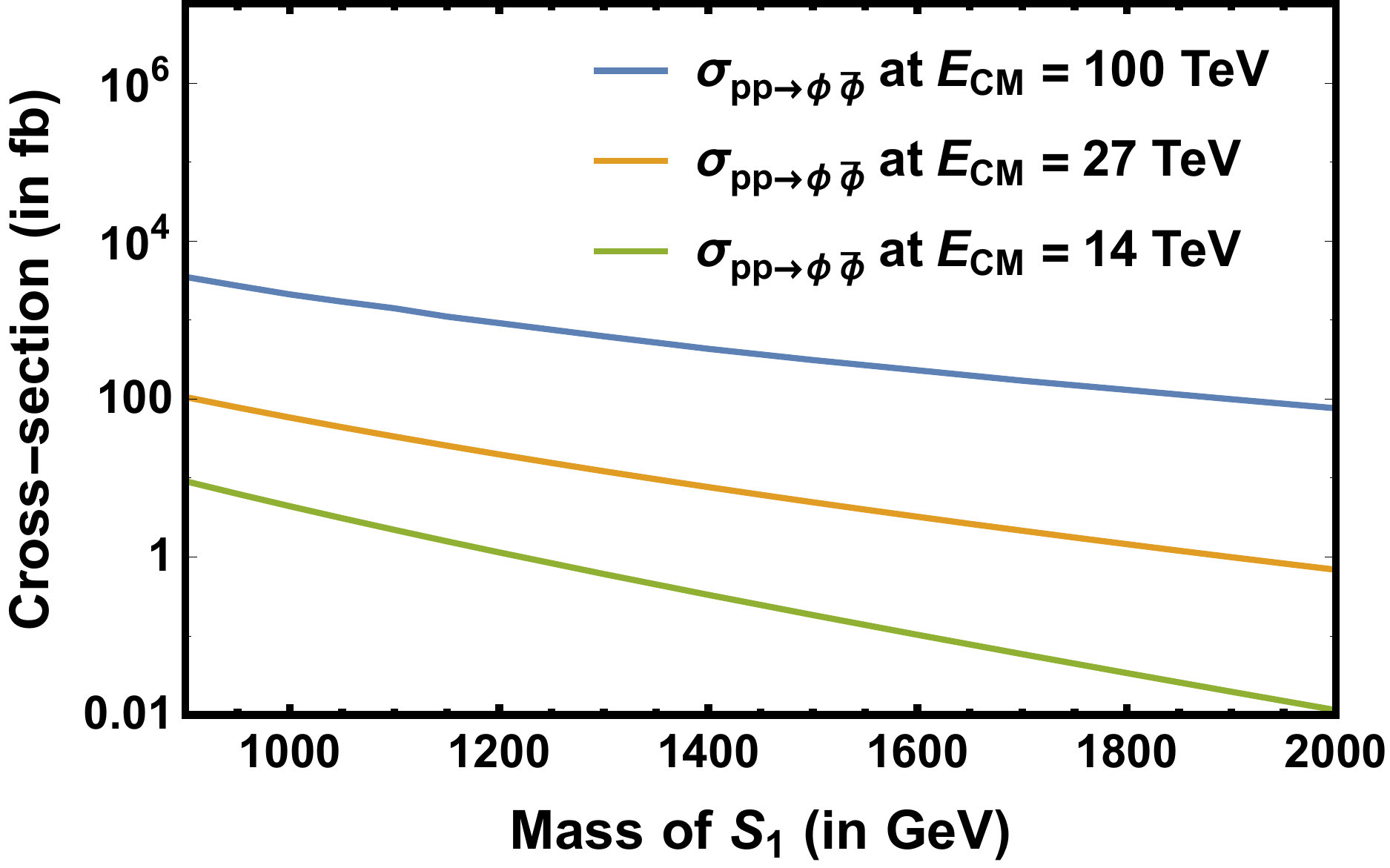}
	\hfil
    \includegraphics[height=0.21\textheight]{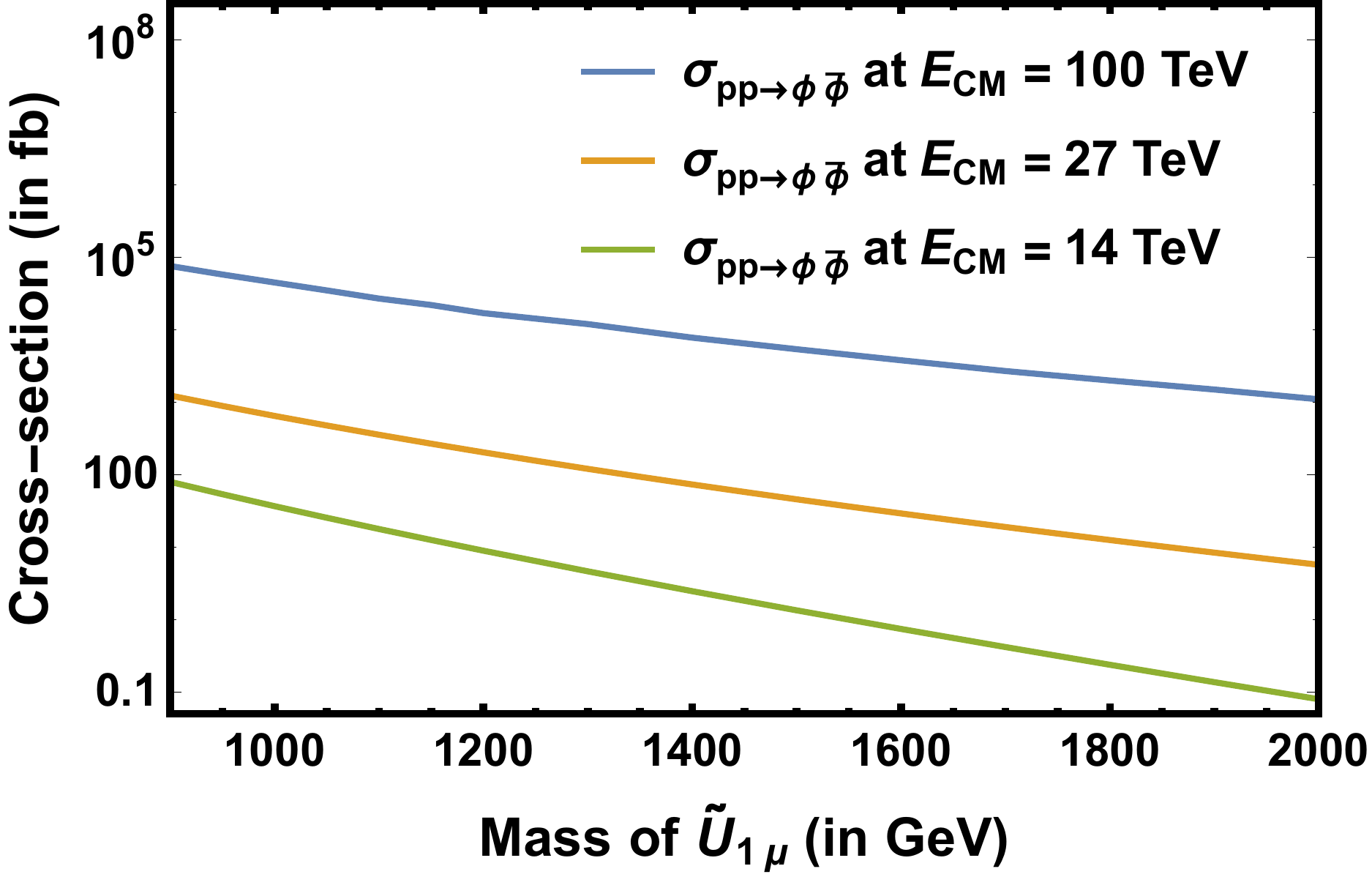}
    \caption{Variation of hard scattering cross-sections for pair production of $S_1$ and $\widetilde U_{1\mu}$ with their masses in proton-proton collider for centre of momentum energies 14 TeV (green), 30 TeV (yellow) and 100 TeV (blue) respectively.}
    \label{fig:hard-cross}
\end{figure}

\begin{table}[h!]
	\renewcommand{\arraystretch}{1.2}
	\centering
		\begin{tabular}{|c|c|c|c|c|c|c|}
			\hline
			&\multicolumn{3}{c|}{\textbf{Leptoquark }$\bm {S_1}$}&\multicolumn{3}{c|}{\textbf{Leptoquark }$\bm{\widetilde{U}_{1\mu}}$}\\
			\cline{2-7}
			Benchmark&\multicolumn{3}{c|}{Production cross-section in fb }&\multicolumn{3}{c|}{Production cross-section in fb}\\
			points&\multicolumn{3}{c|}{ at different $\sqrt s$}&\multicolumn{3}{c|}{ at different $\sqrt s$}\\
			\cline{2-7}
			&14 TeV&27 TeV&100 TeV&14 TeV&27 TeV&100 TeV\\
			\hline
			BP1&4.38&58.03&2183.73&36.80&648.10&44636.30\\ 
			BP2&0.18&4.91&320.40&1.34&45.79&5380.41\\ 
			BP3&0.01&0.69&76.21&0.80&5.72&1116.03\\ 
			\hline
		\end{tabular}
	\caption{Cross-section for pair-production of \LQs $S_1$ and $\widetilde{U}_{1\mu}$ at LHC/FCC with different centre of momentum energies and benchmark points.}\label{tab:cross}
\end{table}

 In Figure \ref{fig:hard-cross}, we show the dependence of hard scattering cross-sections for pair production of $S_1$ (left panel) and $\widetilde{U}_{1\mu}$ (right panel) on their masses in proton-proton collision. The blue, yellow and green curves corresponds to the centre of momentum energy being 100 TeV, 30 TeV and 14 TeV respectively. As expected, the cross-section falls monotonically with increasing mass of Leptoquark and it increases with rise in energy of collision. It is important to notice that at any given $\sqrt s$ and $M_{\phi}$, cross-section for production of vector \LQ is higher than that of scalar \LQ by order of magnitude. This happens because the vector \LQ has three different polarization states which enhance the cross-section for pair production by factor nine relative to the scalar one. It can also be observed from Table \ref{tab:cross} which presents the hard scattering cross-sections for pair production of scalar \LQ $S_1$ and vector \LQ $\widetilde{U}_{1\mu}$ for our chosen energies and benchmark points. For example, the hard scattering cross-sections for pair production of $S_1$ \LQ with mass 1 TeV and coupling 0.2 (BP1) at centre of momentum energies 14 TeV, 27 TeV and 100 TeV are 4.38 fb, 58.03 fb and 2183.73 fb respectively while the same for $\widetilde U_{1\mu}$ \LQ are 36.80 fb, 648.10 fb and 44636.30 fb respectively. Similarly, for BP3 the hard scattering cross-sections at the same centre of momentum energies with \LQ $S_1$ are 0.01 fb, 0.69 fb and 76.21 fb and the same with \LQ $\widetilde U_{1\mu}$ are 0.80 fb, 5.72 fb and 1116.03 fb respectively. So, just looking at the production cross-section, one can easily guess whether the produced \LQ is a scalar or vector one for the same final state topologies. It is worth mentioning that here we have demonstrated the results for vector \LQ in minimal coupling ($\kappa=0$ or $\kappa_G=1$) scenario. For Yang-Mills coupling the hard-scattering cross-section would be even higher \cite{Sirunyan:2018vhk}.

\subsubsection{\LQs with same mass same decay at LHC}

We analyse the $S_1^{1/3}$ and $\widetilde{U}^{5/3}_{1\mu}$ of identical mass, via their $c\mu$ decay modes at the LHC with centre of mass energies of $14,\,27$ and $100$ TeV respectively by simulating the signal and dominant SM background via PYTHIA8 \cite{Pythia8}. We summarize below the steps followed for the generation of events:
\begin{itemize}
\item A detailed simulation requires the models to be written in SARAH \cite{SARAH4}, which is then executed to generate the model files for CalcHEP\cite{CalCHEP3,Belyaev:2005ew}.

\item The ``.lhe'' events were then generated by CalcHEP using NNPDF2.3\cite{pdf} for parton distribution and fed into PYTHIA8 to account for the parton showering, hadronization and jet formation.  The initial state and final state radiations (ISR/FSR) were switched on  for the completeness of the analysis.

\item  We used  Fastjet-3.2.3 \cite{FastJet} with jet radius of $\Delta R=0.5$ using the anti-kT algorithm, constructed from the stable hadrons, and photons originated from the decay of neutral pions.

\begin{itemize}
\item The calorimeter coverage is taken to be  $|\eta| < 4.5, 2.5$ for the jets and leptons respectively.
\item  A taggable lepton needs to be hadronically clean by demanding the hadronic activity within a cone of $\Delta R < 0.3$ around each lepton to be less than $15\%$ of the leptonic transverse momentum ($p_T$).
\item The minimum $p_T$ for the jets and leptons are demanded as $20$ GeV with the respective criteria for the jet-lepton isolation $(\Delta R_{lj} > 0.4)$ and the lepton-lepton isolation $(\Delta R_{ll} > 0.2)$. 
\end{itemize}

\item The dominant SM backgrounds are also taken into account in order to estimate the signal significance  at the LHC. We choose three benchmark points (BPs), with the \LQ masses 1.0 (BP1), 1.5 (BP2) and 2.0 (BP3) TeV respectively and Yukawa coupling 0.2 as mentioned in Table \ref{tab:BP}. To reduce the SM backgrounds we choose \LQ decays to  $c \, \mu$ with respective branching fractions $\mathcal{B}(\phi_{s(v)} \to c \mu)=0.23(0.33)$ for scalar (vector) Leptoquark for the rest of the analysis, as shown in Table \ref{tab:BF}, which are compatible with the LHC bounds \cite{Aaboud:2019bye,Aaboud:2019jcc,Sirunyan:2018btu,Sirunyan:2018kzh,Sirunyan:2018nkj,Sirunyan:2018ryt,Sirunyan:2018vhk}.
\end{itemize}     

\begin{figure}[!htb]
\centering
\mbox {
\subfigure[$S_1$]{\includegraphics[width=0.45\linewidth, height=0.18\textheight]{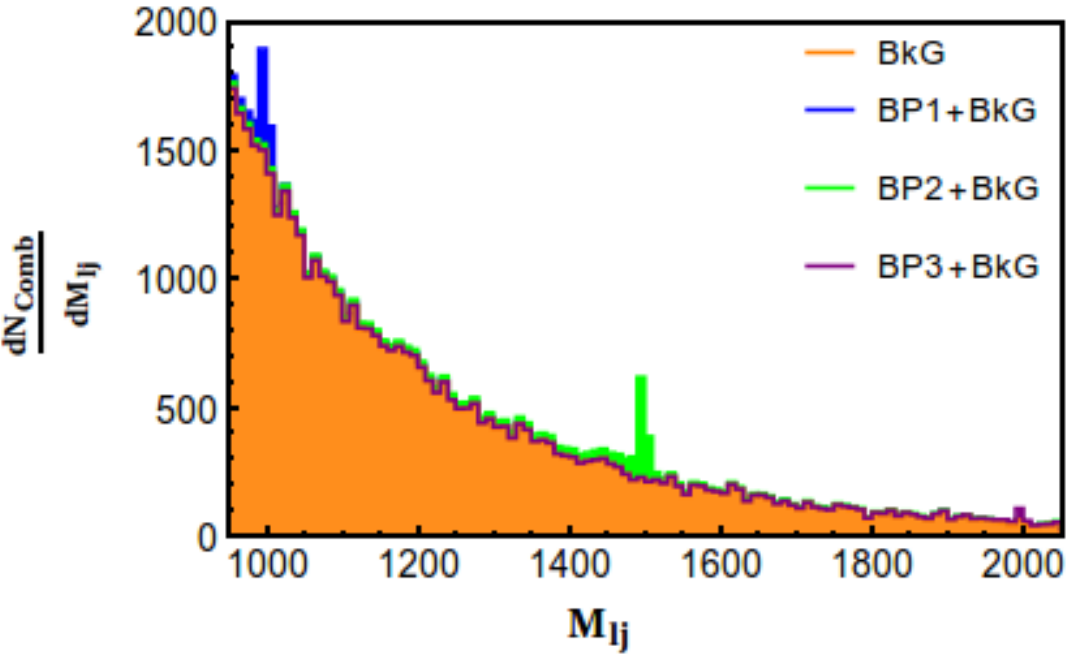} }
\hspace*{1.0cm}
\subfigure[$\widetilde{U}_{1\mu}$]{\includegraphics[width=0.45\linewidth, height=0.18\textheight]{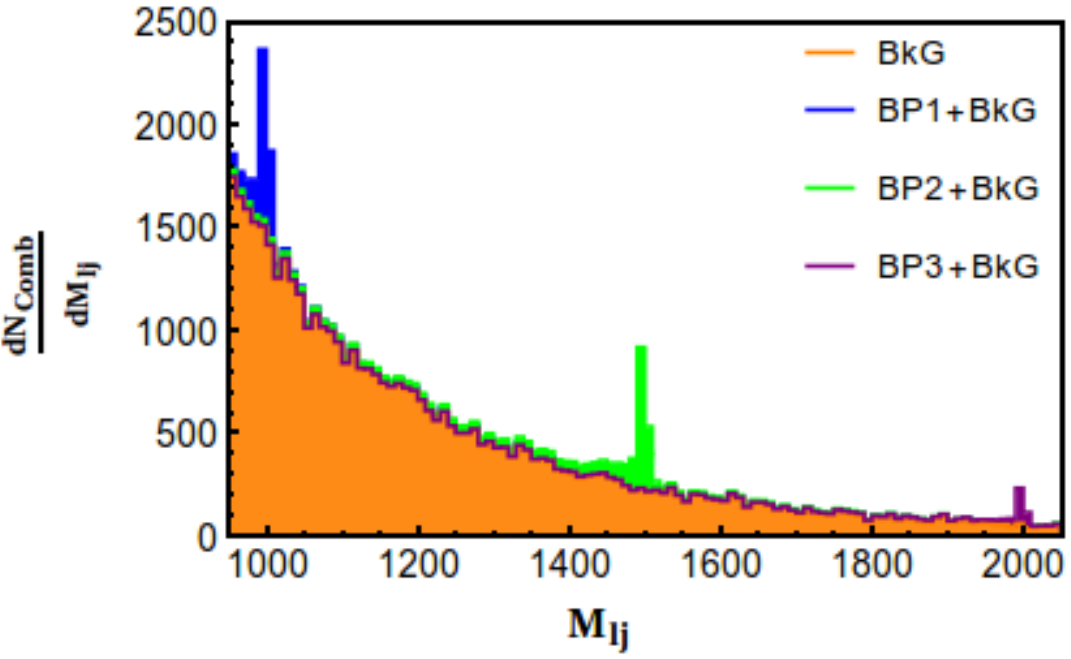} }  }
\caption{Invariant mass distributions of $j \mu$ for both scalar and vector Leptoquarks along with the dominant SM backgrounds at the LHC at 14 TeV.}\label{Invjl1}
\end{figure}

\begin{figure}[!htb]
\centering
\mbox {
\subfigure[$S_1$]{\includegraphics[width=0.45\linewidth, height=0.18\textheight]{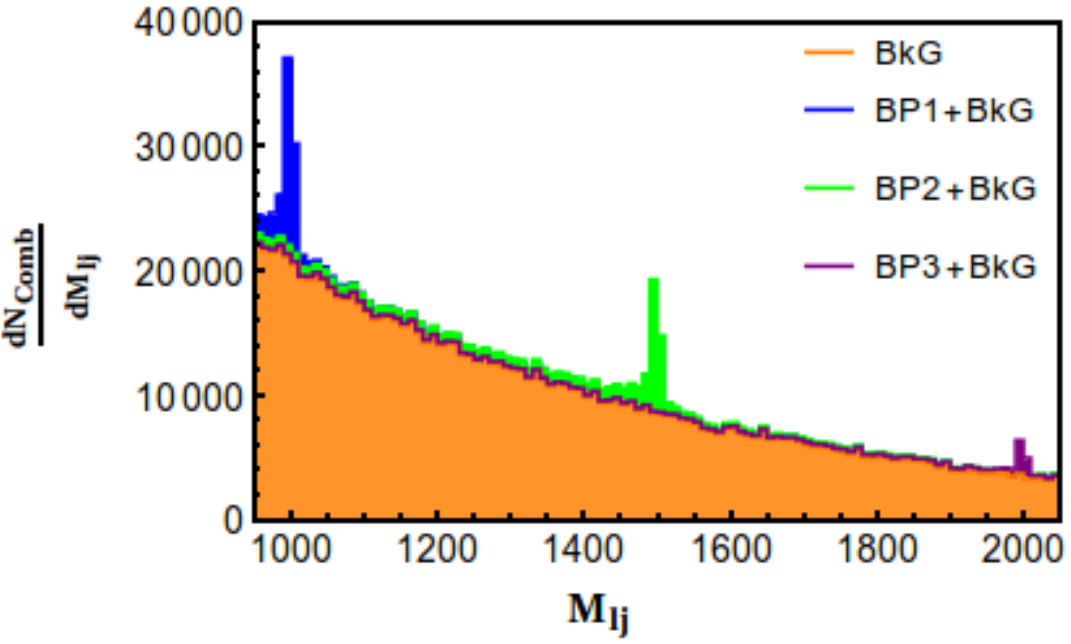} }
\hspace*{1.0cm}
\subfigure[$\widetilde{U}_{1\mu}$]{\includegraphics[width=0.45\linewidth, height=0.18\textheight]{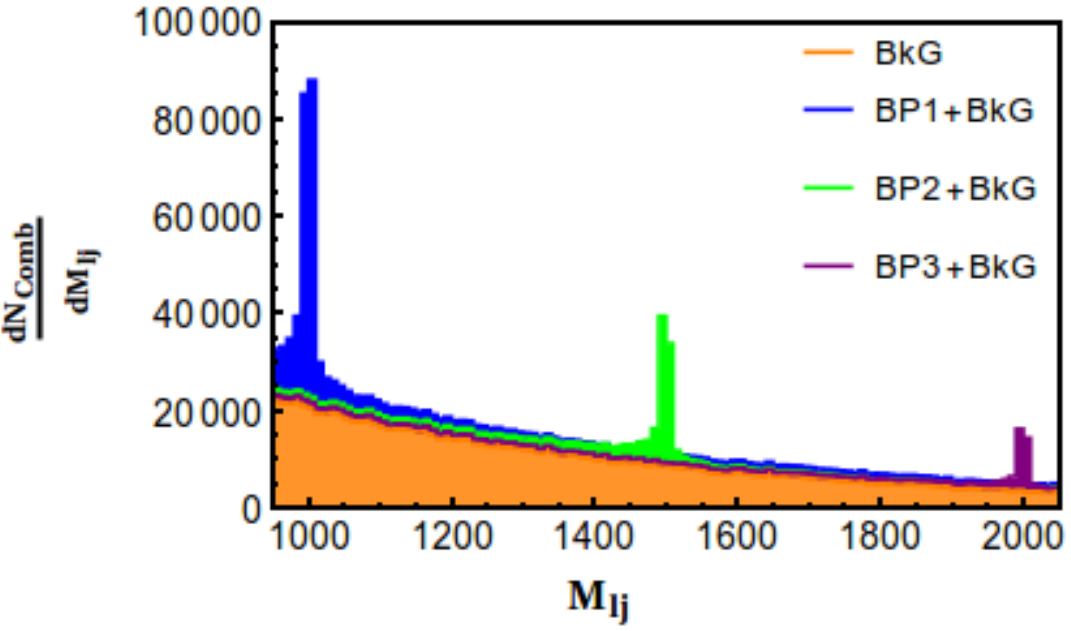} }  }
\caption{Invariant mass distributions of $j \mu$ for both scalar and vector Leptoquarks along with the dominant SM backgrounds at the LHC at 100 TeV.}\label{Invjl2}
\end{figure}

For our purpose, we first boost back the lab frame to CM frame for which the reconstruction of the Leptoquark mass is necessary.  We reconstruct the \LQ mass for each case from the invariant mass of $\rm{jet},\, \mu$ i.e. $M_{\ell j}$ as described in Figures~\ref{Invjl1},~\ref{Invjl2} for the chosen benchmark points (in blue, green and purple) along with the dominant SM backgrounds (in orange) respectively. We consider all possible dominant SM backgrounds for the analysis, viz. $t\bar{t}, \, t\bar{t}V, \, tVV, \, VV, \, VVV$, where $V=Z,\, W^\pm$ and $tVV=tZW^-, \, \bar{t}ZW^+$. In order to obtain more statistics at higher values of jet-lepton invariant mass ($\gtrsim 1.0$ TeV), we imposed following cuts when generating background events: $M_{t\bar{t}} \geqslant 0.95, \, M_{t\bar{t}V} \geqslant 0.95, \, M_{tVV} \geqslant 0.95, M_{VV} \geqslant 0.95 \text{ and }M_{VVV} \geqslant 0.95$ TeV. The tagging of high $p_T$ muons along with a $c$-jet further reduces the Standard Model QCD backgrounds to a negligible level. Since SM backgrounds already depletes for high jet-lepton invariant mass at TeV scale, c-jet tagging and selection of high $p_T$ muons have not been considered in our study.  Due to large energy of interaction for the $pp$ collision, the boost for the interacting partons are mostly longitudinal, thus for the reconstruction of CM frame, transverse boost has been neglected.

We summarize in the following, the criteria set for the selection of the final states:
\begin{itemize}
\item For our simulation, we select each event with $\geq 1\mu^+ + 1\mu^- + 2j$.

\item In order to exclude backgrounds with an on-shell Z boson, we impose every combination of opposite charged leptons, and jets to satisfy $\lvert M_{\ell \ell} - M_Z \rvert > 5,\, \lvert M_{jj} - M_Z \rvert > 10$ GeV.

\item We next take all possible combinations of the jet-lepton pairs and evaluate the invariant mass. The pairs originated from the Leptoquark decay will peak at the invariant mass of the Leptoquarks while the rest will form a continuum, whereas for the SM backgrounds the pattern show an exponential fall with the increase in the jet-lepton invariant mass as obtained in Figures~\ref{Invjl1},~\ref{Invjl2}.

\item Finally in order to obtain signals with a Leptoquark pair, we claim each event with exactly one pair of jet-lepton invariant mass satisfying $\lvert M_{\mu^\pm j} - M_\Phi \rvert \leq 10$ GeV. The sequential impositions of these cuts and their effects on Signal and Backgrounds has been enlisted in Tables~\ref{LQnum1},~\ref{LQnum2},~\ref{LQnum3}.

\item As discussed in Section~\ref{JC}, jet charge is an effective observable to discern different degenerate states of the same $SU(2)$ multiplet, and can optimise signatures of one member over the rest based on different Leptoquark decay modes leading to different event topologies. But in this section, since we focus on distinguishing two $SU(2)$ singlet Leptoquark with different spins having identical decay modes, we did not impose this cut. 
\end{itemize}
   
In order to estimate the significance, the signal and the dominant SM background numbers are determined for all the Benchmark Points of the Signals in Tables~\ref{LQnum1},~\ref{LQnum2},~\ref{LQnum3} at an integrated luminosity of 1000 fb$^{-1}$ at the LHC/FCC with centre of mass energies of 14, 27 and 100 TeV respectively. Obtaining the invariant Leptoquark mass from all possible combinations of invariant mass of the jet-lepton pairs for each event, and subsequent reconstruction of Leptoquark pair imposing the 10 GeV cut around the resonance peak is instrumental in reconstructing the centre of mass frame in which the angular distributions, as shown in Figure~\ref{angdisLQLHC} would exhibit patterns unique to the Leptoquark spins. With judicial imposition of cumulative cuts, we succeed to minimise the SM background to considerable proportion.

\begin{figure}[!htb]
	\hspace*{-1cm}
\centering
\mbox {
\subfigure[BP1]{\includegraphics[width=0.35\linewidth, height=0.15\textheight]{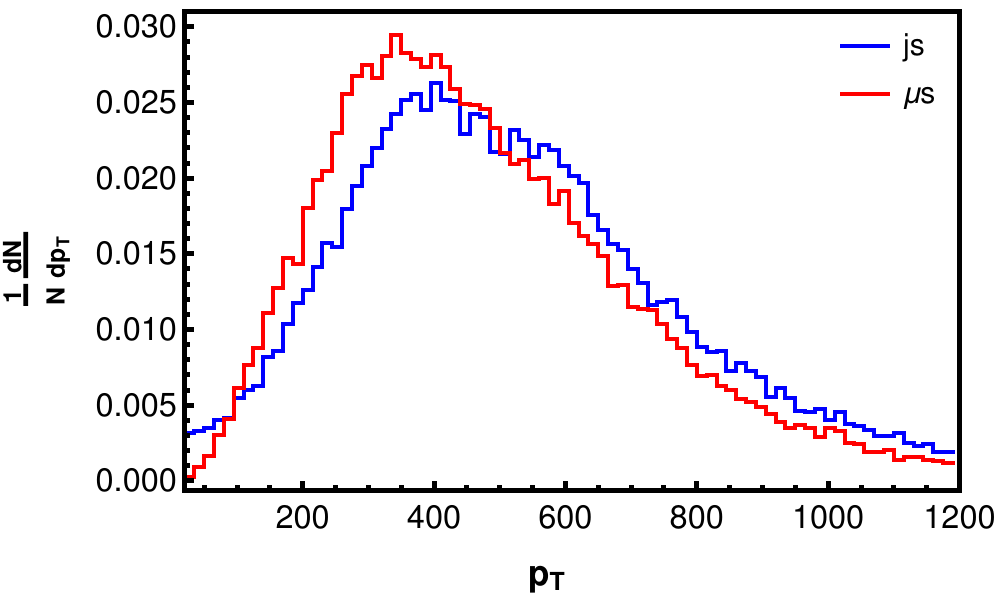} }
\hfill
\subfigure[BP2]{\includegraphics[width=0.35\linewidth, height=0.15\textheight]{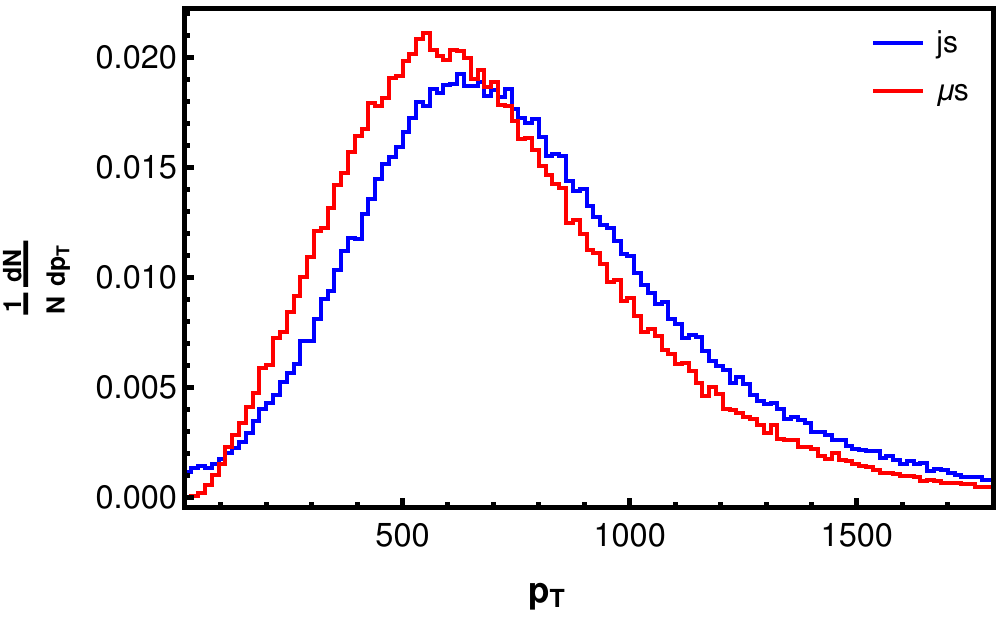} }
\hfill
\subfigure[BP3]{\includegraphics[width=0.35\linewidth, height=0.15\textheight]{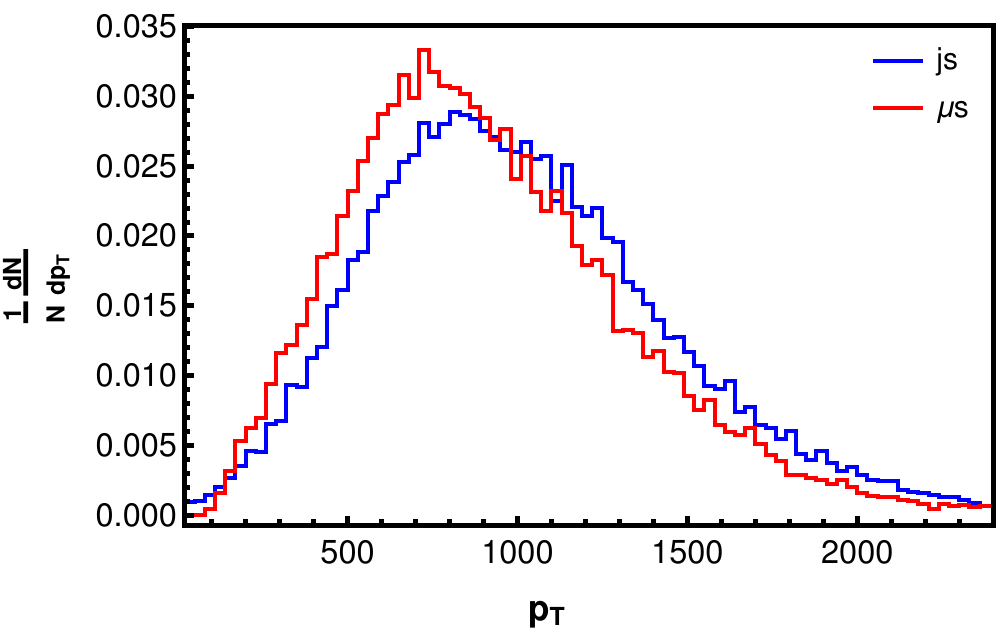} }  }
\caption{Transeverse momenta of the jets and leptons from the decay of the scalar Leptoquark $S_1$ pair produced at 14 TeV.}\label{pT14}
\end{figure}

\begin{figure}[!htb]
		\hspace*{-1cm}
\centering
\mbox {
\subfigure[BP1]{\includegraphics[width=0.35\linewidth, height=0.15\textheight]{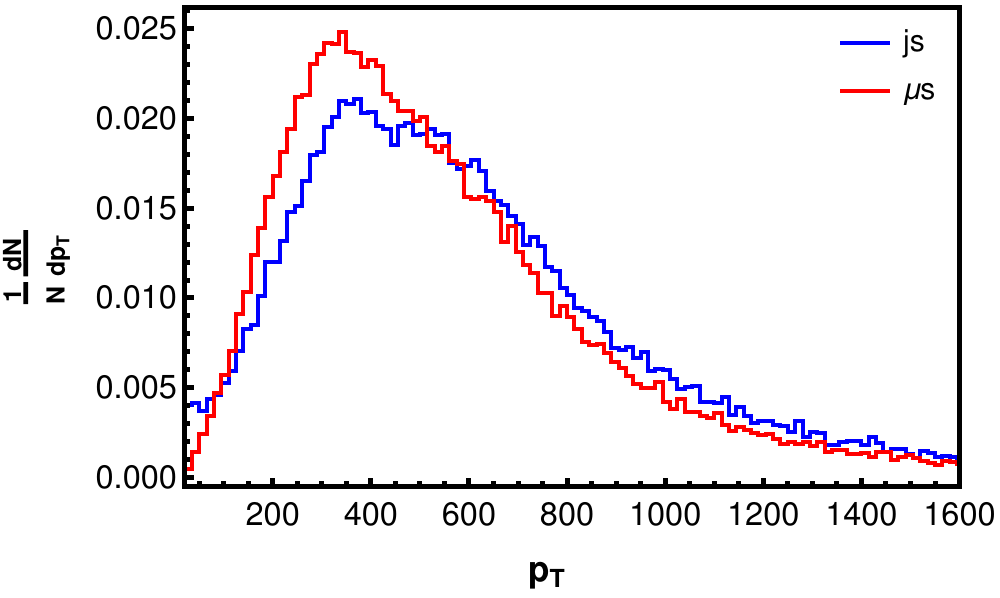} }
\hfill
\subfigure[BP2]{\includegraphics[width=0.35\linewidth, height=0.15\textheight]{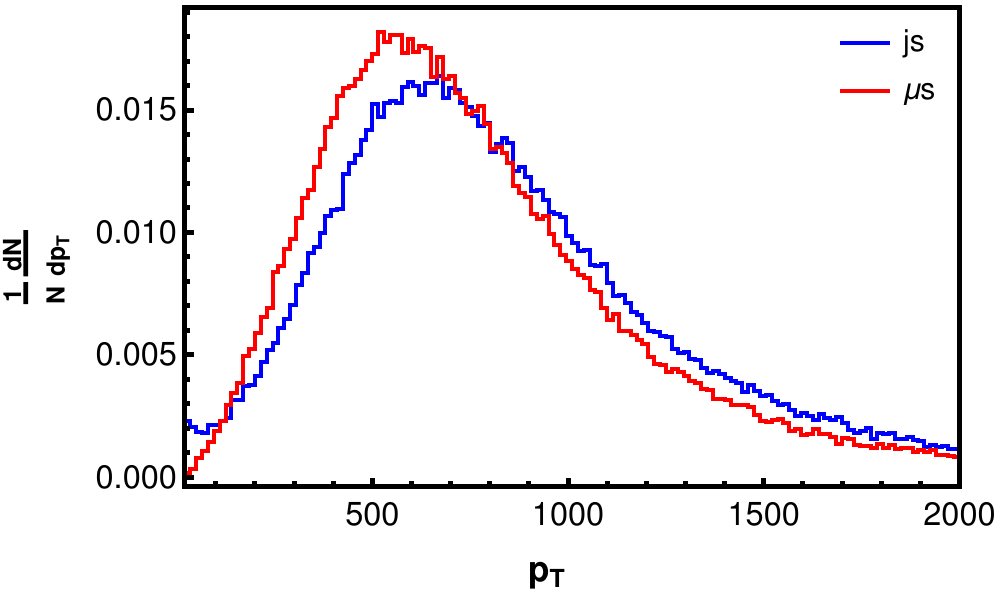} }
\hfill
\subfigure[BP3]{\includegraphics[width=0.35\linewidth, height=0.15\textheight]{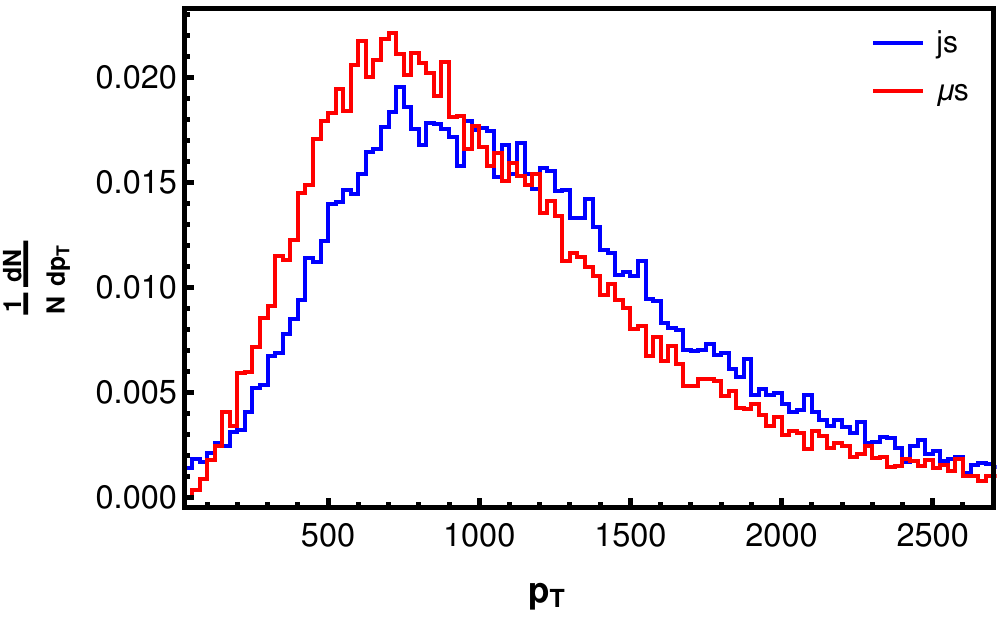} }  }
\caption{Transeverse momenta of the jets and leptons from the decay of the scalar Leptoquark $S_1$ pair produced at 100 TeV.}\label{pT100}
\end{figure}

We begin our analysis with the kinematics of the Leptoquark decays. We select signal events with exactly 1 jet-muon and 1 jet-antimuon invariant masses falling within a 10 GeV window around the Leptoquark resonance peak as shown in Figures~\ref{Invjl1},~\ref{Invjl2}. We next plot the $p_T$s of these jets ($js$), muons and antimuons ($\mu s$) as specified below. We observe that both the jet and muon $p_T$s peak roughly around half the masses of the respective Leptoquarks, and this behaviour is independent of the Leptoquark spins. We present our results in Figures~\ref{pT14},~\ref{pT100} for three benchmark points at 14 and 100 TeV collision energies. We also observe the presence of longer tails for both jet and muon $p_T$s for 100 TeV collisions, compared to the 14 TeV ones.

\begin{figure}[!htb]
		\hspace*{-1cm}
\centering
\mbox {
\subfigure[BP1]{\includegraphics[width=0.35\linewidth, height=0.15\textheight]{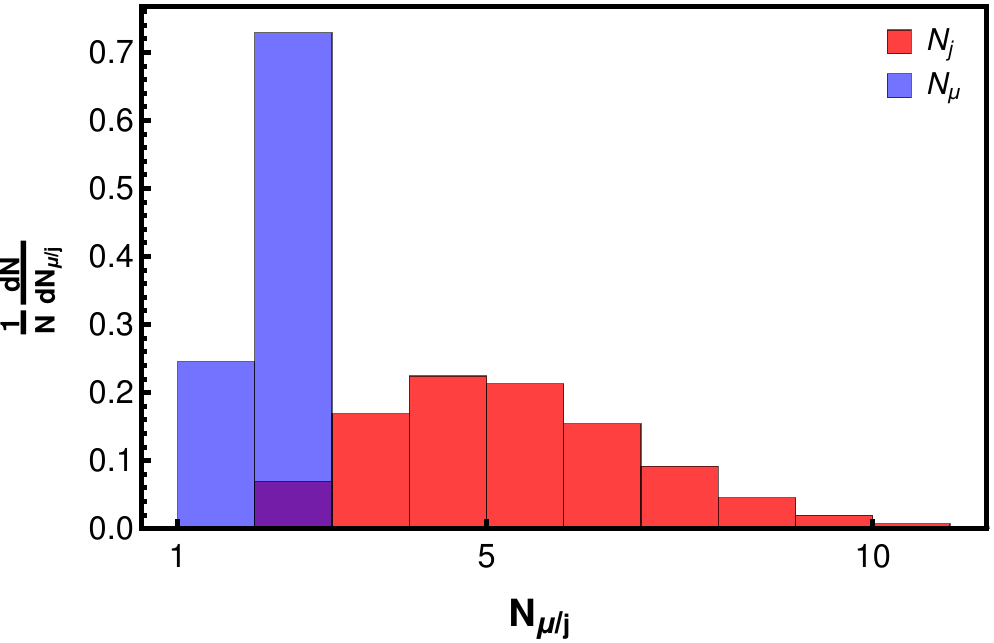} }
\hfill
\subfigure[BP2]{\includegraphics[width=0.35\linewidth, height=0.15\textheight]{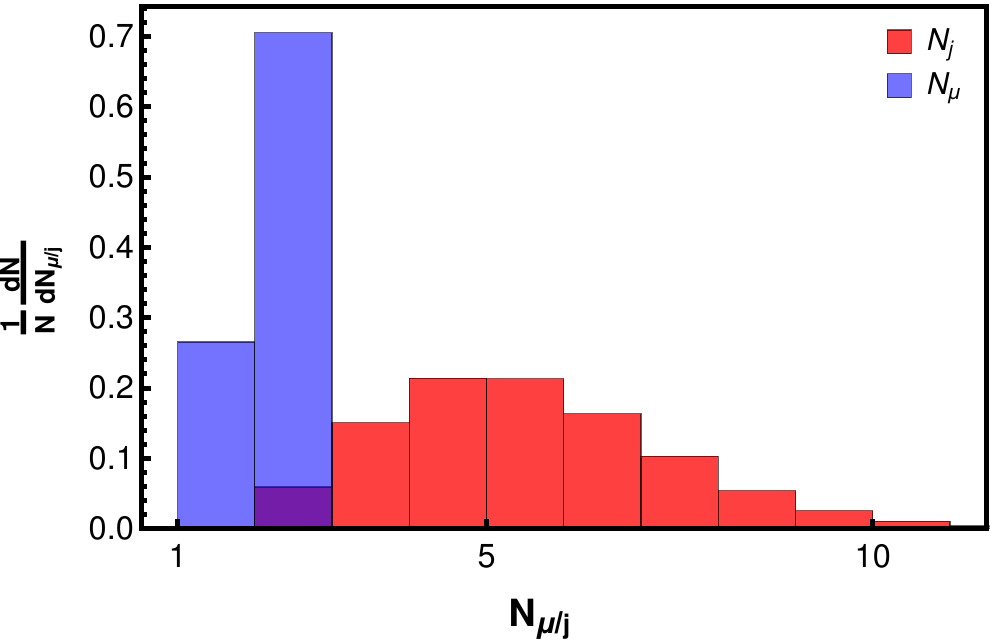} }
\hfill
\subfigure[BP3]{\includegraphics[width=0.35\linewidth, height=0.15\textheight]{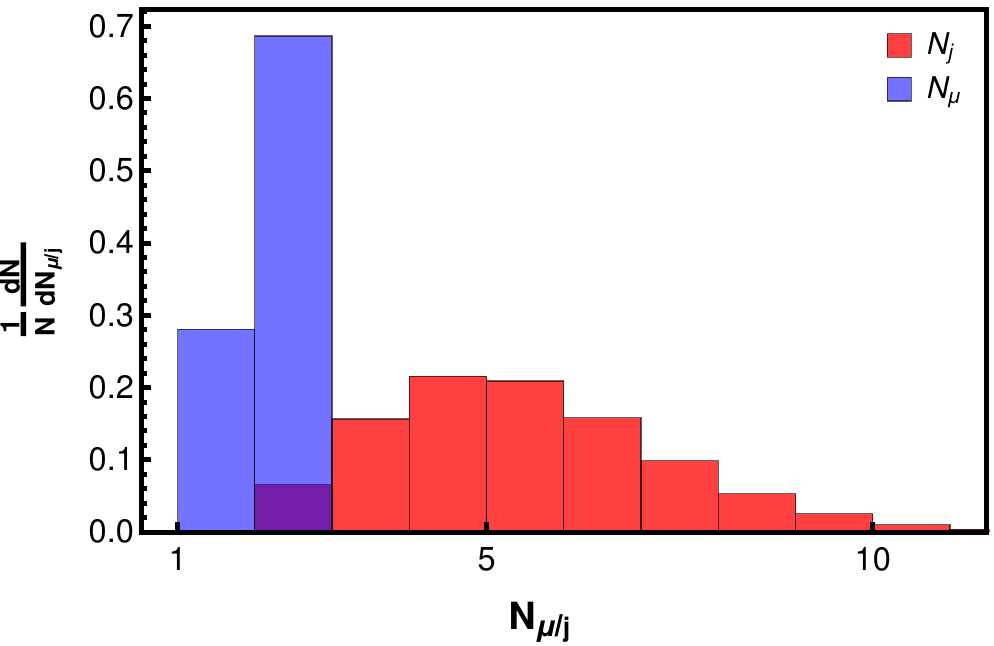} }  }
\caption{Multiplicity of the jets and muons for the pair production of the scalar Leptoquark $S_1$ at 14 TeV.} \label{Mult14}
\end{figure}

We also present, in Figures~\ref{Mult14} and \ref{Mult100}, the jet and muon multiplicities of the signal events for all three different benchmark points and at 14 TeV and 100 TeV collisions. The multiplicities at different collision energies show similar patterns for different Leptoquark spins. We also observe that irrespective of the collision energies, di-muon final states are dominant. Also, due to the radiation effects, jet multiplicities peak roughly at 5 for both collision energies, irrespective of the Leptoquark mass. With our studies on kinematics performed, we next move on to the analysis of the signals and SM backgrounds for different benchmark points, at different collision energies.

\begin{figure}[!htb]
		\hspace*{-1cm}
\centering
\mbox {
\subfigure[BP1]{\includegraphics[width=0.35\linewidth, height=0.15\textheight]{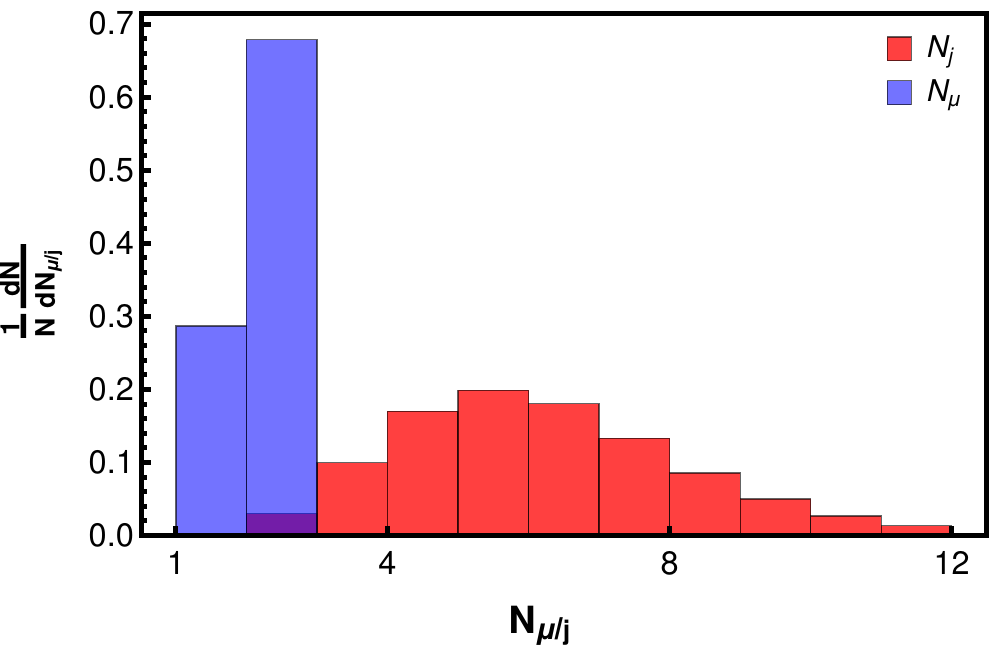} }
\hfill
\subfigure[BP2]{\includegraphics[width=0.35\linewidth, height=0.15\textheight]{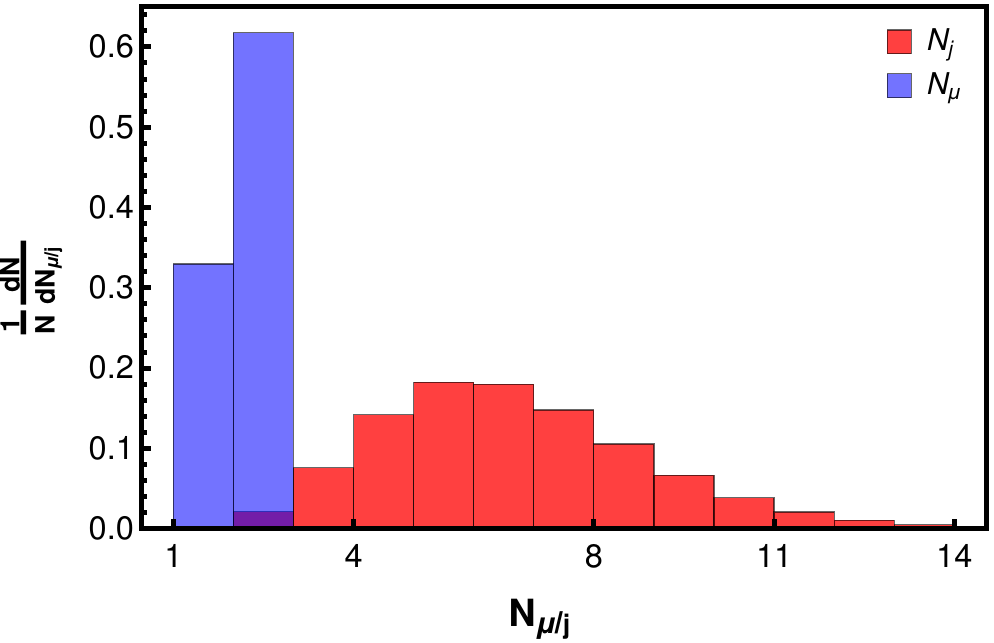} }
\hfill
\subfigure[BP3]{\includegraphics[width=0.35\linewidth, height=0.15\textheight]{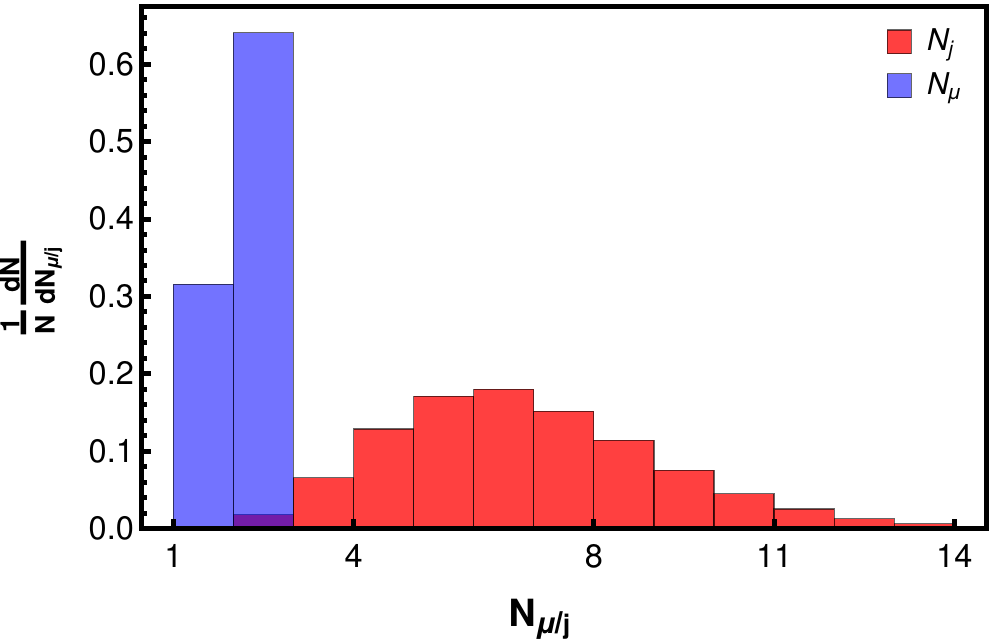} }  }
\caption{Multiplicity of the jets and muons for the pair production of the scalar Leptoquark $S_1$ at 100 TeV.} \label{Mult100}
\end{figure}

%%%%%%%%%%%%%%%%%%%%%%%%%%%%%%%%%%%%%%%%%%%%%%%%%%%%%%%%%%%%%%%%%%

Table~\ref{LQnum1} shows the number of signals and background events for a collision energy of 14 TeV, with an integrated luminosity ($\mathcal{L}_{int}$) of 1000 fb$^{-1}$. The cuts cumulatively imposed, optimise the signal events over the SM background. As discussed in Section~\ref{EvRat}, the vector Leptoquarks have larger cross-section for the pair-production, and thereby greater event rates over the scalar. The greater event rates for the vector Leptoquark $\widetilde{U}_{1\mu}$ over the scalar, $S_1$ are also reinforced by greater decay branching fraction to the second generation quark and muon ($\sim 33 \%$) compared to the scalar one ($\sim 23 \%$), as mentioned in Table~\ref{tab:BF}. With the increase in the Leptoquark mass, the event rates for the signal fall down, along with an exponential decrease in total background events, apparent from the Figures~\ref{Invjl1},~\ref{Invjl2} due to imposition of  high jet-muon invariant mass requirement.

%%%%%%%%% Table for the events the LHC with Ecm: 14 TeV %%%%%%%%%%
\begin{table}[htb!]
	\renewcommand{\arraystretch}{1}
	\centering
	\hspace*{-1.9cm}
	\scalebox{0.9}{
		\begin{tabular*}{1.35\textwidth}{|@{\hspace{3mm}\extracolsep{\fill}}cllllllll|}
			\hline
			%%%%%%%%%%%%%%%%%%%%%%%%
			\multirow{2}{*}{BPs} & \multirow{2}{*}{Cuts} & \multicolumn{2}{c}{Signal} & \multicolumn{5}{c|}{Background} \\  \cline{3-4} \cline{5-9} &&&&&&&&\\[-4mm]
			& & $S_1$ & $\widetilde U_{1\mu}$ & $t\bar{t}$ & $t\bar{t}\,V$ & $t\,VV$ & VV & VVV \\ \cline{1-9}
			&&&&&&&&\\ [-4.6mm] 
			\cline{1-9}
			 &&&&&&&&\\ [-3mm]
			%%%%%%%%%%%%%%%%%%%		
			\multirow{8}{*}{BP1} & $\geq 1\mu^+ + 1\mu^- + 2j$ & 163.91 & 2834.67 & 25526.21 & 243.19 & 26.48 & 543.26  & 116.47 \\ [1mm]
			
			& $ + ~ \begin{cases}
			\lvert M_{\ell\ell} - M_Z \rvert > 5 \text{ GeV} \\
			\lvert M_{jj} - M_Z \rvert > 10 \text{ GeV}
			\end{cases} $   & 119.87 & 2082.75 & 17089.94 & 154.27 & 10.46 & 83.19 & 26.91  \\ [2.5mm]
			
			& $ + ~ \begin{cases}
			\lvert M_{\mu^- j} - M_\Phi \rvert \leq 10 \text{ GeV} \\
			\lvert M_{\mu^+ j} - M_\Phi \rvert \leq 10 \text{ GeV}
			\end{cases} $   & 45.02 & 773.31 & 27.79 & 0.48 & 0.03 & 0.11 & 0.16  \\ [1mm]
			
			& Total & 45.02 & 773.31 & \multicolumn{5}{c|}{28.57} \\ \cline{2-2} \cline{3-4} \cline{5-9} 
			&&&&&&&&\\ [-4mm]
			& ${Sig}$ & 5.25 & 27.31 & & &\multirow{2}{*}{------} & & \\
			& $\mathcal{L}_{5\sigma}$ (in fb$^{-1}$) & 907.03 & 33.52 & & & & & \\ [1mm] \cline{1-9} 
			&&&&&&&&\\ [-4.6mm] 
			\cline{1-9} &&&&&&&&\\ [-3mm]
			
			%%%%%%%%%%%%%%%%%%%%
			
			\multirow{8}{*}{BP2} & $\geq 1\mu^+ + 1\mu^- + 2j$ & 6.99 & 107.55 & 25526.21 & 243.19 & 26.48 & 543.26  & 116.47 \\ [1mm]
			
			& $ + ~ \begin{cases}
			\lvert M_{\ell\ell} - M_Z \rvert > 5 \text{ GeV} \\
			\lvert M_{jj} - M_Z \rvert > 10 \text{ GeV}
			\end{cases} $   & 5.23 & 83.21 & 17089.94 & 154.27 & 10.46  & 83.19 & 26.91   \\ [2.5mm]
			
			& $ + ~ \begin{cases}
			\lvert M_{\mu^- j} - M_\Phi \rvert \leq 10 \text{ GeV} \\
			\lvert M_{\mu^+ j} - M_\Phi \rvert \leq 10 \text{ GeV}
			\end{cases} $   & 1.82 & 27.56 & 1.60 & 0.02 & 0.002 &  0.00 & 0.02  \\ [1mm]
			
			& Total & 1.82 & 27.56 & \multicolumn{5}{c|}{1.64} \\ \cline{2-2} \cline{3-4} \cline{5-9} &&&&&&&&\\ [-4mm]
			& ${Sig}$ & 0.98 & 5.10 & & &\multirow{2}{*}{------} & & \\
			& $\mathcal{L}_{5\sigma}$ (in fb$^{-1}$) & 26030.82 & 961.17 & & & & & \\ [1mm] \cline{1-9} &&&&&&&&\\ [-4.6mm] \cline{1-9} &&&&&&&&\\ [-3mm]
			
			%%%%%%%%%%%%%%%%%%%%
			
			\multirow{8}{*}{BP3} & $\geq 1\mu^+ + 1\mu^- + 2j$ & 0.44 & 6.10 & 25526.21 & 243.19 & 26.48 & 543.26 & 116.47 \\ [1mm]
			
			& $ + ~ \begin{cases}
			\lvert M_{\ell\ell} - M_Z \rvert > 5 \text{ GeV} \\
			\lvert M_{jj} - M_Z \rvert > 10 \text{ GeV}
			\end{cases} $   & 0.34 & 4.79 & 17089.94 & 154.27 & 10.46 & 83.19 & 26.91  \\ [2.5mm]
			
			& $ + ~ \begin{cases}
			\lvert M_{\mu^- j} - M_\Phi \rvert \leq 10 \text{ GeV} \\
			\lvert M_{\mu^+ j} - M_\Phi \rvert \leq 10 \text{ GeV}
			\end{cases} $   & 0.11 & 1.48 & 0.15 & 0.004 & 0.0003 & 0.00 & 0.001  \\ [1mm]
			
			& Total & 0.11 & 1.48 & \multicolumn{5}{c|}{0.16} \\ \cline{2-2} \cline{3-4} \cline{5-9} &&&&&&&&\\ [-4mm]
			& ${Sig}$ & 0.21 & 1.16 & & &\multirow{2}{*}{------} & & \\
			& $\mathcal{L}_{5\sigma}$ (in fb$^{-1}$) & 566893.42 & 18579.07 & & & & & \\ [1mm] \hline
			
			%%%%%%%%%%%%%%%%%%%%
	\end{tabular*}}
	\caption{Table displaying number of signal and background events after cumulative effect of cuts for different Benchmark Points at HL-LHC, for the centre of mass energy of 14 TeV at 1000 fb$^{-1}$ of integrated luminosity.}  \label{LQnum1}
\end{table}
%%%%%%%%%%%%%%%%%%%%%%%%%%%%%%%%%%%%%%%%%%%%%%%%%%%%%%%%%%%%%%%%%%%%%%%%%

As the data suggests, the signal significances for the vector singlet Leptoquark pair production is roughly five times to that for the scalar singlet, in compliance with the factors discussed above. For the probe of both scalar and vector Leptoquarks for different benchmark points at 14 TeV LHC,  a 5$\sigma$ discovery can be achieved for 1 TeV vector Leptoquark $\widetilde{U}_{1\mu}$ at a relatively early stage of high luminosity LHC (HL-LHC) run, for an integrated luminosity of 34 fb$^{-1}$, while 1 TeV scalar, $S_1$ requires $\sim 910$ fb$^{-1}$ to achieve 5$\sigma$ significance.

For 1.5 TeV vector Leptoquark probe at 14TeV, an integrated luminosity of 960 fb$^{-1}$ is required for the 5$\sigma$ significance while, an integrated luminosity of $\sim 26 \times 10^3$ fb$^{-1}$ is required for the 5$\sigma$ significance of $S_1$. 2.0 TeV $\widetilde{U}_{1\mu}$ requires $\sim 18 \times 10^3$ fb$^{-1}$ whereas, $S_1$ with identical mass requires $\sim 566 \times 10^3$ fb$^{-1}$ of dataset to achieve a 5$\sigma$ significance. Certainly their discovery, or ruling out necessitates greater size of the dataset and can therefore might be possible at the later phase of HL-LHC run.

%%%%%%%%% Table for the events the LHC with Ecm: 27 TeV %%%%%%%%%%

\begin{table}[h!]
	\renewcommand{\arraystretch}{1}
\centering
\hspace*{-1.9cm}
\scalebox{0.9}{
	\begin{tabular*}{1.35\textwidth}{|@{\hspace{3mm}\extracolsep{\fill}}cllllllll|}
		\hline
		%%%%%%%%%%%%%%%%%%%%%%%%
		\multirow{2}{*}{BPs} & \multirow{2}{*}{Cuts} & \multicolumn{2}{c}{Signal} & \multicolumn{5}{c|}{Background} \\  \cline{3-4} \cline{5-9} &&&&&&&&\\[-4mm]
		& & $S_1$ & $\widetilde U_{1\mu}$ & $t\bar{t}$ & $t\bar{t}\,V$ & $t\,VV$ & VV & VVV \\ \cline{1-9}
		&&&&&&&&\\ [-4.6mm] 
		\cline{1-9}
		&&&&&&&&\\ [-3mm]
		%%%%%%%%%%%%%%%%%%%		
			\multirow{8}{*}{BP1} & $\geq 1\mu^+ + 1\mu^- + 2j$ & 2225.96 & 49592.87 & 136095.83 & 952.73 & 108.02 & 1282.24 & 1042.79 \\ [1mm]
			& $ + ~ \begin{cases}
			\lvert M_{\ell\ell} - M_Z \rvert > 5 \text{ GeV} \\
			\lvert M_{jj} - M_Z \rvert > 10 \text{ GeV}
			\end{cases} $   & 1555.52 & 34282.62 & 87006.97 & 499.01 & 52.26 & 205.64 & 521.29  \\ [2.5mm]
			& $ + ~ \begin{cases}
			\lvert M_{\mu^- j} - M_\Phi \rvert \leq 10 \text{ GeV} \\
			\lvert M_{\mu^+ j} - M_\Phi \rvert \leq 10 \text{ GeV}
			\end{cases} $   & 576.11 & 12882.17 & 56.56 & 1.40 & 0.85 & 4.03 & 1.77  \\ [1mm]
			& Total & 576.11 & 12882.17 & \multicolumn{5}{c|}{64.61} \\ \cline{2-2}\cline{3-4} \cline{5-9} 
			&&&&&&&&\\ [-4mm]
			& ${Sig}$ & 22.76 & 113.22 & & &\multirow{2}{*}{------} & & \\
			& $\mathcal{L}_{5\sigma}$ (in fb$^{-1}$) & 48.26 & 1.95 & & & & & \\ [1mm] \cline{1-9} &&&&&&&&\\ [-4.5mm] \cline{1-9} &&&&&&&&\\ [-3mm]
			%%%%%%%%%%%%%%%%%%%%
			\multirow{8}{*}{BP2} & $\geq 1\mu^+ + 1\mu^- + 2j$ & 183.94 & 3432.36 & 136095.83 & 952.73 & 108.02 & 1282.24 & 1042.79 \\ [1mm]
			& $ + ~ \begin{cases}
			\lvert M_{\ell\ell} - M_Z \rvert > 5 \text{ GeV} \\
			\lvert M_{jj} - M_Z \rvert > 10 \text{ GeV}
			\end{cases} $   & 131.08 & 2451.09 & 87006.97 & 499.01 & 52.26 & 205.64 & 521.29  \\ [2.5mm]
			& $ + ~ \begin{cases}
			\lvert M_{\mu^- j} - M_\Phi \rvert \leq 10 \text{ GeV} \\
			\lvert M_{\mu^+ j} - M_\Phi \rvert \leq 10 \text{ GeV}
			\end{cases} $   & 45.62 & 882.63 & 4.01 & 0.05 & 0.04 & 0.99 & 0.46  \\ [1mm]
			& Total & 45.62 & 882.63 & \multicolumn{5}{c|}{5.56} \\ \cline{2-2}\cline{3-4} \cline{5-9} &&&&&&&&\\ [-4mm]
			& ${Sig}$ & 6.38 & 29.62 & & &\multirow{2}{*}{------} & & \\
			& $\mathcal{L}_{5\sigma}$ (in fb$^{-1}$) & 614.18 & 28.50 & & & & & \\ [1mm] \cline{1-9} &&&&&&&&\\ [-4.5mm] \cline{1-9} &&&&&&&&\\ [-3mm]
			%%%%%%%%%%%%%%%%%%%%
			\multirow{8}{*}{BP3} & $\geq 1\mu^+ + 1\mu^- + 2j$ & 25.73 & 414.49 & 136095.83 & 952.73 & 108.02 & 1282.24 & 1042.79 \\ [1mm]
			& $ + ~ \begin{cases}
			\lvert M_{\ell\ell} - M_Z \rvert > 5 \text{ GeV} \\
			\lvert M_{jj} - M_Z \rvert > 10 \text{ GeV}
			\end{cases} $   & 18.31 & 298.04 & 87006.97 & 499.01 & 52.26 & 205.64 & 521.29  \\ [2.5mm]
			& $ + ~ \begin{cases}
			\lvert M_{\mu^- j} - M_\Phi \rvert \leq 10 \text{ GeV} \\
			\lvert M_{\mu^+ j} - M_\Phi \rvert \leq 10 \text{ GeV}
			\end{cases} $   & 6.06 & 95.11 & 0.50 & 0.004 & 0.006 & 0.00 & 0.08  \\ [1mm]
			& Total & 6.06 & 95.11 & \multicolumn{5}{c|}{0.59} \\ \cline{2-2}\cline{3-4} \cline{5-9} &&&&&&&&\\ [-4mm]
			& ${Sig}$ & 2.35 & 9.72 & & &\multirow{2}{*}{------} & & \\
			& $\mathcal{L}_{5\sigma}$ (in fb$^{-1}$) & 4526.94 & 264.61 & & & & & \\ [1mm] \hline
			%%%%%%%%%%%%%%%%%%%%
	\end{tabular*}}
	\caption{Table displaying number of signal and background events after cumulative effect of cuts for different Benchmark Points at FCC, for the centre of mass energy of 27 TeV at 1000 fb$^{-1}$ of integrated luminosity.} \label{LQnum2}
\end{table}
%%%%%%%%%%%%%%%%%%%%%%%%%%%%%%%%%%%%%%%%%%%%%%%%%%%%%%%%%%%%%%%%%%%%%%%%%

The next table, Table~\ref{LQnum2} shows the number of signals and background events at 27 TeV collision with an integrated luminosity ($\mathcal{L}_{int}$) of 1000 fb$^{-1}$ for all three benchmark points. Comparing with Table~\ref{LQnum1}, we observe that an increase in collision energy increases the event rates manyfold depending on the leptoquark mass and spins, although the ratio of signal significance of the vector singlet to that of the scalar singlet leptoquarks roughly retains the same value. We observe that, a 5$\sigma$ discovery can be achieved for both scalar and vector leptoquarks of masses 1.0 and 1.5 TeV at fairly earlier stage of the run. An integrated luminosity of $\sim$ 1.95 fb$^{-1}$ is required for 5$\sigma$ discovery of 1.0 TeV $\widetilde{U}_{1\mu}$, while the same for $S_1$ requires $\sim$ 48.26 fb$^{-1}$. For the 1.5 TeV mass, $\widetilde{U}_{1\mu}$ requires $\sim$ 28.5 fb$^{-1}$ and $S_1$ requires $\sim$ 614.2 fb$^{-1}$ of integrated luminosities to be probed with 5$\sigma$ significance. The 2.0 TeV  $\widetilde{U}_{1\mu}$ would require a datset of size $\sim$ 264.61 fb$^{-1}$ and $S_1$ would require that of $\sim$ 4.5 ab$^{-1}$ for 5$\sigma$ discovery.

%%%%%%%%% Table for the events the LHC with Ecm: 100 TeV %%%%%%%%%%
\begin{table}[h!]
	\renewcommand{\arraystretch}{1}
	\centering
	\hspace*{-1.9cm}
	\scalebox{0.9}{
	\begin{tabular*}{1.35\textwidth}{|@{\hspace{3mm}\extracolsep{\fill}}cllllllll|}
		\hline
		%%%%%%%%%%%%%%%%%%%%%%%%
		\multirow{2}{*}{BPs} & \multirow{2}{*}{Cuts} & \multicolumn{2}{c}{Signal} & \multicolumn{5}{c|}{Background} \\  \cline{3-4} \cline{5-9}&&&&&&&&\\[-4mm]
		& & $S_1$ & $\widetilde U_{1\mu}$ & $t\bar{t}$ & $t\bar{t}\,V$ & $t\,VV$ & VV & VVV \\ \cline{1-9} &&&&&&&&\\ [-4.5mm] \cline{1-9} &&&&&&&&\\ [-3mm]
		%%%%%%%%%%%%%%%%%%%		
		\multirow{8}{*}{BP1} & $\geq 1\mu^+ + 1\mu^- + 2j$ & 55706.1 & 2500796.2 & 2500301.8 & 6463.6 & 1412.1 & 19969.4 & 2355.9 \\ [1mm]
		
		& $ + ~ \begin{cases}
		\lvert M_{\ell\ell} - M_Z \rvert > 5 \text{ GeV} \\
		\lvert M_{jj} - M_Z \rvert > 10 \text{ GeV}
		\end{cases} $   & 37971.4 & 1665106.2 & 1553908.9 & 3354.9 & 806.2 & 1190.3 & 630.8  \\ [2.5mm]
		
		& $ + ~ \begin{cases}
		\lvert M_{\mu^- j} - M_\Phi \rvert \leq 10 \text{ GeV} \\
		\lvert M_{\mu^+ j} - M_\Phi \rvert \leq 10 \text{ GeV}
		\end{cases} $   & 13976.6 & 663855.1 & 292.2 & 18.3 & 6.2 & 13.9 & 6.6  \\ [1mm]
		
	& Total & 13976.6 & 663855.1 & \multicolumn{5}{c|}{337.2} \\ \cline{2-2} \cline{3-4}\cline{5-9} &&&&&&&&\\ [-4mm]
		& ${Sig}$ & 116.8 & 814.6 & & &\multirow{2}{*}{------} & & \\
		& $\mathcal{L}_{5\sigma}$ (in fb$^{-1}$) & 1.83 & 0.04 & & & & & \\ [1mm] \cline{1-9} &&&&&&&&\\ [-4.5mm] \cline{1-9} &&&&&&&&\\ [-3mm]
		
		%%%%%%%%%%%%%%%%%%%%
		
		\multirow{8}{*}{BP2} & $\geq 1\mu^+ + 1\mu^- + 2j$ & 10138.3 & 348740.3 & 2500301.8 & 6463.6 & 1412.1 &  19969.4 & 2355.9 \\ [1mm]
		
		& $ + ~ \begin{cases}
		\lvert M_{\ell\ell} - M_Z \rvert > 5 \text{ GeV} \\
		\lvert M_{jj} - M_Z \rvert > 10 \text{ GeV}
		\end{cases} $   & 6923.8 & 237557.8 & 1553908.9 & 3354.9 & 806.2 & 1190.3 & 630.8  \\ [2.5mm]
		
		& $ + ~ \begin{cases}
		\lvert M_{\mu^- j} - M_\Phi \rvert \leq 10 \text{ GeV} \\
		\lvert M_{\mu^+ j} - M_\Phi \rvert \leq 10 \text{ GeV}
		\end{cases} $   & 2364.3 & 82965.4 & 45.2 & 2.2 & 1.5 & 9.3 & 1.9  \\ [1mm]
		
	& Total & 2364.3 & 82965.4 & \multicolumn{5}{c|}{60.1} \\ \cline{2-2} \cline{3-4} \cline{5-9} &&&&&&&&\\ [-4mm]
		& ${Sig}$ & 48.0 & 287.9 & & &\multirow{2}{*}{------} & & \\
		& $\mathcal{L}_{5\sigma}$ (in fb$^{-1}$) & 10.84 & 0.30 & & & & & \\ [1mm] \cline{1-9} &&&&&&&&\\ [-4.5mm] \cline{1-9} &&&&&&&&\\ [-3mm]
		
		%%%%%%%%%%%%%%%%%%%%
		
		\multirow{8}{*}{BP3} & $\geq 1\mu^+ + 1\mu^- + 2j$ & 2088.7 & 130571.7 & 2500301.8 & 6463.6 & 1412.1 & 19969.4 & 2355.9 \\ [1mm]
		
		& $ + ~ \begin{cases}
		\lvert M_{\ell\ell} - M_Z \rvert > 5 \text{ GeV} \\
		\lvert M_{jj} - M_Z \rvert > 10 \text{ GeV}
		\end{cases} $   & 1431.5  & 87408.9 & 1553908.9 & 3354.9 & 806.2 & 1190.3 & 630.8  \\ [2.5mm]
		
		& $ + ~ \begin{cases}
		\lvert M_{\mu^- j} - M_\Phi \rvert \leq 10 \text{ GeV} \\
		\lvert M_{\mu^+ j} - M_\Phi \rvert \leq 10 \text{ GeV}
		\end{cases} $   & 468.7 & 29548.9 & 8.2 & 0.5 & 0.3 & 2.8 & 0.3  \\ [1mm]
		
		& Total & 468.7 & 29548.9 & \multicolumn{5}{c|}{12.3} \\ \cline{2-2} \cline{3-4} \cline{5-9}&&&&&&&&\\[-4mm]
		& ${Sig}$ & 21.3 & 171.9 & & &\multirow{2}{*}{------} & & \\
		& $\mathcal{L}_{5\sigma}$ (in fb$^{-1}$) & 55.10 & 0.85 & & & & & \\ [1mm] \hline
		
		%%%%%%%%%%%%%%%%%%%%
	\end{tabular*}}
	\caption{Table displaying number of signal and background events after cumulative effect of cuts for different Benchmark Points at FCC, for the centre of mass energy of 100 TeV at 1000 fb$^{-1}$ of integrated luminosity.} \label{LQnum3} 
% The numbers are given for $1c$-jet and $2c$-jet tagged also and the corresponding required luminosities for $5\sigma$ are also mentioned.
\end{table}
%%%%%%%%%%%%%%%%%%%%%%%%%%%%%%%%%%%%%%%%%%%%%%%%%%%%%%%%%%%%%%%%%%%%%%%%%

Finally, we present, in Table~\ref{LQnum3}, the event rates for signals and background for a 100 TeV collision with an integrated luminosity ($\mathcal{L}_{int}$) of 1000 fb$^{-1}$ for all three BPs. The ratio of the signal significances of the vector $\widetilde{U}_{1\mu}$ to the scalar $S_1$ for different benchmark points roughly amounts to 7. It is evident that an earlier stage of FCC run will be able to discover or rule out the Leptoquarks of all three BPs, of both spins. An integrated luminosity of $\sim$38 pb$^{-1}$ would lead to a signal significance of 5$\sigma$ for 1.0 TeV $\widetilde{U}_{1\mu}$, $\sim$0.3 fb$^{-1}$ for 1.5 TeV $\widetilde{U}_{1\mu}$ and of $\sim$0.9 fb$^{-1}$ for 2.0 TeV $\widetilde{U}_{1\mu}$. Similarly, an integrated luminosity of $\sim$1.8 fb$^{-1}$ would lead to the same for 1.0 TeV $S_1$, $\sim$10.8 fb$^{-1}$ for 1.5 TeV $S_1$ and of $\sim$55 fb$^{-1}$ for 2.0 TeV $S_1$.

%%%%%%%%%%%%%%%%%%%%%%%%%%%%%%%%%%%%%%%%%%%%%%%%%%%%%%%%%%%%%%%%%%%%%%%%%%%%%%%%%%%%%%%%%%%%%%%
\subsubsection{Angular distribution}

Before going to the simulation of angular distribution at LHC, let us first discuss the theoretical description of the parton level contributions. We investigate the contributions from quarks and gluons in the normalized angular distribution of \LQ pair production at LHC for both scalar and vector \LQs with respect to the angle $(\theta)$ between produced \LQ and the beam axis in the centre of momentum (CM) frame. The QCD contributions to the hard scattering cross-sections and angular distributions for these processes are already discussed in Section \ref{sec:theory}. However, in this section we incorporate the lepton mediated $t-$channel diagrams also. 

\begin{figure*}[!htb]
	\ContinuedFloat*
	\includegraphics[width=0.48\textwidth]{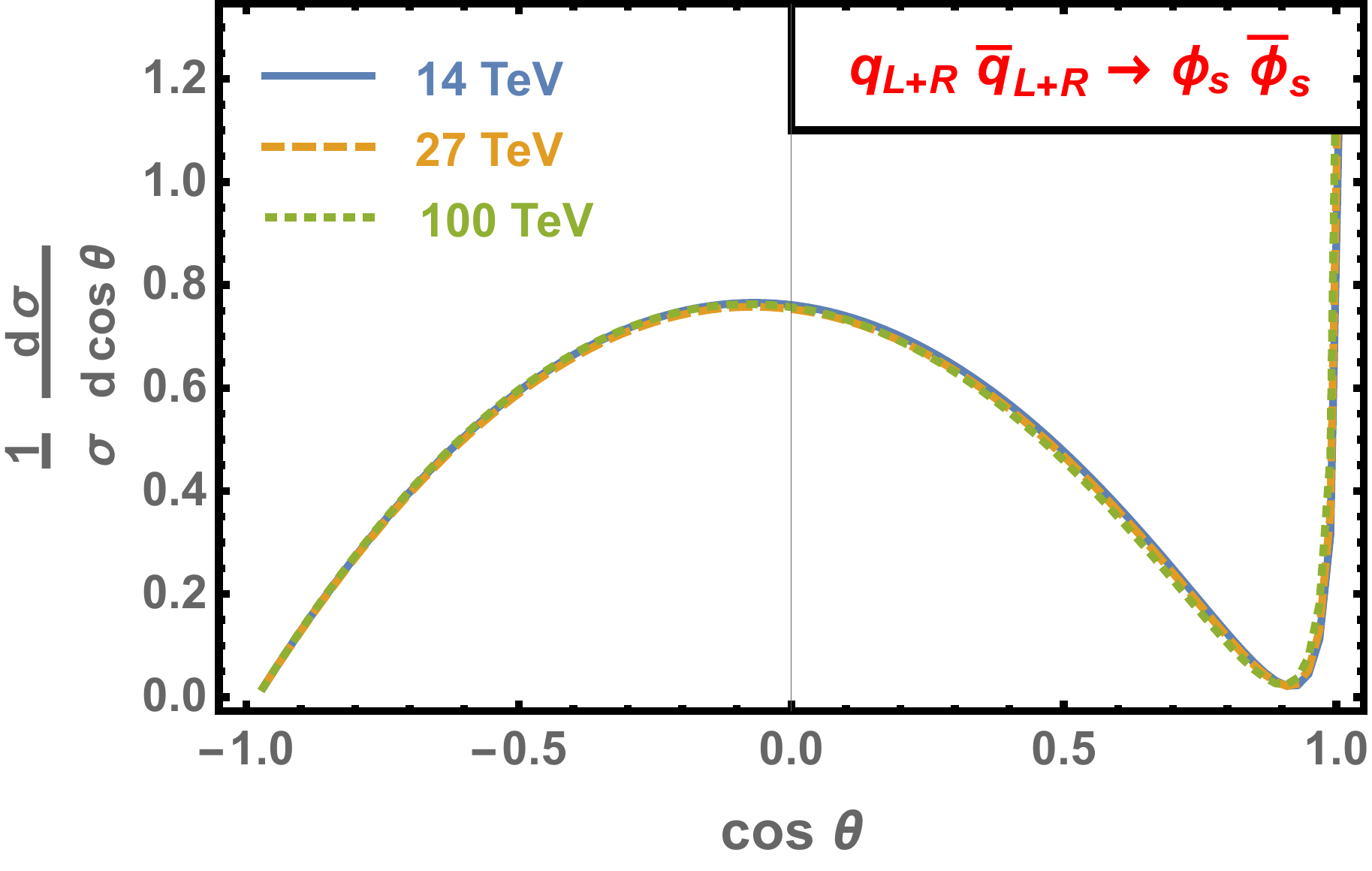}
	\hfil
	\includegraphics[width=0.48\textwidth]{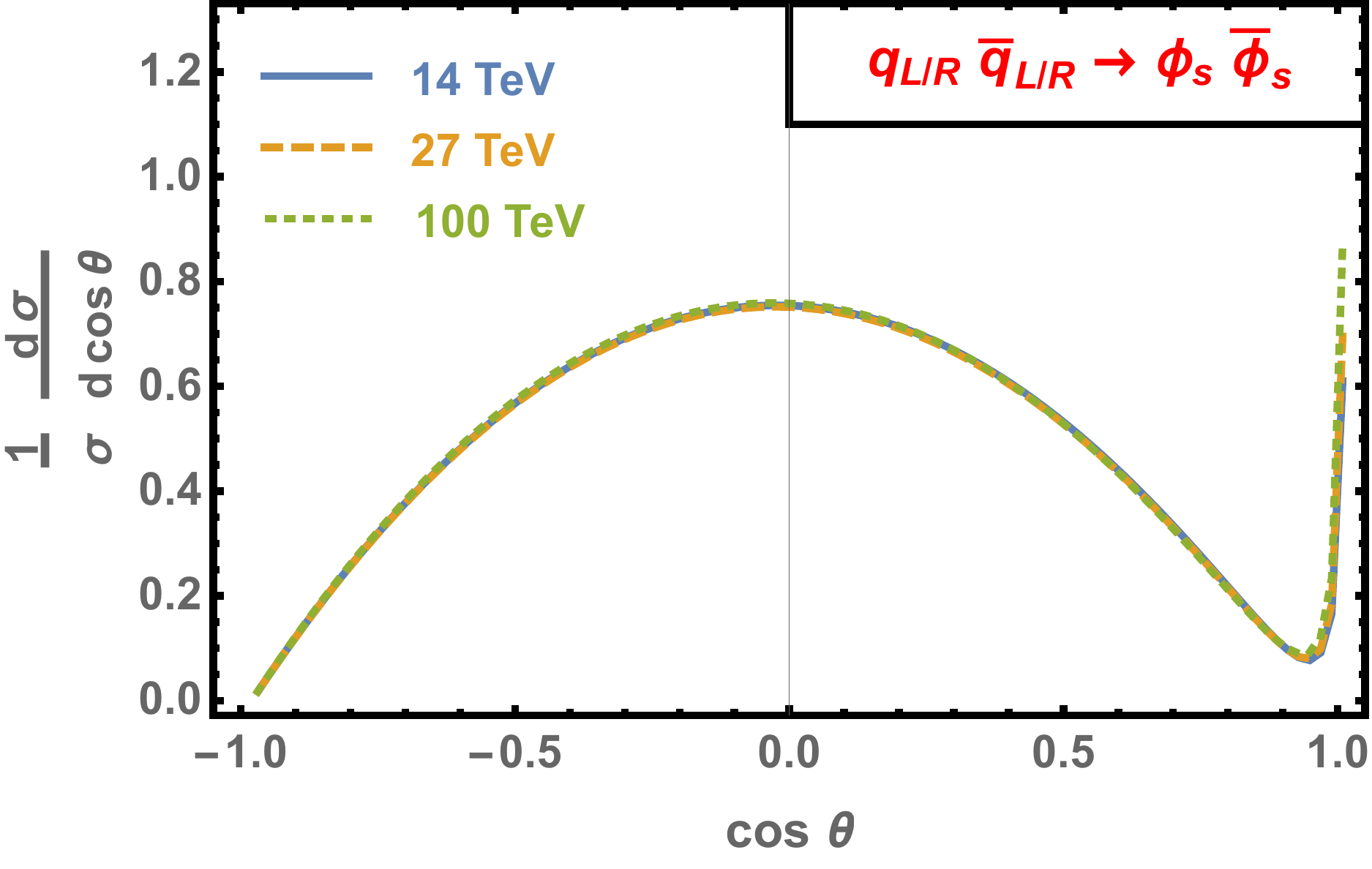}
	
	\includegraphics[width=0.48\textwidth]{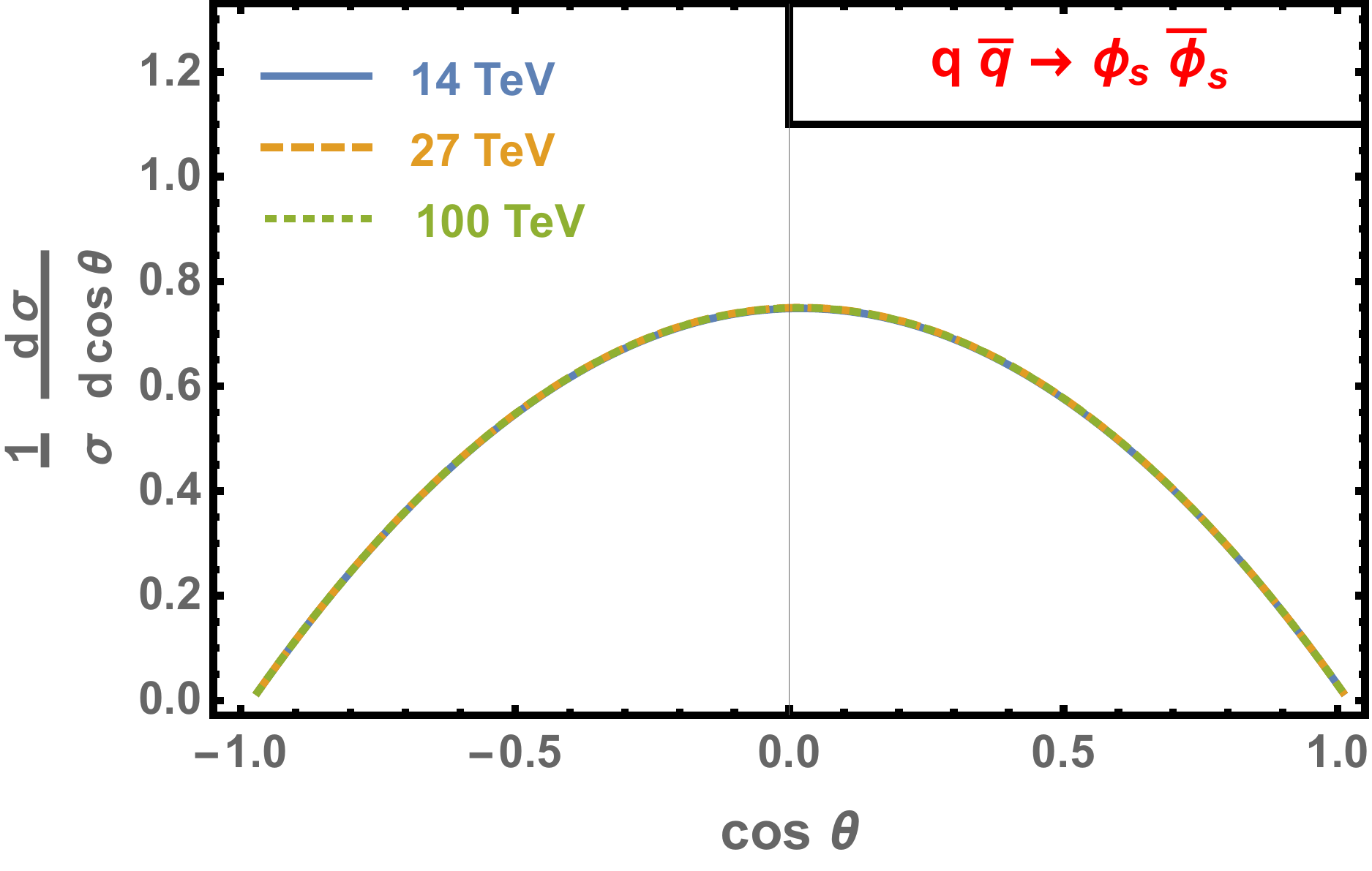}
	\hfil
	\includegraphics[width=0.48\textwidth]{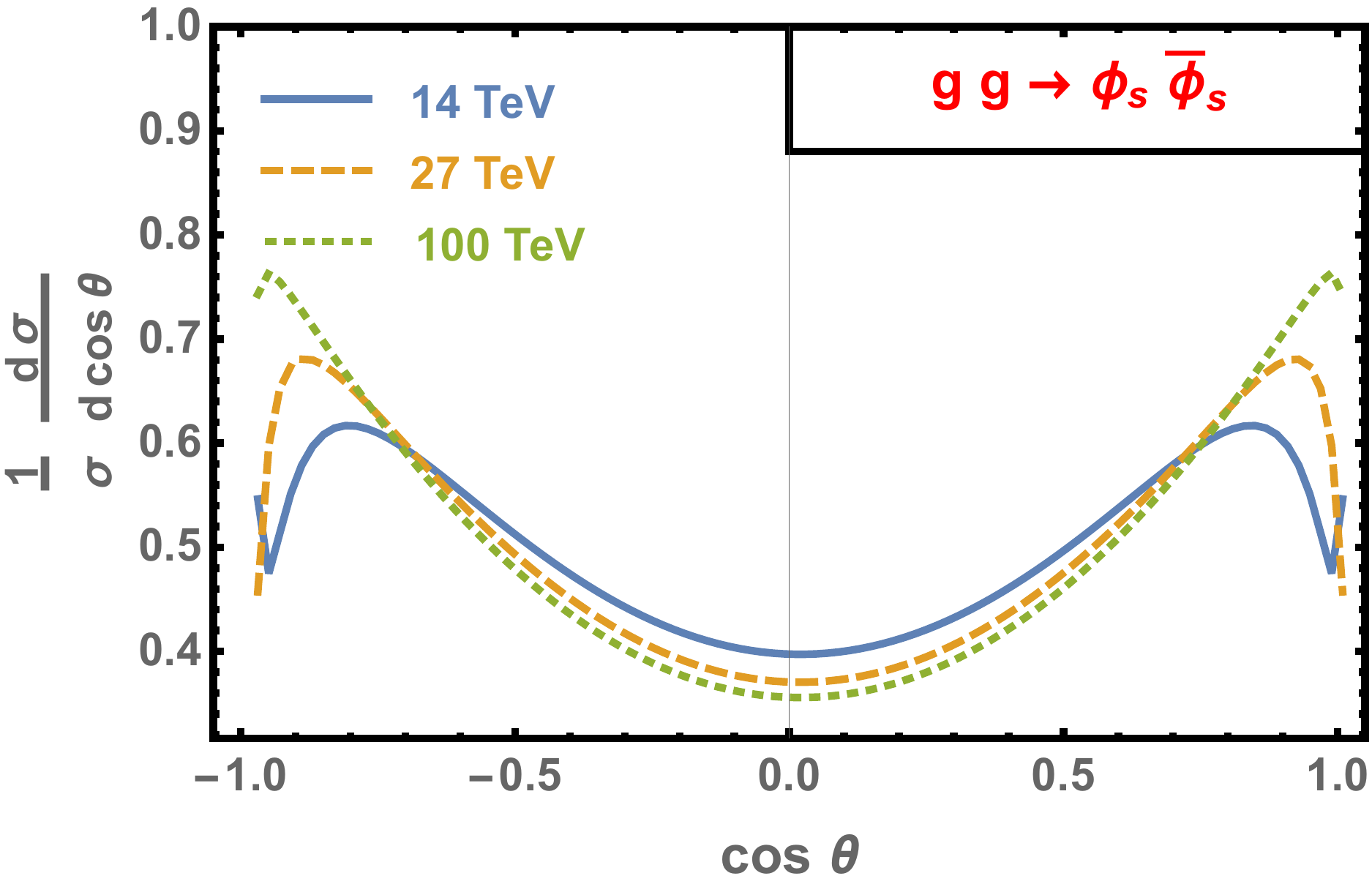}
	\caption{Parton level angular distribution normalized to their cross-sections for pair production of scalar ($\phi_s$) \LQs in the CM frame taking $M_\phi=1.5$ TeV and quark-lepton-Leptoquark coupling to be 0.2. The blue (solid), orange (dashed) and green (dotted) line indicate the distribution at $\sqrt s$ being 14 TeV, 27 TeV and 100 TeV respectively. The first, second and third plots show contributions of quarks under three different scenarios, respectively,: a) when both left and right handed quark couple to \LQ, b) when quark couples to \LQ through one chirality only, c) when \LQ does not couple to a particular quark at all. The fourth one exhibits the distribution for gluon fusion channel.}\label{fig:pairprod1}
\end{figure*}

\begin{figure*}[!htb]
	\ContinuedFloat	
	\vspace*{2mm}
	\includegraphics[width=0.48\textwidth]{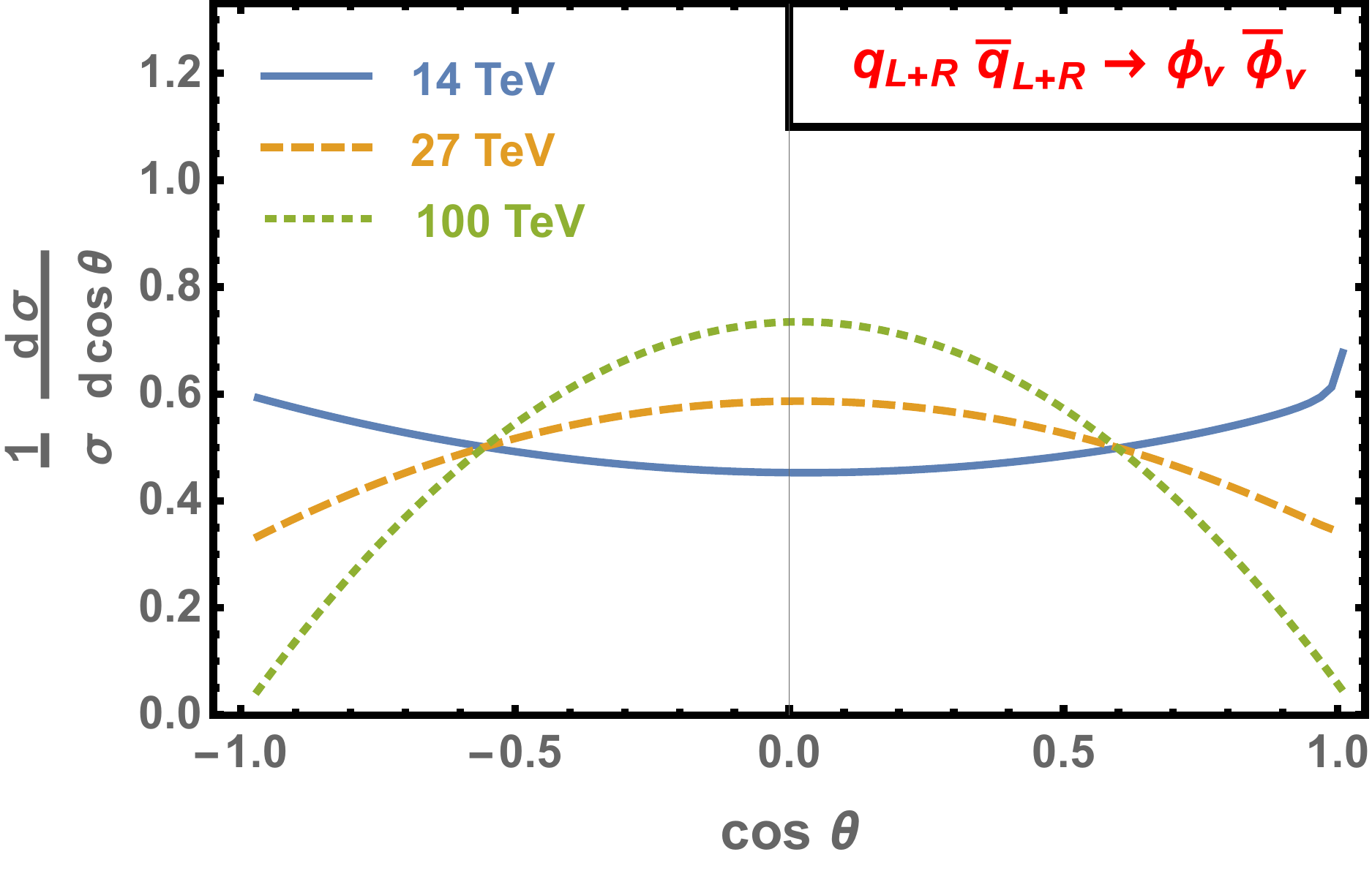}
	\hfil
	\includegraphics[width=0.48\textwidth]{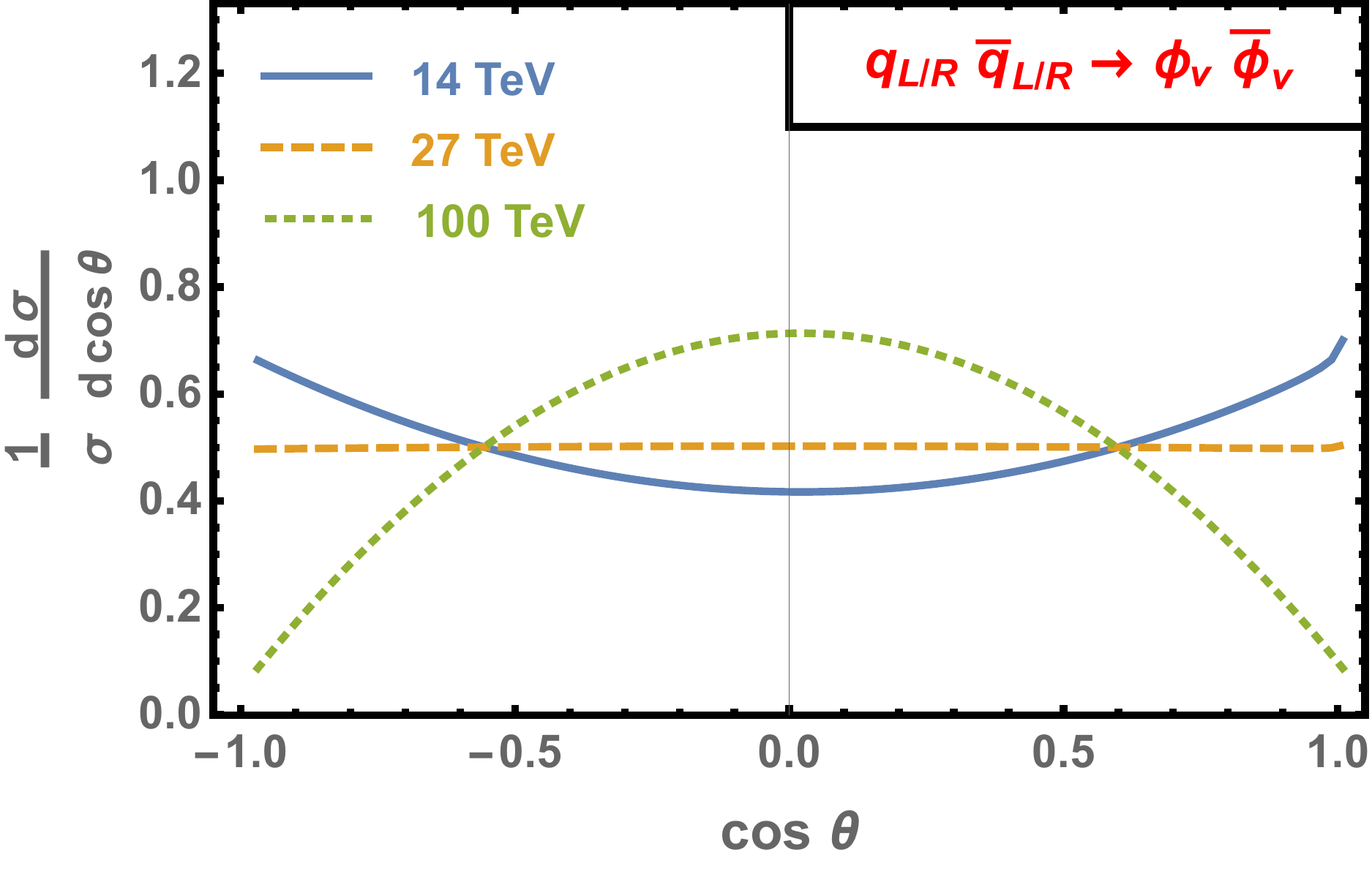}
	
	\includegraphics[width=0.48\textwidth]{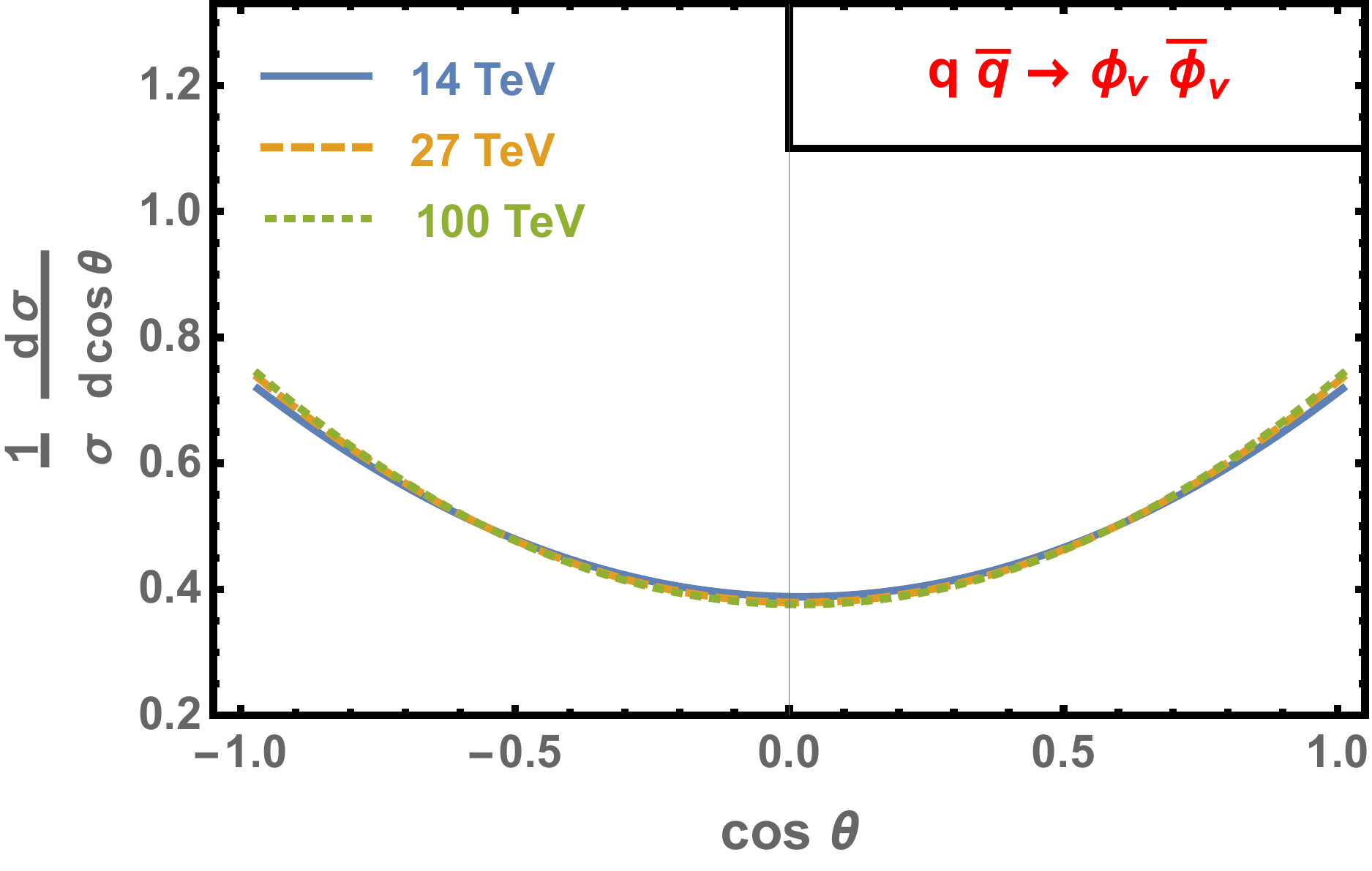}
	\hfil
	\includegraphics[width=0.48\textwidth]{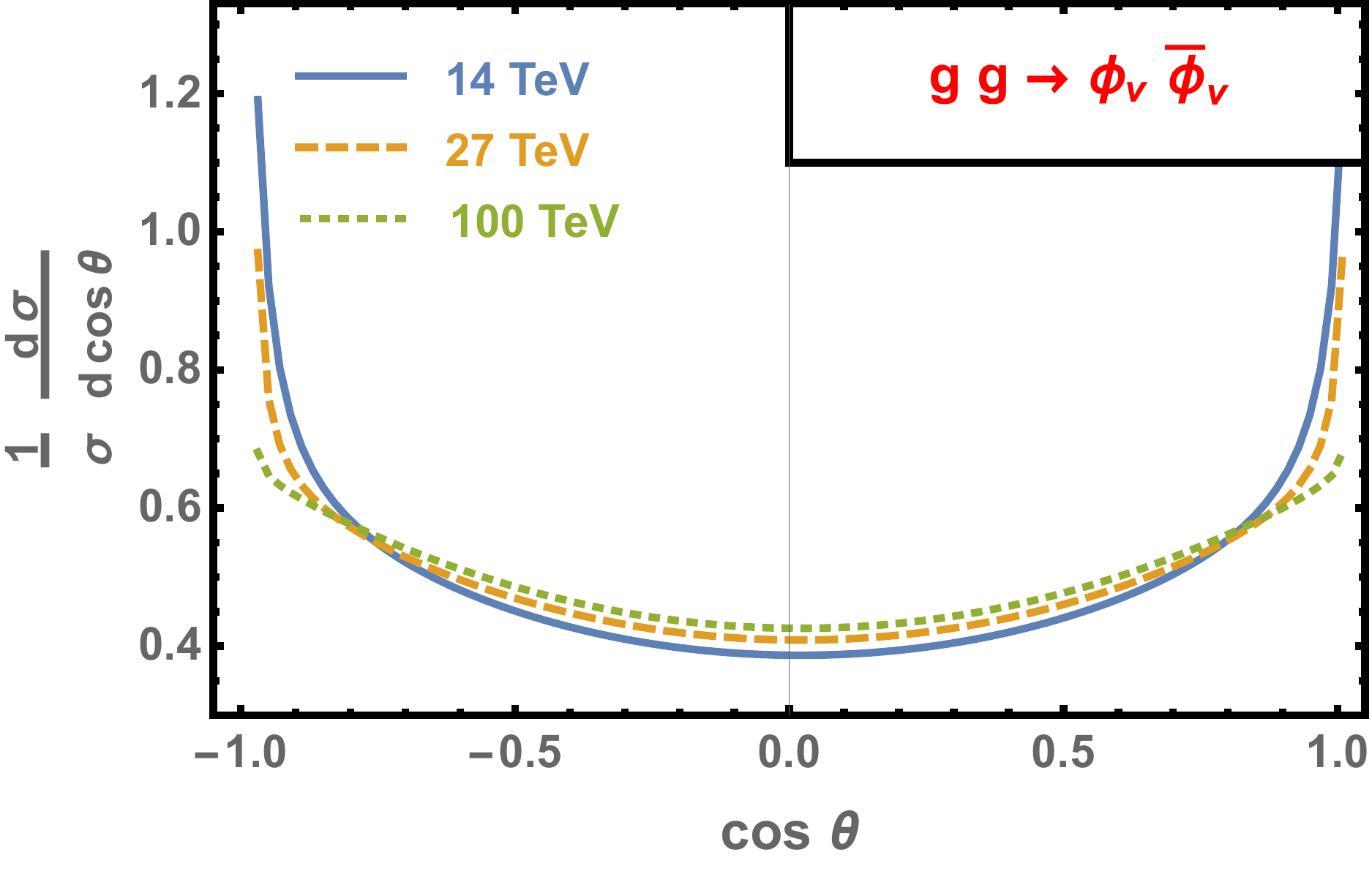}
	\caption{Parton level angular distribution normalized to their cross-sections for pair production of vector ($\phi_v$) \LQs in the CM frame taking $M_\phi=1.5$ TeV and quark-lepton-Leptoquark coupling to be 0.2. The blue (solid), orange (dashed) and green (dotted) line indicate the distribution at $\sqrt s$ being 14 TeV, 27 TeV and 100 TeV respectively. The first, second and third plots show contributions of quarks under three different scenarios, respectively,: a) when both left and right handed quark couple to \LQ, b) when quark couples to \LQ through one chirality only, c) when \LQ does not couple to a particular quark at all. The fourth one exhibits the distribution for gluon fusion channel.}
	\label{fig:pairprod2}
\end{figure*}

 The Feynman Diagrams for pair production of \LQ at LHC are presented in Figure~\ref{FDLQ}. In Figures~\ref{fig:pairprod1} and \ref{fig:pairprod2}, we summarize the parton level angular distributions normalized to the respective cross-sections for scalar and vector \LQs in the CM frame. While Figure \ref{fig:pairprod1} displays the contributions from quarks and gluons in the angular distributions for pair production of scalar \LQs at three different values of $\sqrt s$ for $M_\phi=$1.5 TeV and the quark-lepton-Leptoquark coupling being 0.2 for all the three generations of fermions, Figure \ref{fig:pairprod2} exhibits the same for vector \LQs with minimal coupling $(\kappa_G=1,\,\lambda_G=0)$ \cite{Blumlein:1996qp}. Though bounds from LEP \cite{Abreu:1998fw}, HERA \cite{Chekanov:2003af}, CERN \cite{Alitti:1991dn} and Tevatron \cite{Abe:1995fj,Abe:1996dn} indicate that there is still little room for low mass \LQ with appropriate branching to different generations of fermions \cite{Bandyopadhyay:2020jez,Bandyopadhyay:2020klr}, we take a conservative approach, and considering the constraints from ATLAS \cite{Aaboud:2019bye,Aaboud:2019jcc} and CMS \cite{Sirunyan:2018btu,Sirunyan:2018kzh,Sirunyan:2018nkj,Sirunyan:2018ryt,Sirunyan:2018vhk} only we choose the above mentioned benchmark point. In the first plots of Figure~\ref{fig:pairprod1} and \ref{fig:pairprod2}, we consider the pair production of \LQ from that quark (one generation) only whose left and right both chiral components couple to the Leptoquark, e.g. the contribution of $u$-quark in the pair production of $S_1$ or that of $d$-quark in the pair production of $U_{1\mu}$, and we denote it as $q_{L+R}$. In the second plots of Figure~\ref{fig:pairprod1} and \ref{fig:pairprod2}, we show the effects of the quark (one generation) that couples to the \LQ through one chirality only, for example the pair production of $\widetilde S_1$ from $d\,\bar d$ as $\widetilde S_1$ couples to $d_R^{\,c}$ only or that of $\widetilde U_{1\mu}$ from the fusion of $u$-quark, and we write it as $q_{L/R}$. Similarly, in the third plots of both the figures, we represent the contribution from the quark which does not couple to the \LQ at all and the pair production happens through gluon mediated $s$-channel diagram only, e.g. production of $\widetilde S_1$ from $u\bar u$ channel or production of $\widetilde U_{1\mu}$ from $d\bar d$ mode. One can easily notice the tiny effects of lepton-mediated $t$-channel diagrams around $\cos\theta\sim1$ in first and second columns while comparing them with the third one. It is worth mentioning that we have not considered contributions from photon and $Z^0$ mediated $s$-channel processes since their distributions are quite similar to the gluon-mediated one but with very small magnitude. Finally in the fourth plots we exhibit the angular distributions for \LQ pair production from gluon fusion. It can be seen that the distribution for scalar \LQ production in gluon fusion increases at both sides of $\cos\theta=0$. However, near  $\cos\theta\sim \pm 1$, it  attains maximum and starts decreasing, then it reaches minimum and starts increasing rapidly around the edge of phase space. Though it is difficult to observe the minimum for higher values of $\sqrt s$, since it is too close to $\cos\theta= \pm 1$, the maximum is clearly visible. This effect disappears for $\hat \beta<0.908$ and we get a monotonically increasing curve on both sides of $\cos\theta=0$.

The above effects, however, is bound to change at the real colliders due the effects of parton distribution function as well as the energies. The asymmetric behaviour of angular distribution in quark fusion channels will also be symmetrized in actual proton-proton collision since inside each proton there are quarks and anti-quarks distributed according to  parton distribution function. For our choices of  $\sqrt s$ and \LQ mass  in $pp$ collision at the LHC the tail effects near $\cos{\theta}\sim \pm 1$ are diminished. Now, we shall analyse the total cumulative effects of quarks and gluon fusion leading to \LQ pair production and the angular distribution of the \LQs in CM frame.

\begin{figure}[h!]
	\centering
	\mbox{
		\subfigure[14 TeV]{\includegraphics[width=0.31\textwidth]{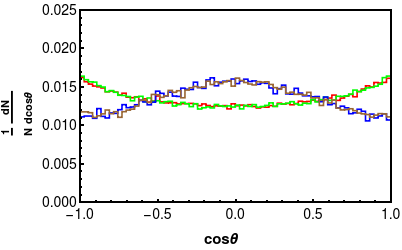}} 
		\hfil
		\subfigure[27 TeV]{\includegraphics[width=0.31\textwidth]{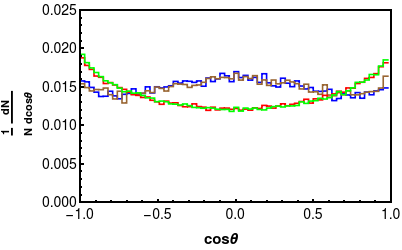}}
		\hfil
		\subfigure[100 TeV]{\includegraphics[width=0.37\textwidth]{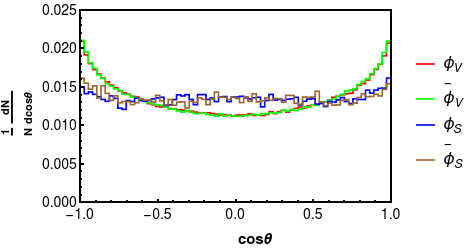}} }
	\caption{Normalised angular distribution of the 1.5 TeV scalar and vector Leptoquark pairs (BP2) for three different collision energies after correction for longitudinal boost effect. The plots are made with the events lying within 10 GeV window of Leptoquark invariant mass peak, \textit{i.e,} after imposing the final cut on signals and backgrounds in Table~\ref{LQnum1}, \ref{LQnum2} and \ref{LQnum3}.} \label{angdisLQLHC}
\end{figure}

Equipped with the CM frame we reconstruct the angular distribution with respect to angle  $\theta$, the angle between the incoming parton and \LQ (anti-\LQ). The \LQ(anti-\LQ) can be identified via the presence of $\mu^-(\mu^+)$ in the final states while reconstructing the \LQs masses. The angular distributions for the scalar $S_1$ and  vector $\widetilde{U}^{5/3}_{1\mu}$ \LQs are shown in Figure~\ref{angdisLQLHC}. The red(green) coloured are for the vector-like Leptoquark (anti-Leptoquark); whereas the blue(brown) are for the scalar \LQ (anti-\LQ) respectively. We note that the subprocess displayed in first two plots of Figure~\ref{fig:pairprod1} and \ref{fig:pairprod1} contribute to the pair production with a strength of fourth power of Yukawa coupling coefficient while, the subprocesses in the last two column is QCD mediated and contribute with a strength of fourth power of strong coupling coefficient. Hence, the effect in angular distribution pattern induced by Yukawa-mediated cross-channels are suppressed in comparison to QCD-mediated s-channel processes. The final angular distribution of the scatter Leptoquark pair is thus dominated by cumulative contribution of s-channel fusions. The gluon fusion contributes to a trough for scattering angles ($\theta$) close to $\nicefrac{\pi}{2}$ and a trough for $\theta \sim 0, \pi$. For quark pair fusion the effect is reverse. We could see that for scalar Leptoquark, though the final angular distribution in pp collision mimics the shape of quark pair annihilation process in the central region (low $|cos\theta|$), it gets more contribution from gluon fusion in the peripheral portion (high $|cos\theta|$) of phase space (see Figure~\ref{fig:pairprod1} and \ref{fig:pairprod2}). For reasons discussed above, the tail-effects in angular distribution of scalar pair are also diminished due to parton distribution function. However, for vector \LQ the angular distribution is mostly dominated by gluon fusion.  Nevertheless, it is clear from  Figure~\ref{angdisLQLHC} that the scalar and vector \LQs can be segregated via their angular distribution in the reconstructed CM frame.  Figure~\ref{angdisLQLHC} (a), (b), (c) describe the angular distributions at the LHC and FCC with centre of mass energies of 14, 27 and 100 TeV respectively. For the chosen \LQ mass of 1.5 TeV, we see such discerning of spins of \LQs are possible even when they generate the similar final state.

\subsubsection{Leptoquark Reaches at LHC/FCC }

As evident from Tables \ref{LQnum1}, \ref{LQnum2} and  \ref{LQnum3}, for similar masses, couplings and collision energies, the vector \LQs have larger significance than their scalar counterparts. As discussed earlier, three polarizations contributing to three degrees of freedom for vector \LQs increase its pair-production cross-section, thereby increasing the signal significance. Hence, comparatively a lower luminosity will be required to achieve $5\sigma$ significance for the vector Leptoquarks. Therefore, they will be discovered or ruled out at much earlier stage of the run compared to the scalar ones.

\begin{figure*}[h!]
	\includegraphics[scale=0.28]{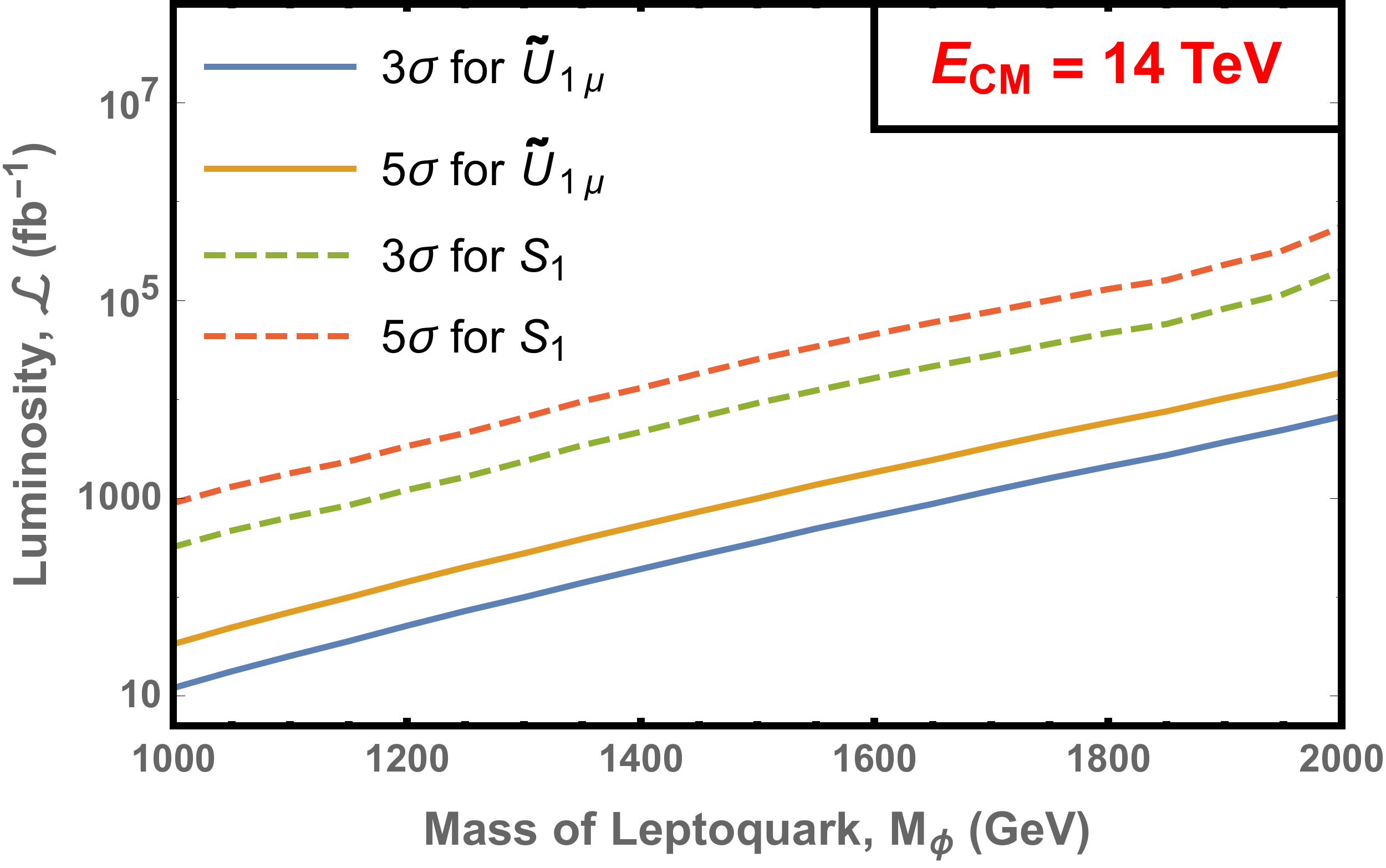}
	\hfil
	\includegraphics[scale=0.28]{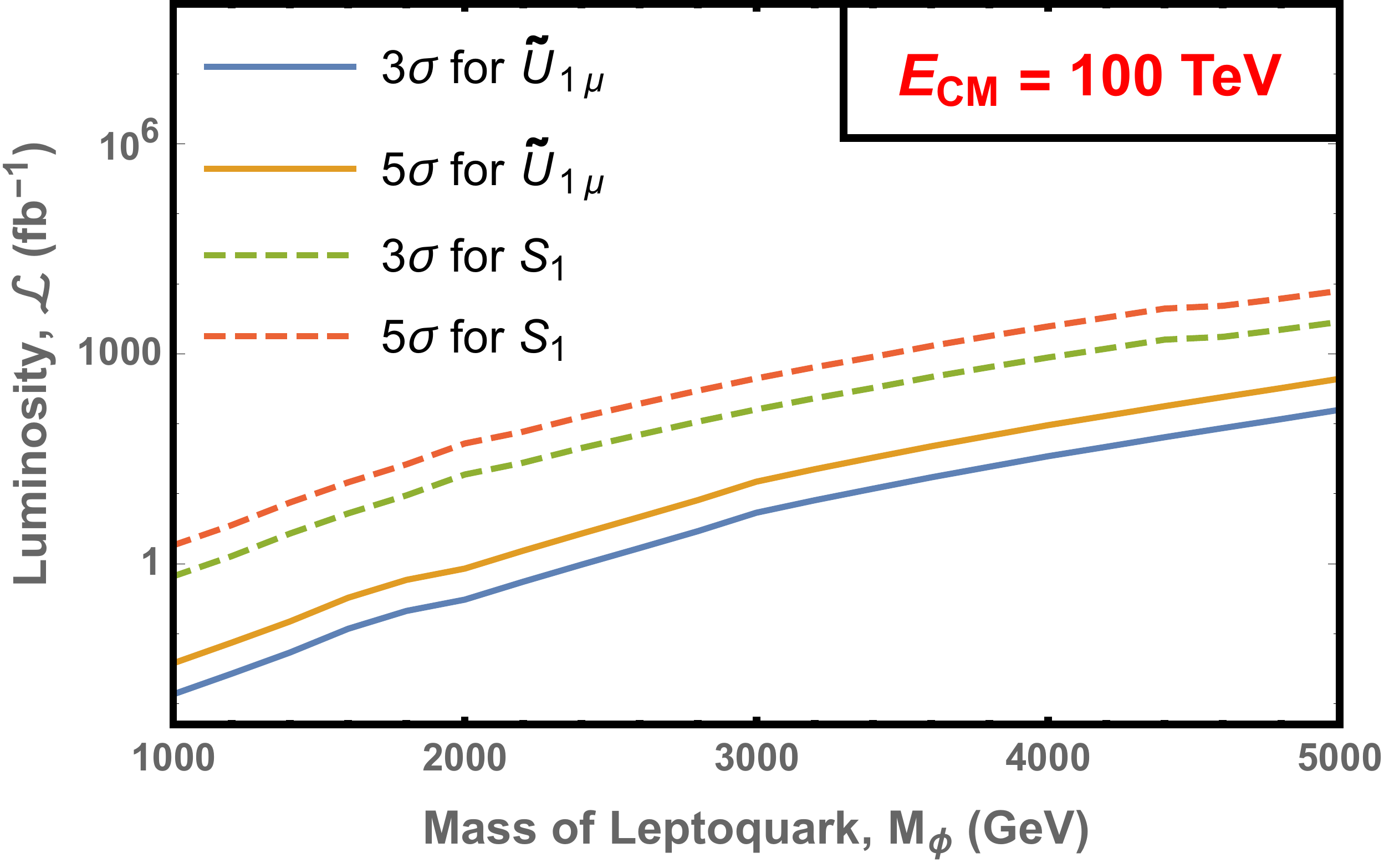}
	
	\caption{Required integrated luminosity for finding \LQs $\widetilde U_{1\mu}$ and $S_1$ with $3\sigma$ and $5\sigma$ significances as function of their mass at centre of momentum energy being 14 TeV (panel) and 100 TeV (right panel). The couplings are taken as described in Table \ref{tab:BP} for all masses of the Leptoquarks. The $3\sigma$ and $5\sigma$ contours for vector singlet \LQ $\widetilde U_{1\mu}$ are indicated by blue and yellow solid lines whereas the same for scalar singlet \LQ $S_1$ are shown by green and red dashed curves respectively.}
	\label{fig:reach}
\end{figure*}

In Figure \ref{fig:reach}, we present plots showing integrated luminosities (in fb$^{-1}$) required for achieving signal significances of 3$\sigma$ and 5$\sigma$ respectively at 14 TeV (at the left) and 100 TeV (at the right) collisions as a function of the Leptoquark mass. We have considered the spin-0 $S_1$ and spin-1 $\widetilde{U}_{1\mu}$ with their decays to the second generation quark and lepton with the branching fraction mentioned in Table \ref{tab:BF}. The blue and yellow solid lines indicate 3$\sigma$ and 5$\sigma$ contours for $\widetilde{U}_{1\mu}$ whereas the green and red dashed lines describe the same for $S_1$. As expected, the scalar \LQ $S_1$ needs much higher luminosity to be probed with an appreciable significance than the vector \LQ $\widetilde{U}_{1\mu}$. Required integrated luminosity to observe pair-production of \LQs $S_1$ and $\widetilde{U}_{1\mu}$ at LHC/FCC with $5\sigma$ significance for different centre of momentum energies and benchmark points are also tabulated in Table \ref{tab:lum}. As can be seen, in order to achieve $5\sigma$ significance for \LQ $S_1$ with BP1 at centre of momentum energies 14 TeV and 100 TeV one needs integrated luminosities of 907 fb$^{-1}$ and 1.83 fb$^{-1}$ respectively whereas for the same with \LQ $\widetilde{U}_{1\mu}$, one requires integrated luminosities of 33.5 fb$^{-1}$ and $0.04$ fb$^{-1}$ respectively. Similarly, to reach $5\sigma$ significance at the same centre of momentum energies for BP3 with \LQ $S_1$, luminosities of $5.67\times 10^5$ fb$^{-1}$ and 55.10 fb$^{-1}$ are needed while for the same with \LQ $\widetilde{U}_{1\mu}$, luminosities of $1.86 \times 10^{4}$ fb$^{-1}$ and $0.85$ fb$^{-1}$ are required respectively. It should also be noticed from Figure \ref{fig:reach} that with 1000 fb$^{-1}$ of integrated luminosity and 14 TeV of centre of momentum energy, one can probe vector \LQ $\widetilde{U}_{1\mu}$ (with minimal coupling) up to mass 1.5 TeV with $5\sigma$ significance while one cannot go much beyond 1 TeV of mass for scalar \LQ $S_1$. On the other hand, at 100 TeV of centre of momentum energy with 1000 fb$^{-1}$ of integrated luminosity, vector \LQ $\widetilde{U}_{1\mu}$ of 5 TeV mass can easily be probed at LHC  with same significance whereas scalar \LQ $S_1$ can be probed till mass 3.5 TeV only.

%\begin{table}[h!]
%	\renewcommand{\arraystretch}{1.2}
%	\centering
%		\begin{tabular}{|c|c|c|c|c|c|c|}
%			\hline
%			&\multicolumn{3}{c|}{\textbf{Leptoquark }$\bm {S_1}$}&\multicolumn{3}{c|}{\textbf{Leptoquark }$\bm{\widetilde{U}_{1\mu}}$}\\
%			\cline{2-7}
%			Benchmark&\multicolumn{3}{c|}{Required luminosity in fb$^{-1}$ for }&\multicolumn{3}{c|}{Required luminosity in fb$^{-1}$ for}\\
%			points&\multicolumn{3}{c|}{$5\sigma$ significance at different $\sqrt s$}&\multicolumn{3}{c|}{ $5\sigma$ significance at different $\sqrt s$}\\
%			\cline{2-7}
%			&14 TeV&27 TeV&100 TeV&14 TeV&27 TeV&100 TeV\\
%			\hline
%			BP1&907.03&48.26&1.83&33.52&1.95&0.04\\ 
%			BP2&26030.82&614.18&10.84&961.17&28.50&0.30\\ 
%			BP3&566893.42&4526.94&55.10&18579.07&264.61&0.85\\ 
%			\hline
%		\end{tabular}
%	\caption{Required integrated luminosity to observe pair-production of \LQs $S_1$ and $\widetilde{U}_{1\mu}$ at LHC/FCC with $5\sigma$ significance for different centre of momentum energies and benchmark points.}\label{tab:lum}
%\end{table}

\begin{table}[h!]
	\renewcommand{\arraystretch}{1.2}
	\centering
	\begin{tabular}{|c|c|c|c|c|c|c|}
		\hline
		&\multicolumn{3}{c|}{\textbf{Leptoquark }$\bm {S_1}$}&\multicolumn{3}{c|}{\textbf{Leptoquark }$\bm{\widetilde{U}_{1\mu}}$}\\
		\cline{2-7}
		Benchmark&\multicolumn{3}{c|}{Required luminosity in fb$^{-1}$ for }&\multicolumn{3}{c|}{Required luminosity in fb$^{-1}$ for}\\
		points&\multicolumn{3}{c|}{$5\sigma$ significance at different $\sqrt s$}&\multicolumn{3}{c|}{ $5\sigma$ significance at different $\sqrt s$}\\
		\cline{2-7}
		&14 TeV&27 TeV&100 TeV&14 TeV&27 TeV&100 TeV\\
		\hline
		BP1&$0.91\times 10^3$&48.26&1.83&33.52&1.95&0.04\\ 
		BP2&$2.60 \times 10^4$&$0.61\times 10^3$&10.84&$0.96 \times 10^3$&28.50&0.30\\ 
		BP3&$5.67\times 10^5$&$4.53 \times 10^3$&55.10&$1.86 \times 10^4$&$0.26\times 10^3$&0.85\\ 
		\hline
	\end{tabular}
	\caption{Required integrated luminosity to observe pair-production of \LQs $S_1$ and $\widetilde{U}_{1\mu}$ at LHC/FCC with $5\sigma$ significance for different centre of momentum energies and benchmark points.}\label{tab:lum}
\end{table}

\subsection{Differentiating \LQs with same spin}

Having discussed the segregation of different Leptoquarks based on spins, which affect uniquely, the distribution of the scattered states at rest frame of interaction, we now concentrate on distinguishing different Leptoquark with the same spin, but \textit{i.e,} with different electromagnetic charges and $SU(2)$ representations.

\subsubsection{Different $SU(2)_L$ or $U(1)$ representation}

Apart from the singlet Leptoquarks, there are also other \LQs in the doublet and triplet representations of $SU(2)_L$ for the cases of scalar ($R_2, \, \widetilde{R}_2, \vec S_3$) and vectors ($V_{2\mu}, \, \widetilde{V}_{2\mu}, \vec U_{3\mu}$) as shown in the Table~\ref{tab:LQ}. All of these \LQs have $SU(2)_L$ partners with same tree-level mass but have different final state topologies. As an illustration, let us consider the example of $\vec S_3 \, (\vec U_{3\mu})$, with components $S_3^{\nicefrac{+4}{3}}, \, S_3^{\nicefrac{+1}{3}}$ and $S_3^{\nicefrac{-2}{3}}$ ($U_{3\mu}^{\nicefrac{+5}{3}}, \, U_{3\mu}^{\nicefrac{+2}{3}}$ and $U_{3\mu}^{\nicefrac{-1}{3}}$). The last component of $SU(2)_L$ multiplet decays only to anti-up (down) type quarks and anti-neutrinos and thus has topology distinct from the first two members. The first component decays only to the charged anti-lepton and anti-down (up) type quark while the second one decays to charged anti-lepton anti-up (down) type quarks and antineutrino anti-down (up) type quarks simultaneously. The first two members of the weak isospin multiplet shows complementary signatures while the third one shows semi-invisible mode. Determination of electromagnetic charge of jet originated from \LQ decay has been shown instrumental in segregating such complimentary jet final states \cite{Bandyopadhyay:2020jez}. Observations of such complementary modes with decay to charged leptons and quarks can eventually distinguish different gauge representation within the same spin group (scalar or vector). Precisely, \LQs (or, anti-\LQs) with electromagnetic charge, $-1 < Q_\phi < 0$ have zeros in their angular distribution in electron-photon collider\cite{Bandyopadhyay:2020klr} while others, with $\lvert Q_\phi \rvert > 1$ can manifest similar phenomena at electron-hadron collider when produced in association with a photon in the final state\cite{Bandyopadhyay:2020jez}.

\subsubsection{Jet charge} \label{JC}

\begin{figure}[!htb]
	\centering
	\mbox {
		\subfigure[Produced with $\mu^+$]{\includegraphics[width=0.50\linewidth, height=0.20\textheight]{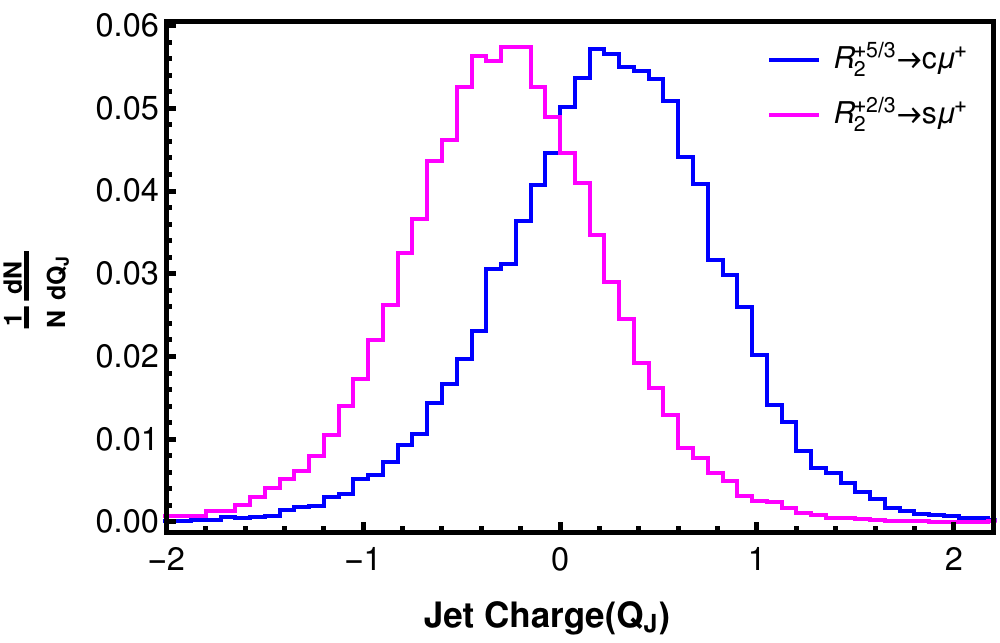} } 
		\hfill
		\subfigure[Produced with $\mu^-$]{\includegraphics[width=0.50\linewidth, height=0.20\textheight]{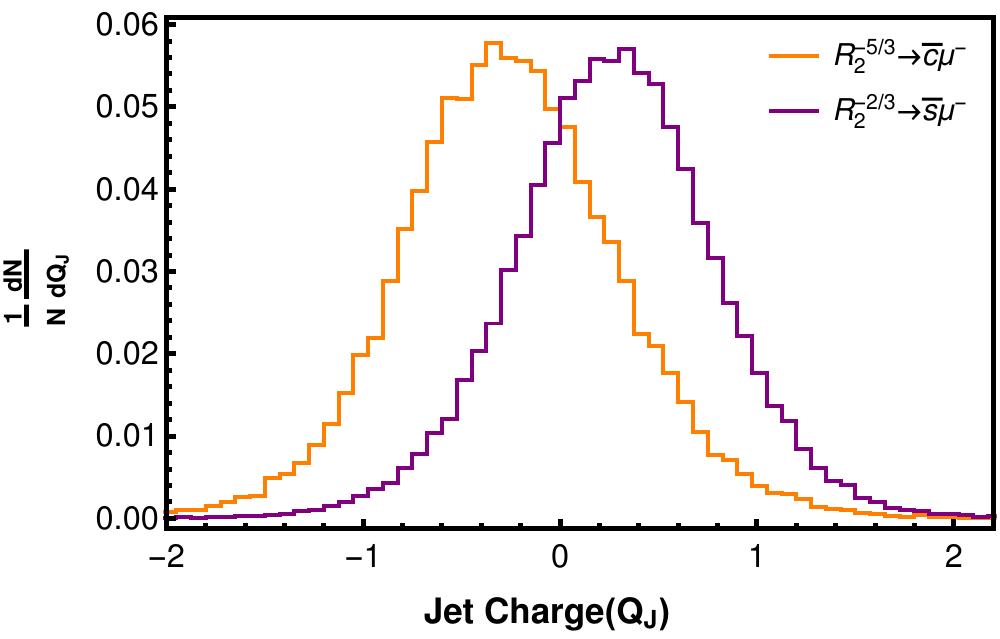} } }
	\caption{Charge of the jets from the decay of the scalar doublet Leptoquarks $R_2$ pair produced at 14 TeV.}
	\label{fig:jcharR2}
\end{figure}

\begin{figure}[!htb]
	\centering
	\mbox {
		\subfigure[Produced with $\mu^+$]{\includegraphics[width=0.50\linewidth, height=0.20\textheight]{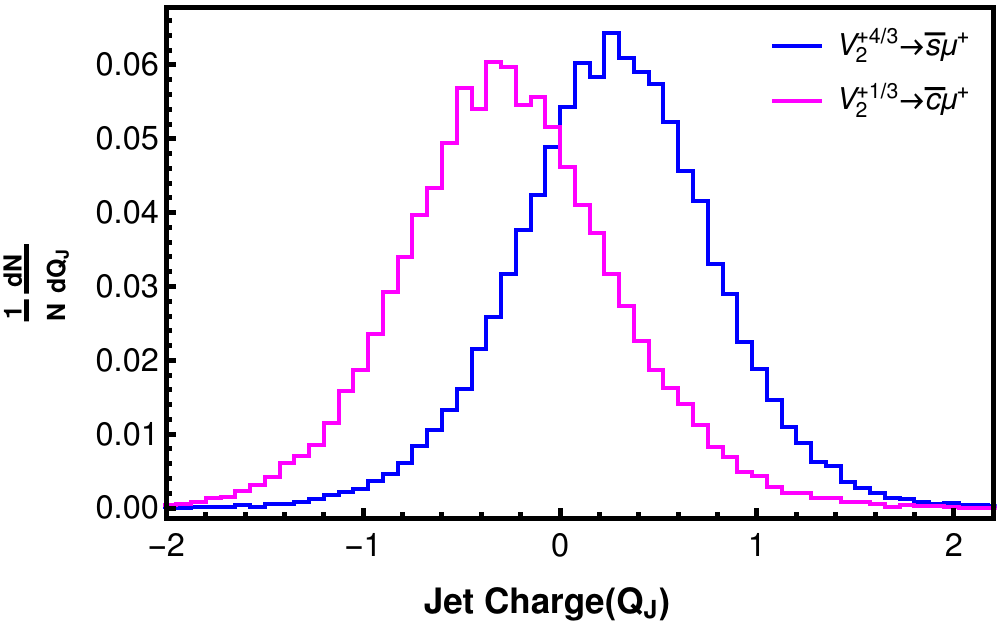} } 
		\hfill
		\subfigure[Produced with $\mu^-$]{\includegraphics[width=0.50\linewidth, height=0.20\textheight]{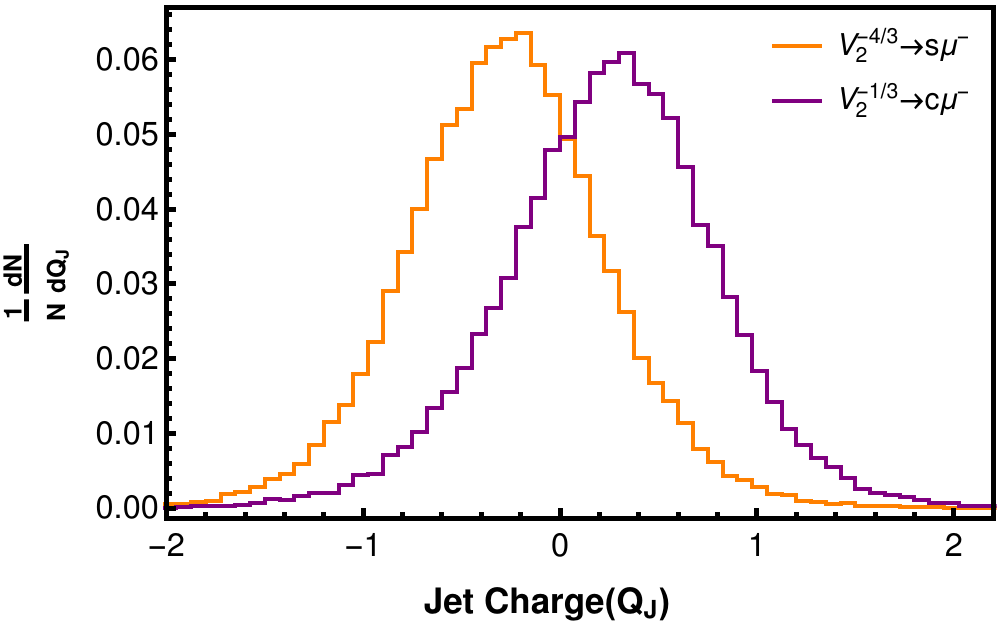} }
	}
	\caption{Charge of the jets from the decay of the scalar doublet Leptoquarks $V_{2\mu}$ pair produced at 14 TeV.}
	\label{fig:jcharV2}
\end{figure}

In case of doublet and triplet Leptoquarks, different components of same multiplet will be produced simultaneously at LHC due to degeneracy of their respective masses. Then it becomes important to distinguish the signatures of different excitations of same multiplet. Determination of the charge for the jets \cite{Krohn:2012fg,Tokar:2017syr,Sirunyan:2017tyr} produced from the decay of the \LQs turns out to be instrumental in this regard. For example, the scalar doublet \LQ $R_2$ consists of two components, $R_2^{+5/3}$ and $R_2^{+2/3}$ which would eventually decay to $c\mu^+$ and $s\mu^+$ respectively (considering decay to second generation of fermions only). Hence, if one tags the $\mu^+$ and determines the charge of the jet produced with it\footnote{The invariant mass of jet-$\mu^+$ pair would peak around the \LQ mass if they come from same Leptoquark.}, it will be seen that the charge of jet from $R_2^{+5/3}$ peaks around $+0.4$ whereas the same from $R_2^{+2/3}$ peaks at $-0.3$ which can be observed from the left panel of Figure \ref{fig:jcharR2}. Similarly, in case of $\bar R_2$ (anti-particle of $R_2$), the jet produced along with $\mu^-$ should be considered for charge determination. However, the results will be opposite to previous case, as shown in the right panel of Figure\ref{fig:jcharR2} , since $R_2^{-5/3}$ produces $\bar c$ while $R_2^{-2/3}$ creates $\bar s$. Same study for the vector doublet \LQ $V_{2\mu}$, produced at proton-proton collision at 14 TeV has been depicted in Figure \ref{fig:jcharV2}. Thus different members of same weak isospin multiplet can be isolated using the technique of jet charge determination.

\section{Conclusion}
\label{sec:conc}
In this paper, we have studied how to distinguish the signatures of different Leptoquarks, if they are produced in proton-proton collision at LHC/FCC. This involves the discrimination of the spin as well as the gauge representations. The information of the spin representations is encoded in their production cross-sections. For example the production cross-sections are a few times higher for vector \LQs compared the scalar ones for the the choices of the same mass at the LHC/FCC, specially for the hadronic collider where the productions are mostly by the strong interactions. Higher degree of freedom of vector \LQs thus will have early signals at the LHC and FCC. 

The spin information can also be probed directly by reconstructing the centre of mass frame as shown in this article. The muon and the jet coming from the \LQ decay can be identified via their invariant mass peak which also enables us to reconstruct the CM frame. It is shown that the angular distributions of the cosine of the angle of \LQs with the beam axis for scalar takes a convex shape while for the vector ones it follows a concave one. The departure from the matrix element calculation to the proton proton collision with the effects of parton distribution functions are also discussed. 

The situation gets even more interesting for higher gauge representations like $SU(2)$ doublets or triplets as they come with more partners within the same mass. However, as pointed out that different dominant decay modes will lead to different final state topologies which ease out the differentiation. In this context we also showed how the reconstruction of the jet charged from the hadronic constituents can pinpoint the decays of the \LQs involved.

Finally we also estimate the required luminosity to probe the scalar and vector \LQs in the TeV range. It is noticed that the LHC with 100 TeV centre of mass energy and with an integrated luminosity of 1000 fb$^{-1}$ can probe the scalar \LQ of mass $\sim 3.5$ TeV. 
For the same LHC specifications mass of $\gsim 5$ TeV can be probed for the vector Leptoquarks. We showed how different such disntiguishers can be instrumental in discerning \LQs with different spin and same gauge representations and vice-versa at the LHC/FCC.

\section*{Acknowledgements}
 PB and AK thank SERB CORE Grant CRG/2018/004971 and MATRICS Grant MTR/2020/000668  for the support. PB wants to thank  Prof. Torbj\"{o}rn Sj\"{o}strand for clarification about transverse boost for the initial state in PYTHIA8. 

%%%%%%%%%%%%%%%%%%%%%%%%%%%%%%%%%%%%%%%%%%%%%%%%%%%%%%%%%%%%%%%%%%%%%%%%%%%%%%%%%%%%%%%%%%%%%%%%%%%%%

\appendix
\section{Relevant functions for pair production of $\phi_v$ under minimal coupling}
\label{app}
\begin{align}
G_0&=1+\frac{1}{16}\Big[\frac{\hat s}{\Mp^2}-(1+3\bh^2)\Big]\,\sin^2\theta~,\\
G_1&=-1-\frac{1}{8}\Big(\frac{\hat s}{\Mp^2}-2\Big)\,\sin^2\theta~,\\
G_3&=\frac{1}{4}+\frac{1}{16}\Big(\frac{\hat s}{\Mp^2}-2\Big)\,\sin^2\theta~,\\
F_0&=(7+9\bh^2\cos\theta^2)\,\Big[19-6\bh^2+6\bh^4+(16-6\bh^2)\,\bh^2\cts+3\bh^4\cos^4\theta\Big]~,\\
F_1&=-4\,\Big(77+143\bh^2\cts+36\bh^4\cos^4\theta\Big)~,\\
F_3&=2\,\Big(117+185\bh^2\cts+18\bh^4\cos^4\theta\Big)\nonumber\\
&\hspace*{1.5cm}+\frac{2\,\hat s}{\Mp^2}\,\Big(8-\bh^2\cts-7\bh^4\cos^4\theta\Big)+\frac{7\,\hat s^2}{4\,\Mp^4}\,(1-\bh^2\cts)^2~,\\
F_6&=-\,61-67\bh^2\cts-\frac{7\,\hat s^2}{4\,\Mp^4}\,(1-\bh^2\cts)^2\nonumber\\
&\hspace{4.5cm}-\frac{\hat s}{2\,\Mp^2}(1-\bh^2\cts)(39+14\bh^2\cts)~,\\
F_{10}&=3+5\bh^2\cts+\frac{5\,\hat s}{4\,\Mp^2}(1-\bh^2\cts)(4-\bh^2\cts)\nonumber\\
&\hspace{4.2cm}+\frac{\hat s^2}{32\,\Mp^4}(1-\bh^2\cts)^2(25+13\bh^2\cts)~.
\end{align}

\end{document}